\documentclass[journal,11pt,onecolumn]{IEEEtran}

\usepackage[latin1]{inputenc}
\usepackage[numbers, square, comma, compress]{natbib}
\usepackage[english]{babel}
\usepackage{dsfont}
\usepackage{setspace}
\usepackage{amsfonts}
\usepackage{amssymb}
\usepackage{amsmath}
\usepackage{mathrsfs}
\usepackage{xcolor}
\usepackage{graphicx}
\usepackage{mathdots}
\usepackage{float}
\usepackage{dblfloatfix}
\usepackage{stmaryrd}
\usepackage{epstopdf}
\usepackage[colorinlistoftodos]{todonotes}
\usepackage[ruled,vlined]{algorithm2e}
\usepackage{verbatim}
\usepackage{tikz}
\usetikzlibrary{shapes,arrows}
\usetikzlibrary{calc}
\usepackage{hyperref}
\usepackage{color}
\usepackage{tikz}
\usetikzlibrary{shapes,arrows}
\usetikzlibrary{calc}
\usepackage{authblk}
\usepackage{paralist}

\newtheorem{teo}{Theorem}
\newtheorem{prop}{Proposition}
\newtheorem{lem}{Lemma}

\newtheorem{oss}{Remark}
\newtheorem{defi}{Definition}
\newtheorem{exe}{Example}

\DeclareMathOperator{\h}{\mathcal H}
\DeclareMathOperator{\V}{\mathcal V}
\DeclareMathOperator{\tv}{\mathbb V}
\DeclareMathOperator{\D}{\mathbb D}

\newcommand{\vv}[1]{``#1''}

\makeatletter
\renewcommand*\env@matrix[1][*\c@MaxMatrixCols c]{%
  \hskip -\arraycolsep
  \let\@ifnextchar\new@ifnextchar
  \array{#1}}
\makeatother

\renewcommand{\IEEEQED}{\IEEEQEDopen}

\title{Strong Coordination of Signals and Actions over Noisy Channels with two-sided State Information}
\author{Giulia Cervia, Laura Luzzi, Ma\"{e}l Le Treust and Matthieu R. Bloch \thanks{ 
The work of M.R. Bloch was supported in part by NSF under grant CIF 1320304. 
The work of M. Le Treust was supported by SRV ENSEA 2014 and INS2I CNRS through projects JCJC CoReDe 2015 and PEPS StrategicCoo 2016. 
The authors thank SRV ENSEA for financial support for the visit of M. R. Bloch  in 2017.\\
This work was presented in part  in \cite{Cervia2017} at the IEEE International Symposium on Information Theory (ISIT 2017), Aachen, Germany 
and in \cite{Cervia2016} at the IEEE Information Theory Workshop (ITW 2016), Cambridge, United Kingdom.\\
G. Cervia, L. Luzzi and  M. Le Treust are with ETIS UMR 8051, Universit\'{e} Paris Seine, Universit\'{e} Cergy-Pontoise, ENSEA, CNRS, Cergy-Pontoise, France (email: \{giulia.cervia, laura.luzzi, mael.le-treust\}@ensea.fr).
M.R. Bloch is with School of Electrical and Computer Engineering, Georgia Institute of Technology, Atlanta, Georgia (email: matthieu.bloch@ece.gatech.edu).}}
\begin{document}

{\let\newpage\relax\maketitle}

\begin{abstract}
  We consider a network of two nodes separated by a noisy channel with two-sided state information, 
in which the input and output signals have to be coordinated with the source and its reconstruction. 
In the case of non-causal encoding and decoding, we propose a joint source-channel coding scheme and develop inner and outer bounds for the strong coordination region. While the inner and outer bounds do not match in general, we provide a complete characterization of the strong coordination region in three particular cases:
\begin{inparaenum}[i)]
\item when the channel is perfect;
\item when the decoder is lossless; and 
\item when the random variables of the channel are independent from the random variables of the source.  
\end{inparaenum}
Through the study of these special cases, we prove that the separation principle does not hold for joint source-channel strong coordination. Finally, in the absence of state information, we show that polar codes achieve the best known inner bound for the strong coordination region, which therefore offers a constructive alternative to random binning and coding proofs.
\end{abstract}


\section{Introduction}
\label{sec:introduction}

While communication networks have traditionally been designed as ``bit pipes'' meant to reliably convey information, 
the anticipated explosion of device-to-device communications, e.g., as part of the Internet of Things, is creating new challenges. 
In fact, more than communication by itself, what is crucial for the next generation of networks is to ensure the \emph{cooperation} and \emph{coordination} of the 
constituent devices, viewed as autonomous decision makers. 
In the present work, coordination is meant in the broad sense of enforcing a joint behavior of the devices through communication. More specifically, we shall quantify this joint behavior in terms of how well we can approximate a target joint distribution between the actions and signals of the devices. Our main objective in the present work is to characterize the amount of communication that is required to achieve coordination for several networks.

A general information-theoretic framework to study coordination in networks was put forward in~\cite{cuff2010}, related to earlier work on 
``Shannon's reverse coding theorem''~\cite{bennet2002entanglement} and the compression of probability distribution sources and mixed quantum states~\cite{Soljanin2002,kramer2007communicating,winter2002compression}. 
This framework also relates to the game-theoretic 
perspective on coordination~\cite{gossner2006optimal} with applications, for instance, power control~\cite{larrousse2015coordination}. Recent extensions of the 
framework have included the possibility of coordination through interactive communication~\cite{yassaee2015channel, haddadpour2017simulation}. 
Two information-theoretic metrics have been proposed to measure the level of coordination:
\emph{empirical coordination}, which requires the joint histogram of the devices' actions to approach a target distribution, and \emph{strong coordination}, which requires the joint distribution of sequences of actions to converge to an i.i.d. target distribution, e.g., in variational distance~\cite{cuff2010,cuff2013distributed}. Empirical coordination captures an ``average behavior'' over multiple repeated actions of the devices; in contrast, strong coordination captures the behavior of sequences. A byproduct of strong coordination is that it enforces some level of ``security,'' in the sense of guaranteeing that sequence of actions will be \emph{unpredictable} to an outside observer beyond what is known about the target joint distribution of sequences. 

Strong coordination in networks was first studied over error free links~\cite{cuff2010} and later extended to noisy communication links~\cite{haddadpour2017simulation}. In the latter setting, the signals that are transmitted and received
over the physical channel become a part of what can be observed, and one can therefore coordinate the actions of the devices with their communication signals~\cite{cuff2011hybrid, treust2017joint}. From a security standpoint, this joint coordination of actions and signals allows one to control the information about the devices' actions that may be inferred from the observations of the communication signals. This ``secure coordination'' was investigated for error-free links in~\cite{satpathy2016secure}.

In the present paper, we address the problem of strong coordination in a two-node network comprised of an
information source and a noisy channel, in which both nodes have access to a common source of randomness. This scenario presents two conflicting goals: the encoder needs to convey a message to the decoder to coordinate the actions,
while simultaneously coordinating the signals coding the message. As in \cite{treust2014correlation,treust2015empirical,larrousse2015coordinating} 
we introduce a random state capturing the effect of the environment, to model actions and channels that change with external factors, and 
we consider a general setting in which state information and side information about the source may or may 
not be available at the decoder. We derive an inner and an outer bound for the strong coordination region by developing a joint source-channel scheme in which an auxiliary codebook allows us to satisfy both goals. 
Since the two bounds do not match, the optimality of our general achievability scheme remains an open question. 
We, however, succeeded to characterize the strong coordination region exactly in some special cases:
\begin{inparaenum}[i)]
\item when the channel is noiseless;
\item  when the decoder is lossless; and
\item when the random variables of the channel are independent from the random variables of the source.
\end{inparaenum}
In all these cases, the set of achievable target distributions is the same as for empirical coordination~\cite{treust2015empirical}, but we show that a positive rate of common randomness is required for strong coordination. We conclude the paper by considering the design of an explicit coordination scheme in this setting. Coding schemes for coordination based on polar codes have already been designed  in~\cite{blasco-serrano2012, bloch2012strong, chou2015coordination, obead2017joint}. Inspired by the binning technique using polar codes in \cite{chou2015polar}, 
we propose an explicit polar coding scheme that achieves the inner bound for the coordination capacity region in \cite{Cervia2017} by extending our coding scheme in \cite{Cervia2016} to strong coordination.  We use a chaining construction as in \cite{hassani2014universal,mondelli2015achieving} to ensure proper alignment of the polarized sets.

The remainder of the paper is organized as follows. $\mbox{Section \ref{sec: prel}}$ introduces the notation and some preliminary results.
$\mbox{Section \ref{sec: isit2017}}$  describes a simple model in which there is no state and no side information and derives an inner and an outer bound for the strong coordination region.
The information-theoretic modeling of coordination problems relevant to this work is best illustrated in 
this simplified scenario. 
$\mbox{Section \ref{sec: innerouter}}$ extends the inner and outer bounds to the general case of a noisy channel with state and side information at the decoder.
In particular, the inner bound is proved by proposing a random binning scheme and a random coding scheme that have the same statistics.
$\mbox{Section \ref{sec: special cases}}$ characterizes the strong coordination region for three special cases and 
shows that the separation principle does not hold for strong coordination. 
$\mbox{Section \ref{sec: polarcoding}}$ presents an explicit polar coding scheme for the simpler setting where there is no state and no side information.
Finally, $\mbox{Section \ref{sec: conclusions}}$ presents some conclusions and open problems.

\section{Preliminaries}\label{sec: prel}

We define the integer interval $\llbracket a,b \rrbracket$ as the set of integers between $a$ and $b$.
Given a random vector $X^{n}:=$ $(X_1, \ldots, X_{n})$, we note $X^{i}$ the first $i$ components of $X^{n}$, 
$X_{\sim i}$ the vector $(X_j)_{j \neq i}$, $j\in \llbracket 1,n \rrbracket $, where the component $X_i$ has been removed and $X[A]$ the vector $(X_j)_{j \in A}$, 
$A \subseteq \llbracket 1,n \rrbracket $.
The total variation between two probability mass
functions $P$ and $Q$ on $\mathcal A$ is  given by
$$\mathbb V (P, Q):= \frac{1}{2} \sum_{a \in \mathcal{A}} \lvert P(a)-Q(a) \rvert.$$
The Kullback-Leibler 
divergence between two discrete distributions $P$ and $Q$ is 
$$\mathbb D (P \Arrowvert Q):=\sum_{a} P(a) \log{\frac{P(a)}{Q(a)}}.$$
We use the notation $f(\varepsilon)$ to denote a function which tends to zero as $\varepsilon$ does, 
and the notation $\delta(n)$ to denote a function which tends to zero exponentially as $n$ goes to infinity.

We now state some useful results. 
First, we recall well-known properties of the variational distance and Kullback-Leibler divergence. 

\begin{lem}[$\mbox{\cite[Lemma 1]{csiszar1996almost}}$]\label{lem1csi}
Given a pair of random variables $(A,B)$ with joint distribution $P_{AB}$, marginals  $P_A$ and $P_B$ and 
$\lvert \mathcal A \rvert \geq 4$, we have
 \begin{equation*}
 \frac{1}{2 \log2} { \mathbb V(P_{AB},P_{A}P_{B}) }^2 \leq I(A;B) \leq \mathbb V(P_{AB},P_{A}P_{B}) \log {\frac{\lvert \mathcal A\rvert }{\mathbb V(P_{AB},P_{A}P_{B})}}.
 \end{equation*}
\end{lem}

\vspace{0,2cm}
\begin{lem}[$\mbox{\cite[Lemma 16]{cuff2009thesis}}$]\label{cuff16}
For any two joint distributions $P_{AB}$ and
$\widehat P_{AB}$, the total variation distance between them can
only be reduced when attention is restricted to $P_{A}$ and $\widehat P_{A}$.
That is,
$$\tv (P_{A}, \widehat P_{A}) \leq \tv (P_{AB}, \widehat P_{AB}).$$
\end{lem}

\vspace{0,2cm}
\begin{lem}[$\mbox{\cite[Lemma 17]{cuff2009thesis}}$]\label{cuff17}
When two random variables are passed through the same channel, the total variation between the resulting input-output joint distributions is the same as the total variation between the input distributions. Specifically,
$$\tv (P_A, \widehat P_A)= \tv (P_AP_{B|A}, \widehat P_A P_{B|A}).$$
\end{lem}

\vspace{0,2cm}
\begin{lem}\label{lemkl}
When two random variables are passed
through the same channel, the Kullback-Leibler divergence between the resulting input-output joint distributions is the same as the Kullback-Leibler divergence between the input distributions. Specifically,
$$\D \left(P_A \Arrowvert \widehat P_A\right)= \D \left(P_AP_{B|A} \Arrowvert \widehat P_AP_{B|A}\right).$$
\end{lem}
\vspace{0,2cm}

\begin{lem}[$\mbox{\cite[Lemma 4]{yassaee2014achievability}}$]\label{lem4}
If $ \tv \left(P_{Y^{n}} P_{X^{n}|Y^{n}},P'_{Y^{n}} P'_{X^{n} |Y^{n}} \right) = \varepsilon$  then there exists $\mathbf y \in \mathcal Y^{n}$ such that
 \begin{equation*}
 \tv \left(P_{X^{n}|Y^{n}= \mathbf y},P'_{X^{n} |Y^{n}= \mathbf y} \right) = 2 \varepsilon.
 \end{equation*}
\end{lem}

The proofs  of the following results are in Appendix \ref{appendix prel}. 
The following lemma is in the same spirit as  \cite[Lemma VI.3]{cuff2013distributed}. 
We state a slightly different version which is more convenient for our proofs. 
\vspace{0,2cm}

\begin{lem}\label{lemmit}
Let $P_{A^{n}}$ such that 
$\tv\left(P_{A^{n}}, \bar P_{A}^{\otimes n}\right) \leq \varepsilon,$
then we have
\begin{equation*}
\sum_{t=1}^{n} I(A_t;A_{\sim t}) \leq n f(\varepsilon). 
\end{equation*}
In particular, if $P_{AB}$ is such that $\tv\left(P_{AB}, \bar P_{A}\bar P_{B}\right) \leq \varepsilon$, 
then $ I(A;B) \leq  f(\varepsilon) $.
\end{lem}

\vspace{0,2cm}

\begin{lem}\label{lemab}
 Let $P_{A^{n} B^{n}}$ such that 
$\tv\left(P_{A^{n} B^{n}}, \bar P_{AB}^{\otimes n}\right) \leq \varepsilon$.
Then we have
\begin{equation}\label{lemab1}
 \sum_{t=1}^{n} I(A_t;A^{t-1} B_{\sim t}|B_t) \leq n f(\varepsilon).
\end{equation}
Let the variable $T$ serve as a random time index, for any random variable $C$ we have
\begin{equation}\label{lemab2}
  H(C|B^{n}) \geq  n I(A_T;CB_{\sim T} T|B_T ) -n I(A_T  B_T;T)  -n f(\varepsilon).
\end{equation}
\end{lem}

\section{Inner and outer bounds for the strong coordination region} \label{sec: isit2017}

\subsection{System model} \label{sec: sys}

\begin{center}
\begin{figure}[ht]
 \centering
 \includegraphics[scale=0.21]{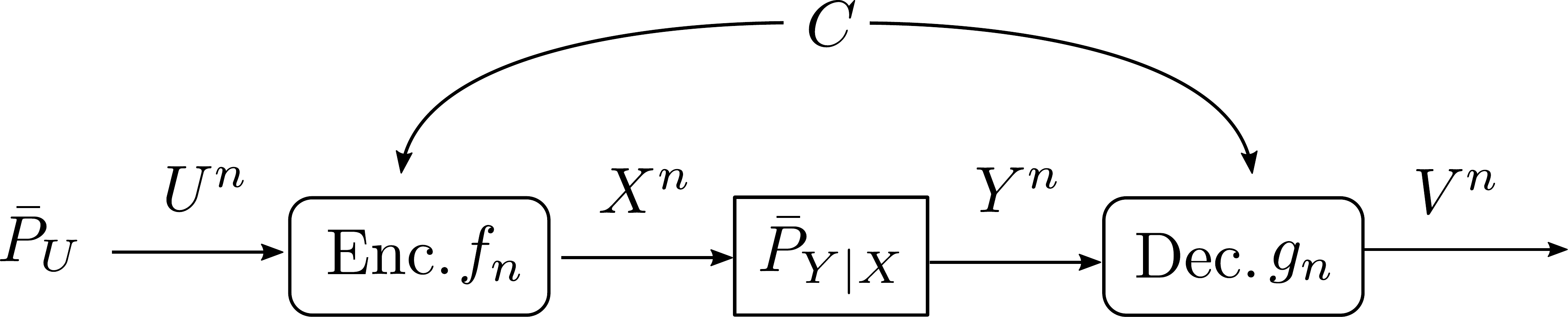}
\caption{Coordination of signals and actions for a two-node network with a noisy channel with non-causal encoder and decoder.}
\label{fig: coordisit}
\end{figure}
\end{center}

\vspace{-0,7cm}
Before we study the general model with a state in detail, it is  helpful 
to  consider a simpler model depicted in Figure \ref{fig: coordisit} to understand the nature of the problem. 
Two agents, the encoder and the decoder, wish to coordinate their behaviors: 
the stochastic actions of the agents should follow a known and fixed joint distribution.

We suppose that the encoder and the decoder have access to a shared source of uniform randomness $C \in \llbracket 1,2^{nR_0} \rrbracket$.
Let $U^{n} \in \mathcal U^n $ be an i.i.d. source with distribution $\bar P_{U}$.
The encoder observes the sequence  $U^{n} \in \mathcal U^n$  and selects a signal 
$X^{n}= f_n(U^{n},  C)$,  $f_n: \mathcal U^n  \times  \llbracket 1,2^{nR_0} \rrbracket \rightarrow \mathcal X^n$.
The signal $X^{n}$ is transmitted over a discrete memoryless channel parametrized by the conditional distribution $\bar P_{Y |X}$.
Upon  observing  $Y^{n}$ and common randomness $C$, 
the decoder selects an action $V^{n} = g_n(Y^{n}, C)$, where
$g_n: \mathcal Y^n  \times \llbracket 1,2^{nR_0} \rrbracket \rightarrow \mathcal V^n$ is a stochastic  map. 
For block length $n$, the pair $(f_n , g_n )$  constitutes a code.

We recall the definitions of achievability and of the coordination region for empirical and strong coordination \cite{cuff2010,cuff2009thesis}.

\begin{defi}
A distribution $\bar{P}_{UXYV}$ is \emph{achievable} for \emph{empirical coordination} if for all $\varepsilon>0$ there exists a sequence $(f_n,g_n)$ of encoders-decoders such that
$$ \mathbb P \left\{ \tv \left( T_{U^{n}  X^{n} Y^{n}  V^{n}}, \bar{P}_{UXYV} \right) > \varepsilon \right\} < \varepsilon$$
where $T_{U^{n}  X^{n} Y^{n}  V^{n}}$  is the joint histogram of the actions induced by the code.
The \emph{empirical coordination region} $\mathcal R_e$ is the closure of the set of achievable distributions $\bar{P}_{UXYV}$.
\end{defi}

\begin{defi}
A pair $(\bar{P}_{UXYV}, R_0)$ is \emph{achievable} for \emph{strong coordination} if there exists a sequence $(f_n,g_n)$ of encoders-decoders with rate of common randomness $R_0$, such that
$$\lim_{n \to \infty} \tv \left( P_{U^{n}  X^{n} Y^{n}  V^{n}}, \bar{P}_{UXYV}^{\otimes n} \right)=0$$
where $P_{U^{n} X^{n} Y^{n}  V^{n}}$ is the joint distribution induced by the code.
The \emph{strong coordination region} $\mathcal{R}$ is the closure of the set of achievable pairs 
$(\bar P_{UXYV}, R_0)$\footnote{As in \cite{cuff2010}, 
we define the achievable region as the closure of the set of achievable rates and distributions. This definition
allows to avoid boundary complications.  
For a thorough discussion on the boundaries of the achievable region when $\mathcal{R}$ is 
defined as the closure of the set of rates for a given distribution, see \cite[Section VI.D]{cuff2013distributed}.}.
\end{defi}

Our first result is an inner and outer bound for the strong coordination region $\mathcal{R}$ \cite{Cervia2017}.
\begin{teo} \label{teoisit}
Let $\bar P_{U}$ and $\bar P_{Y|X}$ be the given source and channel parameters, then 
$\mathcal R'_{\text{in}} \subseteq \mathcal{R} \subseteq \mathcal R'_{\text{out}}$ where:
{\allowdisplaybreaks
\begin{align}
\mathcal R'_{\text{in}} &:= \begin{Bmatrix}[c|l]
& \bar P_{UXYV}= \bar P_{U} \bar P_{X|U} \bar P_{Y|X} \bar P_{V|UXY}  \\
&\exists \mbox{ } W \mbox{ taking values in $\mathcal W$}\\ 
(\bar P_{UXYV}, R_0)  & \bar P_{UXYWV}= \bar P_{U} \bar P_{W|U} \bar P_{X|UW} \bar P_{Y|X} \bar P_{V|WY}\\
& I(W;U) \leq I(W;Y) \\
& R_0 \geq I(W;UXV|Y)\\
\end{Bmatrix} \label{eq: regionisit}\\
\quad
\mathcal R'_{\text{out}} &:= \begin{Bmatrix}[c|l]
& \bar P_{UXYV}= \bar P_{U} \bar P_{X|U} \bar P_{Y|X} \bar P_{V|UXY}  \\
&\exists \mbox{ } W \mbox{ taking values in $\mathcal W$}\\ 
(\bar P_{UXYV}, R_0) &\bar P_{UXYWV}= \bar P_{U} \bar P_{W|U} \bar P_{X|UW} \bar P_{Y|X} \bar P_{V|WY}\\
& I(W;U) \leq I(X;Y)\\
& R_0 \geq I(W;UXV|Y)\\
&\lvert \mathcal W \rvert \leq \lvert \mathcal U \times  \mathcal X \times \mathcal Y \times {\mathcal V} \rvert+4 \\
\end{Bmatrix}. \label{eq: regionisit2}
\end{align}}
\end{teo}

\begin{oss}
 Observe that the decomposition of the joint distributions  $\bar P_{UXYV}$  and  $\bar P_{UWXYV} $
 is equivalently characterized in terms of Markov chains: 
{\allowdisplaybreaks
\begin{equation}\label{markov chain isit}
 Y-X-U,
\quad \quad
 \begin{cases}
Y-X-(U,W)\\
V-(Y,W)-(X,U)
   \end{cases}.
\end{equation}
}
\end{oss}

\begin{oss}
The empirical coordination region for the setting of Figure \ref{fig: coordisit} was investigated in \cite{cuff2011hybrid}, in which the authors derived an inner and outer bound. 
Note that the information constraint $I(W;U) \leq$ $I(W;Y)$ and 
the decomposition of the joint probability distribution $\bar P_{U} \bar P_{W|U} \bar P_{X|UW} \bar P_{Y|X} \bar P_{V|WY}$
are the same for empirical coordination \cite[Theorem 1]{cuff2011hybrid}. 
The main difference is that strong coordination requires a positive rate of common randomness $R_0 > I(W;UXV|Y)$.
\end{oss}

\subsection{Proof of Theorem \ref{teoisit}: inner bound}
We postpone the achievability proof because it is a corollary of the inner bound in the general setting of Theorem \ref{teouv} proven in Section \ref{inner}.
A stand-alone proof can be found in the conference version of the present paper \cite{Cervia2017}.

\begin{oss}
With the same random binning techniques, \cite{haddadpour2017simulation} characterizes an inner bound for the strong coordination region 
in the slightly different scenario in which only $U^n$ and $V^n$ need to be coordinated. 
Given the source  and channel parameters $\bar P_{U}$ and $\bar P_{Y|X}$ respectively, the inner bound in \cite{haddadpour2017simulation} is: 
{\allowdisplaybreaks
\begin{equation}\label{eq: regionhaddad}
 \mathcal{R}_{\text{Hadd,in}}:= \begin{Bmatrix}[c|l]
& \bar P_{UXYV}= \bar P_{U} \bar P_{X|U} \bar P_{Y|X} \bar P_{V|UXY}  \\
&\exists \mbox{ } W \mbox{ taking values in $\mathcal W$}\\ 
(\bar P_{UV}, R_0)  & \bar P_{UXYWV}= \bar P_{U} \bar P_{WX|U}  \bar P_{Y|X} \bar P_{V|WY}\\
& I(W;U) \leq I(W;Y) \\
& R_0 \geq I(W;UV)-I(W;Y)\\
\end{Bmatrix}.
\end{equation}}

Note that the joint distribution and the information constraints are the same as in \eqref{eq: regionisit} 
but in  \eqref{eq: regionisit} the rate of common randomness is larger since
\begin{equation*}
 I(W;UXV|Y)=I(W;UXYV)-I(W;Y) \geq I(W;UV)-I(W;Y).
\end{equation*}
The difference in common randomness rate  $I(W;XY|UV)$ stems from the requirement in \cite{haddadpour2017simulation},
which coordinates $U^n$ and $V^n$ but not necessarly $(U^{n}, X^{n}, Y^n, V^{n})$.
\end{oss}

\subsection{Proof of Theorem \ref{teoisit}: outer bound}
Consider a code $(f_n,g_n)$ that induces a distribution $P_{U^{n} X^{n} Y^{n} V^{n}}$ 
that is $\varepsilon$-close in total variational distance to the i.i.d. distribution $\bar P_{U X Y V}^{\otimes n}$.
Let the random variable $T$ be uniformly distributed over the
set $\llbracket 1,n\rrbracket$ and independent of sequence
$(U^{n}, X^{n}, Y^{n}, V^{n}, C)$. The variable $T$ will serve as a random time index. 
The variable $U_T$ is independent of $T$ because $U^{n}$ is an i.i.d. source sequence \cite{cuff2010}.

\subsubsection{Bound on $R_0$}\label{isitconvpart1}
We apply Lemma \ref{lemab} to $A^{n}:= U^{n} X^{n} V^{n}$, $B^{n}:=Y^{n}$ and $C$ and, using \eqref{lemab2}, we have
{\allowdisplaybreaks
\begin{align*}
& nR_0 \geq H(C) \geq H(C|B^{n}) \\ 
& \overset{(a)}{\geq}  n I(A_T;CB_{\sim T} T|B_T ) -n I(A_T  B_T;T)  -n f(\varepsilon) \stepcounter{equation}\tag{\theequation}\label{boundr0}\\
& \overset{(b)}{\geq} n I(A_T;CB_{\sim T} T|B_T ) -2n f(\varepsilon)=n I(U_T  X_T V_T;CY_{\sim T}  T|Y_T) -2n f(\varepsilon)
\end{align*}}where $(a)$ follows from  Lemma \ref{lemab} and $(b)$ comes from \cite[Lemma VI.3]{cuff2013distributed}.

\subsubsection{Information constraint}\label{isitconvpart1}
We have
{\allowdisplaybreaks
\begin{align*}
& 0  \overset{(a)}{\leq} I(X^{n};Y^{n})-I(C,U^{n};Y^{n}) \\
&\leq I(X^{n};Y^{n})-I(U^{n};Y^{n}|C)\\
& = H(Y^{n})-H(Y^{n}|X^{n}) +H(U^{n}|Y^{n} C)-H(U^{n}|C)\\
& \overset{(b)}{\leq} \sum_{t=1}^{n} H(Y_t) \!-\! \sum_{t=1}^{n} H(Y_t|X_t) + \! \sum_{t=1}^{n} H(U_t|U^{t-1} Y_t Y_{\sim t} C)\!- \! \sum_{t=1}^{n} H(U_t)\\
& \overset{(c)}{\leq} \sum_{t=1}^{n} \left(H(Y_t) -H(Y_t|X_t)+ H(U_t|Y_{\sim t} C)- H(U_t)\right)\\
& \overset{(d)}{\leq}  n H(Y_T)  - n  H(Y_T|X_T  T)+  n H(U_T|Y_{\sim T} C T) - n  H(U_T|T) \\
& \overset{(e)}{=} n H(Y_T) -n H(Y_T|X_T) + nH(U_T|Y_{\sim T} C T)-n H(U_T) \\
&= n I(X_T; Y_T)-n I( U_T;Y_{\sim T}, C, T )
\end{align*}}where $(a)$ comes from the Markov chain $Y^{n}-X^{n}-(C,U^{n})$ and $(b)$ comes from the chain rule for the conditional entropy and the fact that
$U^{n}$ is an i.i.d. source independent of $C$. The inequalities $(c)$ and $(d)$ come 
from the fact that 
conditioning does not increase entropy and 
$(e)$ from the memoryless nature of the channel $\bar P_{Y|X}$ and the i.i.d. nature of the source $\bar P_{U}$.

\subsubsection{Identification of the auxiliary random variable}\label{identification}
We identify the auxiliary random variables $W_t$ with $(C,Y_{\sim t})$ for each $t \in \llbracket 1,n\rrbracket$ and $W$ with
$(W_T,T)=(C,Y_{\sim T}, T)$. For each $t \in \llbracket 1,n\rrbracket$ the following two Markov chains hold: 
{\allowdisplaybreaks
\begin{align}
 Y_t-X_t-(C, Y_{\sim t}, U_t)  \quad &\Longleftrightarrow \quad Y_t-X_t-(W_t, U_t) \label{mc1t}\\
 V_t-(C,Y_{\sim t},Y_t)-(U_t,X_t) \quad &\Longleftrightarrow \quad V_t-(W_t,Y_t)-(U_t,X_t)\label{mc2t}
\end{align}}where \eqref{mc1t} comes from the fact that the channel is memoryless and \eqref{mc2t} from the fact that the decoder is non-causal 
and for each $t \in \llbracket 1,n\rrbracket$ the decoder generates $V_t$ from $Y^{n}$ and common randomness $C$. 
Then, we have 
{\allowdisplaybreaks
\begin{align}
 Y_T-X_T-(C, Y_{\sim T}, U_T, T)  \quad &\Longleftrightarrow \quad  Y_T-X_T-(W_T, U_T, T) \label{mc1T}\\
 V_T-(C,Y_{\sim T},Y_T, T)-(U_T,X_T) \quad &\Longleftrightarrow \quad V_T-(W_T,Y_T, T)-(U_T,X_T)  \label{mc2T}
\end{align}}where \eqref{mc1T} holds because 
\begin{align*}
 \mathbb P \{Y_T=y| X_T=x, Y_{\sim T}=\tilde{\mathbf y} , U_T=u, T=t, C\}=  \mathbb P \{Y_T=y| X_T=x\}
\end{align*}
since the channel is memoryless. Then by \eqref{mc2t}, \eqref{mc2T} holds because 
{\allowdisplaybreaks
\begin{align*}
I(V_T;U_T X_T| C Y^{n} T)= \sum_{i=1}^n \frac{1}{n} I(V_t;U_t X_t| C Y^{n} T=t)=0.
\end{align*}
}
Since $W=W_t$ when $T=t$, we also have $(U,X)-(W,Y)-V$ and $Y-X-(U,W)$. 
The cardinality bound is proved in $\mbox{Appendix \ref{appendix bounds}}$.

\section{Inner and outer bounds for the strong coordination region with state and side information}\label{sec: innerouter}
\begin{center}
\begin{figure}[ht]
\centering
\includegraphics[scale=0.21]{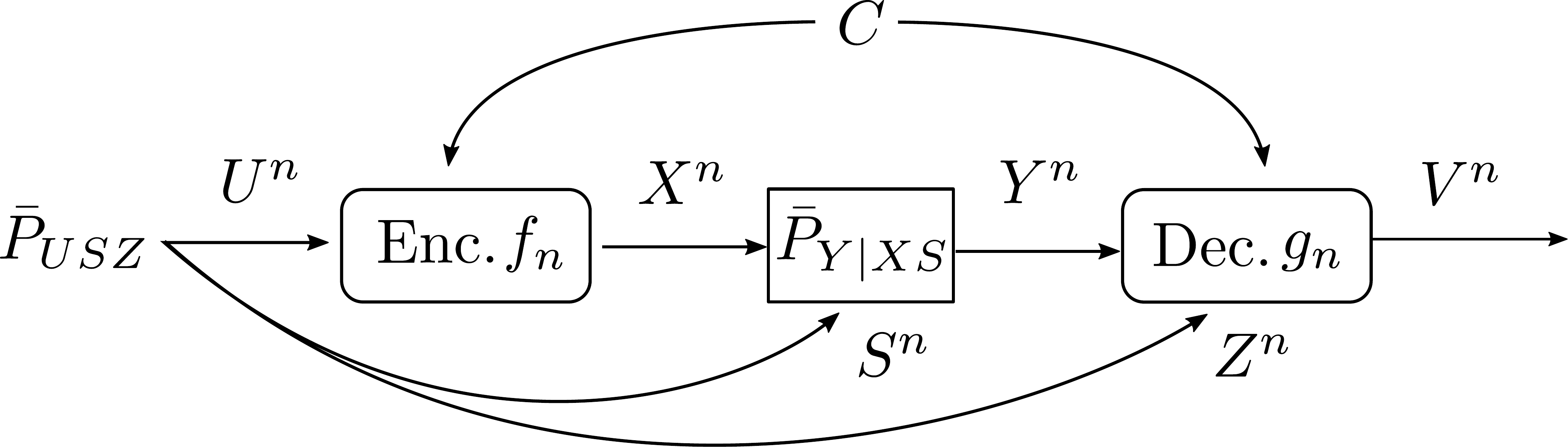}
\caption{Coordination of signals and actions for a two-node network with a noisy channel with state and side information at the decoder.}
\label{fig: generalsetting}
\end{figure}
\end{center}
\vspace{-0,8cm}

In this section we consider the model depicted in Figure~\ref{fig: generalsetting}. 
It is a generalization of the simpler setting of Figure~\ref{fig: coordisit}, where
the noisy channel depends on a state $S^n$, and the decoder has access to non-causal side information $Z^n$. 
The encoder selects a signal $X^{n}= f_n(U^{n},  C)$, with $f_n: \mathcal U^n  \times  \llbracket 1,2^{nR_0} \rrbracket \rightarrow \mathcal X^n$ 
and transmits it  over the discrete memoryless channel $\bar P_{Y |XS}$ where $S$ represents the state.
The decoder then selects an action $V^{n} = g_n(Y^{n}, Z^{n}, C)$,  where 
$g_n: \mathcal Y^n  \times \mathcal Z^n \times \llbracket 1,2^{nR_0} \rrbracket \rightarrow \mathcal V^n$ is a stochastic map and $Z^{n}$ 
represents the side information available at the decoder.

\begin{oss}
 Note that the channel state information and side information at the decoder are represented explicitly by the random 
 variables $S^{n}$ and $Z^{n}$ respectively, but the model is quite general and includes scenarios where 
 partial or perfect channel state information is available at the encoder as well, 
 since the variables $U^{n}$ and $S^{n}$ are possibly correlated.
\end{oss}

We recall the notions of achievability and of the coordination region for empirical and strong coordination \cite{cuff2010,cuff2009thesis} in this setting.

\begin{defi}
A distribution $\bar{P}_{USZXYV}$ is \emph{achievable} for \emph{empirical coordination} if for all $\varepsilon>0$ there exists a sequence $(f_n,g_n)$ of encoders-decoders such that
$$ \mathbb P \left\{ \tv \left( T_{U^{n} S^{n} Z^n X^{n} Y^{n}  V^{n}}, \bar{P}_{USZXYV} \right) > \varepsilon \right\} < \varepsilon$$
where $T_{U^{n} S^{n} Z^n X^{n} Y^{n}  V^{n}}$  is the joint histogram of the actions induced by the code.
The \emph{empirical coordination region} $\mathcal R_e$ is the closure of the set of achievable distributions $\bar{P}_{USZXYV}$.\\
A pair $(\bar{P}_{USZXYV}, R_0)$ is \emph{achievable} for \emph{strong coordination} if there exists a sequence $(f_n,g_n)$ of encoders-decoders with rate of common randomness $R_0$, such that
$$\lim_{n \to \infty} \tv \left( P_{U^{n} S^{n} Z^n X^{n} Y^{n}  V^{n}}, \bar{P}_{USZXYV}^{\otimes n} \right)=0$$
where $P_{U^{n} S^{n} Z^n X^{n} Y^{n}  V^{n}}$ is the joint distribution induced by the code.
The \emph{strong coordination region} $\mathcal{R}$ is the closure of the set of achievable pairs $(\bar P_{USZXYV}, R_0)$.
\end{defi}
\vspace{0,2cm}

In the case of non-causal encoder and decoder, the problem of characterizing the strong
coordination region for the system model in Figure \ref{fig: generalsetting} is still open, but we establish the following inner and outer bounds.

\begin{teo} \label{teouv}
Let $\bar P_{USZ}$ and $\bar P_{Y|XS}$ be the given source and channel parameters, then 
$\mathcal R_{\text{in}} \subseteq \mathcal{R}\subseteq \mathcal R_{\text{out}}$ where:
{\allowdisplaybreaks
\begin{align}
\mathcal R_{\text{in}}  &:= \begin{Bmatrix}[c|l]
& \bar P_{USZXYV}= \bar P_{USZ} \bar P_{X|U} \bar P_{Y|XS} \bar P_{V|UXYSZ}  \\
&\exists \mbox{ } W \mbox{ taking values in $\mathcal W$}\\ 
(\bar P_{USZXYV}, R_0)&\bar P_{USZWXYV}=\bar P_{USZ} \bar P_{W|U} \bar P_{X|UW} \bar P_{Y|XS} \bar P_{V|WYZ}\\
&I(W;U) \leq I(W;YZ)\\
& R_0 \geq I(W;USXV|YZ)\\
\end{Bmatrix} \label{eq: region inn} \\
\mathcal R_{\text{out}} &:=   \begin{Bmatrix}[c|l]
 & \bar P_{USZXYV}= \bar P_{USZ} \bar P_{X|U} \bar P_{Y|XS} \bar P_{V|UXYSZ}  \\
&\exists \mbox{ } W \mbox{ taking values in $\mathcal W$}\\ 
(\bar P_{USZXYV}, R_0) &\bar P_{USZWXYV}=\bar P_{USZ} \bar P_{W|U} \bar P_{X|UW} \bar P_{Y|XS} \bar P_{V|WYZ}\\
& I(W;U) \leq \min \{I(XUS;YZ),I(XS;Y)+I(U;Z)\} \\
& R_0 \geq I(W;USXV|YZ)\\
& \lvert \mathcal W \rvert \leq \lvert \mathcal U \times \mathcal S  \times \mathcal Z \times \mathcal X \times \mathcal Y \times {\mathcal V} \rvert+5\\
\end{Bmatrix} \label{eq: region out}
\end{align}
}
\end{teo}

\begin{oss} 
As in Theorem \ref{teoisit}, even if inner and outer bound do not match, they only differ on the upper bound on $I(W;U)$. 
Note that we cannot compare $I(XUS;YZ)$ and $I(XS;Y)+I(U;Z)$ (for more details, see the discussion in Appendix \ref{appendix compare}).
Hence, in $\mathcal R_{\text{out}}$ the upper bound on the mutual information $I(W;U)$ is the minimum of the two.
\end{oss}

\begin{oss}
 Observe that the decomposition of the joint distributions  $\bar P_{USZXYV}$  and $\bar P_{USZWXYV}$
 is equivalently characterized in terms of Markov chains: 
{\allowdisplaybreaks
\begin{equation}\label{markov chain general}
 \begin{cases}
   Z-(U,S)-(X,Y)\\
   Y-(X,S)-U
   \end{cases},\quad
 \begin{cases}
Z-(U,S)-(X,Y,W)\\
Y-(X,S)-(U,W)\\
V-(Y,Z,W)-(X,S,U)
   \end{cases}.
\end{equation}
}

\end{oss}

\subsection{Proof of Theorem \ref{teouv}: inner bound}\label{inner}
The achievability proof uses the same techniques as in \cite{haddadpour2017simulation} 
inspired by \cite{yassaee2014achievability}.
The key idea of the proof is to define a random binning and a random coding scheme, each of
which induces a joint distribution, and to prove that the two schemes have the same statistics. 

Before defining the coding schemes, we state the results that we will use to prove the inner bound.

The following lemma is a consequence of the Slepian-Wolf Theorem  .  
\begin{lem}[Source coding with side information at the decoder $\mbox{\cite[Theorem 10.1]{elgamal2011nit}}$ ]\label{lem1}
 Given a discrete memoryless source $(A^{n},B^{n})$, where $B^{n}$ is side information available at the decoder,  
 we define a stochastic encoder $\psi_n: \mathcal A^n \to \llbracket 1,2^{nR} \rrbracket$,  where
$C:= \varphi_n(A^{n})$ is a binning of $A^{n}$.
If $R>H(A|B)$, the decoder recovers $A^{n}$ from $C$ and $B^{n}$ with arbitrarily small error probability.
\end{lem}

\begin{lem}[Channel randomness extraction $\mbox{\cite[Lemma 3.1]{ahlswede1998common}}$ and $\mbox{\cite[Theorem 1]{yassaee2014achievability}}$]\label{cor1}  
 Given a discrete memoryless source $(A^{n},B^{n})$, 
 we define a stochastic encoder $\varphi_n: \mathcal B^n \to \llbracket 1, J_n \rrbracket$,  where
$K:= \varphi_n(B^{n})$ is a binning of $B^{n}$ with $J_n$ values chosen independently and uniformly at random.
if $R  < H(B|A)$, then we have
$$ \lim_{n \to \infty} \mathbb E_{\varphi_n} \left[\mathbb V \left( P_{A^{n} K}^{\varphi},  Q_K P_{A^{n}}\right)\right]=0,$$
where $\mathbb{E}_{\varphi_n}$ denotes the average over the random binnings and $Q_K$ is the uniform distribution on $\llbracket 1,J_n\rrbracket$. 
\end{lem}
 
Although Lemma \ref{cor1} ensures the convergence in total variational distance and is therefore enough to prove strong coordination,
it does not bring any insight on the speed of convergence. 
For this reason, throughout the proof we will use the following lemma instead.
We omit the proof as it follows directly from the discussion in \cite[Section III.A]{pierrot2013joint}.

 \begin{lem}[Channel randomness extraction for discrete memoryless sources and channels]\label{1.4.2} 
  Let $A^n$ with distribution $P_{A^n}$ be a discrete memoryless source and $P_{B^n|A^n}$ a discrete memoryless channel. 
Then for every $\varepsilon >0$, there exists a sequence of $(J_n , n)$ codes 
$\varphi_n: \mathcal B^n \to \llbracket 1, J_n \rrbracket$ and a constant $\alpha > 0$ such that for $K:= \varphi_n(B^{n})$ we have
  \begin{equation}\label{eq1lem1.4.2}
   \liminf_{n \to \infty} \frac{\log{J_n}}{n} \leq (1-\varepsilon)H(B|A) \quad \mbox{and} \quad \mathbb D\left(P_{A^nK} \Arrowvert P_{A^n} Q_K \right) \leq 2^{-\alpha n}.
  \end{equation}
 \end{lem}

\subsubsection{Random binning scheme}\label{rb gen}
Assume that the sequences $U^{n}$, $S^{n}$, $Z^{n}$, $X^{n}$, $W^{n}$, $Y^{n}$ and $V^{n}$ 
are jointly i.i.d. with distribution 
 {\allowdisplaybreaks
\begin{align*}
& \bar P_{U^{n}S^{n}Z^{n}} \bar P_{W^{n}| U^{n}} \bar P_{X^{n}| W^{n} U^{n}} \bar P_{Y^{n}|X^{n} S^{n}} \bar P_{V^{n}|W^{n} Y^{n} Z^{n}}.
\end{align*}}

We consider two uniform random binnings for $W^{n}$:
\begin{itemize}
\item first binning $C = \varphi_1(W^{n})$, where  $\varphi_1: \mathcal{W}^{n} \to \llbracket 1,2^{nR_0} \rrbracket$ is an encoder which maps each sequence of $\mathcal{W}^{n}$ uniformly and independently to the set $\llbracket 1,2^{nR_0} \rrbracket$;
\item second binning $F = \varphi_2(W^{n})$, where $\varphi_2: \mathcal{W}^{n} \to \llbracket 1,2^{n \tilde R} \rrbracket$ is an encoder.
\end{itemize}

 \begin{center}
\begin{figure}[h]
 \centering
 \includegraphics[scale=0.23]{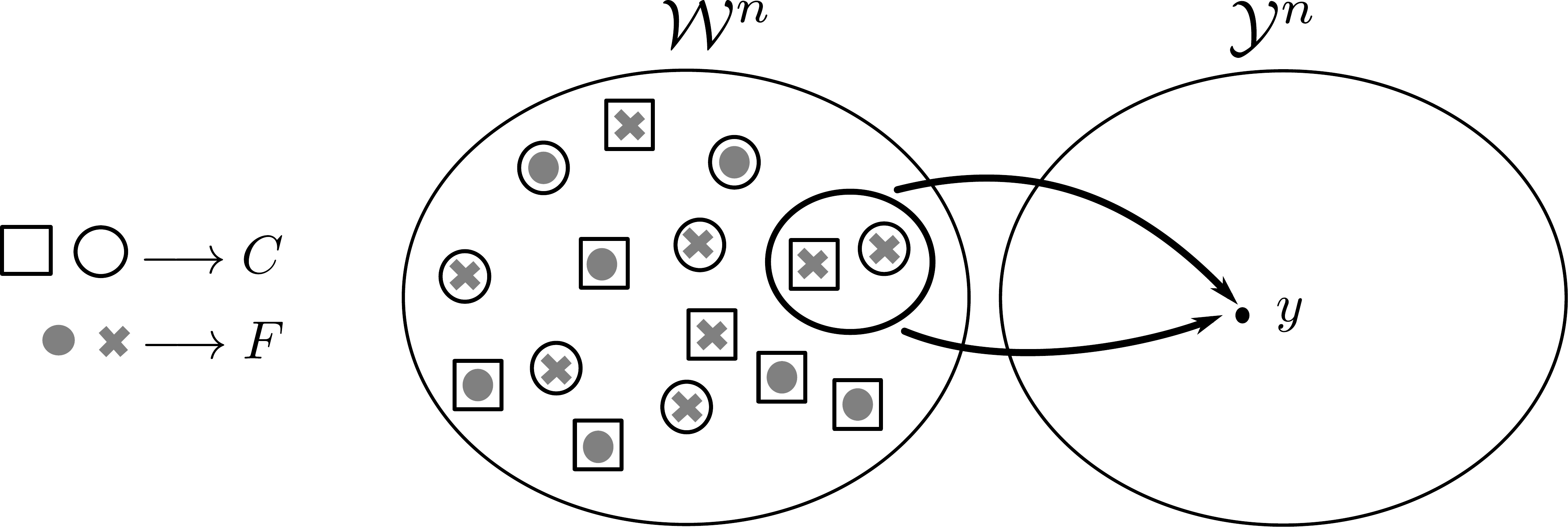}
\caption{The square and the circle represent the possible outputs $C$ of the first binning and the dot and the cross the outputs $F$ of the second binning. Given $\mathbf y$ and the realizations of $C$ and $F$, it is possible to recover $\mathbf w$. }
\label{fig: db}
\end{figure}
\end{center}
\vspace*{-0.6cm}

Note that if $\tilde R+R_0 >H(W|YZ)$, by Lemma \ref{lem1}, it is possible to recover $W^{n}$ from $Y^{n}$, $Z^{n}$ and $(C, F)$ with 
high probability using a Slepian-Wolf decoder via the conditional distribution $P^{\text{SW}}_{\widehat W^{n}|CF Y^{n} Z^{n}}$. 
This defines a joint distribution:
 {\allowdisplaybreaks
\begin{align*}
&\bar P_{U^{n} S^{n}Z^{n}W^{n} \widehat W^{n} X^{n} Y^{n} C F V^{n}}=  \bar P_{U^{n} S^{n}Z^{n}} \bar P_{W^{n}|U^{n}} \bar P_{X^{n}|W^{n}  U^{n}}  \bar P_{C|W^{n}} \bar P_{F|W^{n}}  \bar P_{Y^{n}|X^{n} S^{n}} \bar P_{V^{n}|W^{n}  Y^{n} Z^{n}} P^{\text{SW}}_{\widehat W^{n}|C F  Y^{n} Z^{n}}.
\end{align*}
}In particular, $\bar P_{W^{n}|CFU^{n}}$ is well defined.

\subsubsection{Random coding scheme}\label{rc gen}
In this section we follow the approach in \cite[Section IV.E]{yassaee2014achievability}.
Suppose that in the setting of Figure \ref{fig: generalsetting}, encoder and decoder have access not only to common randomness $C$ 
but also to extra randomness $F$, where $C$ is generated uniformly at random 
in $\llbracket 1,2^{nR_0} \rrbracket$ with distribution $Q_C$ and $F$ is generated uniformly at
random in $\llbracket 1,2^{n \tilde R} \rrbracket$ with distribution $Q_F$ independently of $C$. 
Then, the encoder generates $W^{n}$ according to $\bar P_{W^{n}|CFU^{n}}$ defined above and $X^{n}$ according to $\bar P_{X^{n}|U^{n} W^{n}}$.
The encoder sends $X^{n}$ through the channel.
The decoder obtains $(Y^{n}, Z^{n})$ and $(C,F)$ and reconstructs $W^{n}$ 
via the conditional distribution  $P^{\text{SW}}_{\widehat W^{n}|CF Y^{n} Z^{n}}$. 
The decoder then generates $V^{n}$ letter by letter according to the distribution $P_{V^{n}|\widehat W^{n} Y^{n}Z^{n}}$
(more precisely $\bar P_{V^{n}| W^{n} Y^{n} Z^{n}}(\widehat{\mathbf u}|\widehat{\mathbf w}, \mathbf y ,\mathbf z)$, where $\widehat{\mathbf w}$ is the output of the Slepian-Wolf decoder).
This defines a joint distribution:
{\allowdisplaybreaks
\begin{align*}
& P_{U^{n}S^{n}Z^{n} W^{n} \widehat W^{n} X^{n} Y^{n} C F V^{n}} =Q_C Q_F P_{U^{n}S^{n}Z^{n}} \bar P_{ W^{n}|CFU^{n}} \bar P_{X^{n}|W^{n} U^{n}}  \bar P_{Y^{n}|X^{n} S^{n}} P^{\text{SW}}_{\widehat W^{n}|CF Y^{n} Z^{n}} P_{V^{n}|\widehat W^{n} Y^{n} Z^{n}}.
\end{align*}
}
We want to show that the distribution $\bar P$ is achievable  for strong coordination:
 {\allowdisplaybreaks
\begin{align*}
& \lim_{n \to \infty} \tv (\bar P_{U^{n} S^{n}Z^{n} X^{n} W^{n} \widehat W^{n} Y^{n} V^{n}}, P_{U^{n}S^{n}Z^{n} X^{n} W^{n} \widehat W^{n} Y^{n} V^{n}} )=0.  \stepcounter{equation}\tag{\theequation}\label{shat}
\end{align*}}We prove that the random coding scheme possesses all the properties of the initial source coding 
scheme stated in Section \ref{rb gen}. Note that
{\allowdisplaybreaks
\begin{align*}
& \mathbb D (\bar P_{U^{n} S^{n}Z^{n} W^{n} \widehat W^{n} X^{n} Y^{n} CF} \Arrowvert P_{U^{n} S^{n}Z^{n} W^{n} \widehat W^{n} X^{n} Y^{n} C F}) \\
& = \mathbb D (\bar P_{U^{n}S^{n}Z^{n}} \bar P_{W^{n}|U^{n}} \bar P_{X^{n}|W^{n} U^{n}} \bar P_{C|W^{n}}  \bar P_{F|W^{n}} \bar P_{Y^{n}|X^{n} S^{n}} P^{\text{SW}}_{\widehat W^{n}|CF Y^{n} Z^{n}}   \\
&  \phantom{= \mathbb D |} \Arrowvert Q_C Q_F  P_{U^{n} S^{n}Z^{n}} \bar P_{ W^{n}|CFU^{n}} \bar P_{X^{n}|W^{n} U^{n}} \bar P_{Y^{n}|X^{n} S^{n}} P^{\text{SW}}_{\widehat W^{n}|CF Y^{n} Z^{n}})  \stepcounter{equation}\tag{\theequation}\label{chain1} \\ 
& {\overset{{(a)}}{=}}  \mathbb D ( \bar P_{U^{n}S^{n}Z^{n}} \bar P_{W^{n}|U^{n}} \bar P_{C|W^{n}} \bar P_{F|W^{n}} \Arrowvert Q_C Q_F P_{U^{n}S^{n}Z^{n}} \bar P_{W^{n}|CFU^{n}} )\\
& {\overset{{(b)}}{=}} \mathbb D (\bar P_{U^{n} S^{n}Z^{n}CF} \Arrowvert P_{U^{n}S^{n}Z^{n}}  Q_C Q_F )
\end{align*}
}where $(a)$ comes from Lemma \ref{lemkl}. Note that $(b)$  follows from Lemma \ref{lemkl} as well,
since $W^{n}$ is generated according to $\bar P_{W^{n}|CFU^{n}}$ and because of the Markov chain $W-U-ZS$, $W^{n}$ is conditionally independent of $(Z^{n},S^{n})$ given $U^{n}$. 
Then if $R_0 + \widetilde R < H(W|USZ)=H(W|U)$, we apply Lemma \ref{1.4.2} where $B^{n}=W^{n}$,  $K=(C,F)$, 
and claim that there exists a fixed binning $\varphi':=(\varphi'_1, \varphi'_2)$ such that, if we denote with  $\bar P^{\varphi'}$ and $P^{\varphi'}$  
the distributions $\bar P$ and $P$ with respect to the choice of a binning $\varphi'$, we have
\begin{align*}
 \mathbb D \left(\bar P_{U^{n}S^{n}Z^{n}CF}^{\varphi'} \Arrowvert P_{U^{n}S^{n}Z^{n}}Q_C Q_F  \right)=\delta(n),
\end{align*}
which by \eqref{chain1} implies 
 {\allowdisplaybreaks
\begin{align*}
& \mathbb D \left(\bar P^{\varphi'}_{U^{n} S^{n}Z^{n}W^{n} \widehat W^{n} X^{n} Y^{n} CF} \Arrowvert P^{\varphi'}_{U^{n} S^{n}Z^{n} W^{n} \widehat W^{n} X^{n} Y^{n} C F}\right) =\delta(n).
\end{align*}
}
Then, by Lemma \ref{lem1csi} we have
 {\allowdisplaybreaks
\begin{align*}
& \mathbb V \left(\bar P^{\varphi'}_{U^{n} S^{n}Z^{n}W^{n} \widehat W^{n} X^{n} Y^{n} CF}, P^{\varphi'}_{U^{n} S^{n}Z^{n} W^{n} \widehat W^{n} X^{n} Y^{n} C F}\right) =\delta(n). \stepcounter{equation}\tag{\theequation}\label{suxy}  
\end{align*}
}From now on, we will omit  ${\varphi'}$ to simplify the notation.

Now we would like to show that we have strong coordination for $V^{n}$ as well, but
in the second coding scheme $V^{n}$ is generated 
using the output of the Slepian-Wolf decoder 
$\widehat W^{n}$ and not $W^{n}$ as in the first scheme. 
Because of Lemma \ref{lem1}, the inequality $\widetilde{R}+R_0>H(W|YZ)$  implies that $\widehat W^{n}$ is equal to $W^{n}$ with high probability and we will use this fact to show that
the distributions are close in total variational distance.

First,  we recall the definition of coupling and the basic coupling inequality for two random variables \cite{Lindvall1992coupling}.
\begin{defi}\label{defcoup}
A coupling of two probability distributions $P_A$ and $P_{A'}$ on the same measurable space $\mathcal A$ is any probability distribution 
$\widehat P_{AA'}$ on the product measurable space $\mathcal{A} \times \mathcal{A}$  whose marginals are $P_A$ and $P_{A'}$.
\end{defi}

\begin{prop}[$\mbox{\cite[I.2.6]{Lindvall1992coupling}}$]\label{theocoup}
Given two random variables $A$, $A'$ with probability distributions $P_{A}$, $P_{A'}$, any coupling 
$\widehat P_{AA'}$ of $P_{A}$, $P_{A'}$ satisfies
\begin{equation*}
\tv(P_A, P_{A'})\leq 2 \mathbb P_{\widehat P_{AA'}}\{A \neq A'\} 
\end{equation*}
\end{prop}

Then, we apply Proposition \ref{theocoup} to 
{\allowdisplaybreaks
\begin{align*}
&\begin{matrix}[ll]
 A = U^{n} S^{n}Z^{n}W^{n} X^{n} Y^{n} CF   & A'= U^{n} S^{n}Z^{n} \widehat W^{n} X^{n} Y^{n} CF\\
 P_{A} =\bar P_{U^{n} S^{n}Z^{n}W^{n}  X^{n} Y^{n} CF} & P_{A'}=\bar P_{U^{n} S^{n}Z^{n} \widehat W^{n} X^{n} Y^{n} CF} \\
\end{matrix}\\
&  \mathcal A= \mathcal U \times \mathcal S \times \mathcal Z \times \mathcal W \times \mathcal X \times \mathcal Y \times \llbracket 1,2^{nR_0} \rrbracket \times  \llbracket 1,2^{n \tilde R} \rrbracket. 
\end{align*}
}Since $\widehat W^{n}$ is equal to $W^{n}$ with high probability by Lemma \ref{lem1}, 
and the probability of error goes to zero exponentially in the Slepian-Wolf Theorem  \cite[Theorem 10.1]{elgamal2011nit}, 
we find that for the random binning scheme
{\allowdisplaybreaks
\begin{align*}
 \tv(\bar P_{U^{n} S^{n}Z^{n}W^{n} X^{n} Y^{n} CF}, \bar P_{U^{n} S^{n}Z^{n} \widehat W^{n} X^{n} Y^{n} CF} )= \delta(n).
\end{align*}
}This implies that:
{\allowdisplaybreaks
\begin{align*}
  \tv(\bar P_{U^{n} S^{n}Z^{n}W^{n} \widehat W^{n} X^{n} Y^{n} CF}, \bar P_{U^{n} S^{n}Z^{n}W^{n} X^{n} Y^{n} CF}  \mathds 1_{ \widehat W^{n}| W^{n}})=\delta(n). \stepcounter{equation}\tag{\theequation}\label{34yas'}
\end{align*}
}

Similarly, we apply Proposition \ref{theocoup} again to the random coding scheme and we have
{\allowdisplaybreaks
\begin{align*}
&\tv(P_{U^{n} S^{n}Z^{n}W^{n} \widehat W^{n} X^{n} Y^{n} CF}, P_{U^{n} S^{n}Z^{n}W^{n} X^{n} Y^{n} CF}  \mathds 1_{ \widehat W^{n}| W^{n}})=\delta(n).\stepcounter{equation}\tag{\theequation}\label{35yas'}
\end{align*}
}

Then using the triangle inequality, we find that 
{\allowdisplaybreaks
\begin{align*}
& \tv (\bar P_{U^{n} S^{n}Z^{n}W^{n} \widehat W^{n} X^{n} Y^{n} CF V^{n}},P_{U^{n} S^{n}Z^{n} W^{n} \widehat W^{n} X^{n} Y^{n} C F V^{n}}) \\
& = \tv ( \bar P_{U^{n} S^{n}Z^{n}W^{n} \widehat W^{n} X^{n} Y^{n} CF} \bar P_{V^{n} | W^{n} Y^{n} Z^{n}}, P_{U^{n} S^{n}Z^{n} W^{n} \widehat W^{n} X^{n} Y^{n} C F} P_{V^{n} | \widehat W^{n} Y^{n} Z^{n}} ) \stepcounter{equation}\tag{\theequation}\label{triu}  \\
&\leq  \tv (\bar P_{U^{n} S^{n}Z^{n}W^{n} \widehat W^{n} X^{n} Y^{n} CF}  \bar P_{V^{n} | W^{n} Y^{n} Z^{n}} ,  \bar P_{U^{n} S^{n}Z^{n}W^{n} X^{n} Y^{n} CF} \mathds 1_{\widehat W^{n}| W^{n}} \bar P_{V^{n} | W^{n} Y^{n} Z^{n}} ) \\
&\quad + \tv  (\bar P_{U^{n} S^{n}Z^{n}W^{n} X^{n} Y^{n} CF}   \mathds 1_{\widehat W^{n}| W^{n}}  \bar P_{V^{n} | W^{n} Y^{n} Z^{n}},   P_{U^{n} S^{n}Z^{n}W^{n} X^{n} Y^{n} CF}  \mathds 1_{\widehat W^{n}| W^{n}} P_{V^{n} |\widehat W^{n} Y^{n} Z^{n}})\\
&\quad +  \tv (P_{U^{n} S^{n}Z^{n}W^{n} X^{n} Y^{n} CF} \mathds 1_{\widehat W^{n}| W^{n}}  P_{V^{n} | W^{n} Y^{n} Z^{n}},  P_{U^{n} S^{n}Z^{n}W^{n} \widehat W^{n} X^{n} Y^{n} CF} P_{V^{n} | \widehat W^{n} Y^{n} Z^{n}}) .
\end{align*}
}The first and the third term go to zero exponentially by applying Lemma \ref{cuff17} to \eqref{34yas'} and \eqref{35yas'} respectively.
Now observe that 
\begin{equation*}
 \mathds 1_{\widehat W^{n}| W^{n}}  \bar P_{V^{n} | W^{n} Y^{n} Z^{n}}\!\!= \mathds 1_{\widehat W^{n}| W^{n}}  P_{V^{n} | \widehat W^{n} Y^{n} Z^{n}}
\end{equation*}by definition of $P_{V^{n} | \widehat W^{n} Y^{n} Z^{n}}$. 
Then by using Lemma \ref{cuff17} again the second term is equal to 
\begin{equation*}
 \tv \left(\bar P_{U^{n} S^{n}Z^{n}W^{n}  X^{n} Y^{n} CF}, P_{U^{n} S^{n}Z^{n}W^{n}  X^{n} Y^{n} CF}\right)
\end{equation*}and goes to zero by \eqref{suxy} and Lemma \ref{cuff16}. 
Hence, we have
 {\allowdisplaybreaks
\begin{align*}
 \mathbb V (\bar P_{U^{n} S^{n}Z^{n}W^{n} \widehat W^{n} X^{n} Y^{n} CF V^{n}}, P_{U^{n} S^{n}Z^{n} W^{n} \widehat W^{n} X^{n} Y^{n} C F V^{n}}) =\delta(n).  \stepcounter{equation}\tag{\theequation}\label{suuxycf} 
\end{align*}
}Using Lemma \ref{cuff16}, we conclude that 
$$ \tv (\bar P_{U^{n} S^{n}Z^{n} X^{n} W^{n} \widehat W^{n} Y^{n} V^{n}}, P_{U^{n}S^{n}Z^{n} X^{n} W^{n} \widehat W^{n} Y^{n} V^{n}} )=\delta(n).$$

\subsubsection{Remove the extra randomness F}\label{rf gen}

Even though the extra common randomness $F$ is required to coordinate $\left(U^{n}\right.$, $S^{n}$, $Z^{n}$, ${X}^{n}$, $Y^{n}$, $V^{n}$, $\left.W^{n}\right)$ 
we will show that we do not need it in order to coordinate only $(U^{n}, S^{n},Z^{n},{X}^{n},Y^{n},V^{n})$.
Observe that by Lemma \ref{cuff16}, equation \eqref{suuxycf} implies that 
 {\allowdisplaybreaks
\begin{align*}
\mathbb V (\bar P_{U^{n} S^{n}Z^{n} X^{n} Y^{n}  V^{n} F}, P_{U^{n} S^{n}Z^{n} X^{n} Y^{n}  V^{n} F}) =\delta(n). \stepcounter{equation}\tag{\theequation}\label{convf2} 
\end{align*}}
As in \cite{yassaee2014achievability}, we would like to reduce the amount of common randomness by having the two nodes agree on an instance $F=f$.
To do so, we apply Lemma \ref{1.4.2} again where $B^{n}=W^{n}$,  
$K=F$, $\varphi= \varphi''_2$ and $A^{n}= U^{n} S^{n} Z^{n} X^{n} Y^{n} V^{n}$.
If $\tilde R < H(W| SUZXY V)$, there exists a fixed binning such that
 {\allowdisplaybreaks
\begin{align*}
\tv (\bar P_{U^{n} S^{n}Z^{n} X^{n} Y^{n} V^{n} F}, Q_F \bar P_{U^{n} S^{n}Z^{n} X^{n} Y^{n} V^{n}})=\delta(n). \stepcounter{equation}\tag{\theequation}\label{bin3}  
\end{align*}}
\vspace{-0,3cm}
\begin{oss}\label{binnings2}
Note that in Section \ref{rc gen} we had already chosen a specific binning $\varphi'_2$. In Appendix \ref{appendix bin} we prove that there exists a binning which works for both conditions.
\end{oss}

Because of \eqref{convf2}, \eqref{bin3} implies
 {\allowdisplaybreaks
\begin{align*}
 &  \tv (P_{U^{n} S^{n}Z^{n}X^{n} Y^{n} V^{n} F}, Q_F \bar P_{U^{n} S^{n}Z^{n}X^{n} Y^{n} V^{n}})=\delta(n). \stepcounter{equation}\tag{\theequation}\label{bin4}  
\end{align*}}Hence, we can fix $f \in F$ such that $(U^{n},S^{n},Z^{n},X^{n},Y^{n}, \hat U^{n})$ is almost independent of $F$ according to $P$.
To conclude, if $f \in F$ is fixed, the distribution $P_{ U^{n} S^{n}Z^{n} X^{n} Y^{n} V^{n}}$ changes to $P_{U^{n} S^{n}Z^{n} X^{n} Y^{n} V^{n}|F=f}$
and by  Lemma \ref{lem4} we have
 {\allowdisplaybreaks
\begin{align*}
& \mathbb V (\bar P_{U^{n} S^{n}Z^{n}X^{n} Y^{n} V^{n}|F=f},P_{U^{n} S^{n}Z^{n}X^{n} Y^{n} V^{n}|F=f}) =\delta(n).
\end{align*}}Since $\bar P_{U^{n} S^{n}Z^{n}X^{n} Y^{n} V^{n}|F=f}$ is close to $\bar P_{U^{n}S^{n}Z^{n} X^{n}Y^{n}V^{n}}$ because of \eqref{bin3}, we have
 {\allowdisplaybreaks
\begin{equation}\label{finaleq}
\mathbb V (\bar P_{U^{n} S^{n} Z^{n} X^{n} Y^{n} V^{n}}, P_{U^{n} S^{n} Z^{n} X^{n} Y^{n} V^{n}}) =\delta(n).
\end{equation}}

\subsubsection{Rate constraints}
We have imposed the following rate constraints:
{\allowdisplaybreaks
\begin{align*}
H(W|YZ) &< \widetilde R+R_0 < H(W|U),\\
 \widetilde R & < H(W|USZXYV).
\end{align*}
}
Therefore we obtain:
{\allowdisplaybreaks
\begin{align*}
 & R_0 > H(W|YZ)-H(W|USZXYV) = I(W;USXV|YZ),\\
 & I(W;U)  < I(W;YZ). 
\end{align*}
}

\vspace{-0,7cm}
\subsection{Proof of Theorem \ref{teouv}: outer bound}\label{outer}
Consider a code $(f_n,g_n)$ that induces a distribution $P_{U^{n} S^{n} Z^{n} X^{n} Y^{n} V^{n}}$ that is $\varepsilon$-close in total variational distance to the i.i.d. distribution $\bar P_{U S Z  X Y V}^{\otimes n}$.
Let the random variable $T$ be uniformly distributed over the
set $\llbracket 1,n\rrbracket$ and independent of the sequence
$(U^{n}, S^{n}, Z^{n}, X^{n}, Y^{n}, V^{n}, C)$. 
The variable $(U_T, S_T, {Z}_T)$ is independent of $T$ because $(U^{n},S^{n}, Z^{n})$ is an i.i.d. source sequence \cite{cuff2010}.

\subsubsection{Bound on $R_0$}\label{genconvpart1}
The proof is the same as in Section \ref{isitconvpart1}, using $A^{n}:= U^{n} S^{n} X^{n} V^{n}$ and $B^{n}:=Y^{n} Z^{n}$ 
in \eqref{boundr0}. Then, we obtain $R_0 \geq I(W;USXV|YZ)$.

\subsubsection{Information constraint}\label{genconvpart2}
As proved in Appendix \ref{appendix compare},
in the general case we are not able to compare $I(XUS;YZ)$ and $I(XS;Y)+I(U;Z)$. Then, we show separately that:
\begin{align}
& I(W;U) \leq I(XUS; YZ), \label{outer 1} \\
& I(W;U) \leq I(XS;Y)+I(U;Z).\label{outer 2} 
\end{align}

\paragraph{Proof of \eqref{outer 1}}

We have
{\allowdisplaybreaks
\begin{align*}
&0  \overset{(a)}{\leq} I(X^{n} S^{n};Y^{n})-I(C U^{n}; Y^{n})
=H(Y^{n}|CU^{n})-H(Y^{n}|X^{n}S^{n})\\
&\overset{(b)}{=} H(Y^{n}|CU^{n})-H(Y^{n}|CU^{n}X^{n}S^{n})
 = I(Y^{n};X^{n} S^{n}|C U^{n} ) \leq I(Y^{n} Z^{n};X^{n} S^{n}|C U^{n} )  \stepcounter{equation}\tag{\theequation}\label{convpart2}\\
&= I(Y^{n} Z^{n};X^{n} S^{n} U^{n}| C)-I(Y^{n} Z^{n};U^{n}| C)\overset{(c)}{\leq}  n I(Y_T {Z}_T;X_T S_T U_T|T) -n I( U_T;Y_{\sim T}{Z}_{\sim T} C T )
\end{align*}}where $(a)$ and $(b)$ come from the Markov chain $Y^{n}-(X^{n}, S^{n})-(C, U^{n})$. 
To prove $(c)$, we show separately that:
\begin{itemize}
 \item[(i)] $I(Y^{n} Z^{n};U^{n}| C) \geq n I( U_T;Y_{\sim T}{Z}_{\sim T} C T )$,
 \item[(ii)] $I(Y^{n} Z^{n};X^{n} S^{n} U^{n}| C) \leq  n I(Y_T {Z}_T;X_T S_T U_T |T)$.
\end{itemize}

\paragraph*{Proof of (i)}

Observe that 
{\allowdisplaybreaks
\begin{align*}
I(Y^{n} Z^{n};U^{n}| C)&=H(U^{n}|C)- H(U^{n}|Y^{n} Z^{n} C) \overset{(d)}{=} H(U^{n})- H(U^{n}|Y^{n} Z^{n} C)\\
&  \overset{(e)}{=} \sum_{t=1}^{n} \left( H(U_t)- H(U_t|U^{t-1} Y_t {Z}_t Y_{\sim t}{Z}_{\sim t} C)\right) \geq \sum_{t=1}^{n} \left(H(U_t)- H(U_t|Y_{\sim t}{Z}_{\sim t} C)\right) \\
&= n  H(U_T|T)- n H(U_T|Y_{\sim T}{Z}_{\sim T} C T)  \overset{(f)}{=}   n  H(U_T)  -  n H(U_T|Y_{\sim T}{Z}_{\sim T} C T)  \\
&=     n I( U_T;Y_{\sim T}{Z}_{\sim T} C T )
\end{align*}}where $(d)$ comes from the independence between $U^{n}$ and $C$ and $(e)$ and $(f)$ follow from the i.i.d. nature of $U^{n}$.

\paragraph*{Proof of (ii)}
First, we need the following result (proved in Appendix \ref{appendix ossmc1}.
\begin{lem}\label{ossmc1}
 For every $t\in \llbracket 1,n\rrbracket$ the following Markov chain holds:
 \begin{equation}\label{markovc1}
  (Y_t,{Z}_t) - (X_t,U_t, S_t) - (C, X_{\sim t},U_{\sim t}, S_{\sim t}, Y_{\sim t},{Z}_{\sim t}).
 \end{equation}
\end{lem}
Then, observe that
{\allowdisplaybreaks
\begin{align*}
I(Y^{n} Z^{n};X^{n} S^{n} U^{n}| C)&\leq I(Y^{n} Z^{n};X^{n} S^{n} U^{n} C)\\
&=\sum_{t=1}^{n}  I(Y_t {Z}_t ;X^{n} S^{n} U^{n} C|Y^{t-1} Z^{t-1})  \leq \sum_{t=1}^{n}  I(Y_t {Z}_t; X^{n} S^{n} U^{n} C Y^{t-1} Z^{t-1})\\
&= \sum_{t=1}^{n} I(Y_t {Z}_t;X_t S_t U_t ) +\sum_{t=1}^{n}  I(Y_t {Z}_t ;X_{\sim t} S_{\sim t} U_{\sim t} C Y^{t-1} Z^{t-1}|X_t S_t U_t)\\
& \overset{(g)}{=}  \sum_{t=1}^{n} I(Y_t {Z}_t;X_t S_t U_t)=n I(Y_T {Z}_T;X_T S_T U_T |T)
\end{align*}}where $(g)$  follows from Lemma \ref{ossmc1}. 
Moreover, since the distributions are $\varepsilon$-close to i.i.d. by hypothesis, the last term is close to $n I(Y {Z};X S U)$.
In fact, we have
{\allowdisplaybreaks
\begin{align*}
 I(Y_T {Z}_T;X_T S_T U_T |T)& =H(Y_T {Z}_T|T)+H(X_T S_T U_T |T)-H(Y_T {Z}_T X_T S_T U_T |T)\\
 &=\sum_{t=1}^{n} \frac{1}{n} H(Y_t {Z}_t)+ \sum_{t=1}^{n} \frac{1}{n} H(X_t S_t U_t ) -\sum_{t=1}^{n} \frac{1}{n} H(Y_t {Z}_t X_t S_t U_t).
\end{align*}
}Then, as in the proof of \cite[Lemma VI.3]{cuff2013distributed}, 
{\allowdisplaybreaks
\begin{align*}
 &\lvert H(Y_t {Z}_t)-H(Y {Z}) \rvert \leq 2\varepsilon \log{\left(\frac{\lvert \mathcal Y \times \mathcal Z \rvert}{\varepsilon}\right)}:= \varepsilon_1,\\
 &\lvert H(X_t S_t U_t)-H(X S U) \rvert \leq 2\varepsilon \log{\left(\frac{\lvert \mathcal X  \times \mathcal S  \times \mathcal U \rvert}{\varepsilon}\right)}:= \varepsilon_2,\\
 &\lvert H(Y_t {Z}_t X_t S_t U_t)-H(Y {Z}X S U) \rvert \leq 2\varepsilon \log{\left(\frac{\lvert \mathcal Y  \times \mathcal Z  \times \mathcal X  \times \mathcal S  \times \mathcal U \rvert}{\varepsilon}\right)}:= \varepsilon_3.
\end{align*}
}This implies that 
\begin{equation}\label{g epsilon}
\lvert I(Y_T {Z}_T;X_T S_T U_T |T)-  I(Y {Z};X S U)\rvert \leq g(\varepsilon),
\end{equation}where $g(\varepsilon):= (\varepsilon_1+ \varepsilon_2+ \varepsilon_3)$. 
Then, \eqref{convpart2} becomes $0 \leq n I(Y {Z};X S U) -n I( U_T;Y_{\sim T}{Z}_{\sim T} C T )+g(\varepsilon)$.

\vspace{0,2cm}
\paragraph{Proof of \eqref{outer 2}}

In this case, for the second part of the converse, we have
{\allowdisplaybreaks
\begin{align*}
 0 & \overset{(a)}{\leq} I(X^{n} S^{n};Y^{n})-I(CZ^{n} U^{n};Y^{n}) \overset{(b)}{\leq} I(X^{n} S^{n};Y^{n})-I( U^{n};Y^{n} C|Z^{n}) \\
& = H(Y^{n})-H(Y^{n}|X^{n} S^{n})-H(U^{n}) + I(U^{n};Z^{n} ) + H(U^{n}|Y^{n} Z^{n} C)\\
& \overset{(c)}{\leq} \sum_{t=1}^{n} H(Y_t)  -  \sum_{t=1}^{n} H(Y_t|X_t S_t) -  \sum_{t=1}^{n} H(U_t)  +  \sum_{t=1}^{n} I(U_t;Z_t)+ \sum_{t=1}^{n} H(U_t|U^{t-1} Y_t {Z}_t Y_{\sim t}{Z}_{\sim t} C)\\
&\overset{(d)}{\leq}  n H(Y_T)  - n  H(Y_T|X_T S_T T)- n  H(U_T|T)  +    n I(U_T;Z_T|T) +  n H(U_T|Y_{\sim T}{Z}_{\sim T} C T)  \\
& \overset{(e)}{=} n H(Y_T) -n H(Y_T|X_T S_T) - n  H(U_T)  +    n I(U_T;Z_T)  + nH(U_T|Y_{\sim T} {Z}_{\sim T} C T)\\
&= n I(X_T, S_T; Y_T)-n I( U_T;Y_{\sim T} {Z}_{\sim T} C T) +  n I(U_T;Z_T) 
\end{align*}}where $(a)$ comes from the Markov chain $Y^{n}-(X^{n}, S^{n})-(C,Z^{n}, U^{n})$, $(b)$ 
from the fact that 
\begin{align*}
 I(CZ^{n} U^{n};Y^{n}) \geq I(Z^{n} U^{n};Y^{n}|C)=I(Z^{n} U^{n};Y^{n} C) \geq I( U^{n};Y^{n} C|Z^{n})
\end{align*}
by the chain rule and the fact that $U^{n}$ and $ Z^{n} $ are independent of $C$.
Then $(c)$ comes from the chain rule for the conditional entropy. 
The inequalities $(d)$ comes from the fact that conditioning does not increase entropy (in particular $H(Y_T|T) \leq H(Y_T)$) and 
$(e)$ from the memoryless channel $\bar P_{Y|XS}$ and the i.i.d. source $\bar P_{UZ}$.
Finally, since the source is i.i.d. the last term is $n I(U;Z)$.

\begin{oss}
Note that if $U$ is independent of $Z$ the upper bound for $I(U;W)$ is $I(XS;Y)$.
\end{oss}

\vspace*{0.2cm}
\subsubsection{Identification of the auxiliary random variable}\label{identification2}
For each $t \in \llbracket 1,n\rrbracket$ we identify the auxiliary random variables $W_t$ with $(C,Y_{\sim t}, {Z}_{\sim t})$  and $W$ with
$(W_T, T)=(C,Y_{\sim T},{Z}_{\sim T}, T)$. 

The following Markov chains hold for each $t \in \llbracket 1,n\rrbracket$: 
 {\allowdisplaybreaks
\begin{align}
 {Z}_t -(U_t, S_t)-(C, X_t, Y_t, Y_{\sim t}, Z_{\sim t}) \quad &\Longleftrightarrow \quad {Z}_t -(U_t, S_t)-(X_t, Y_t, W_t), \label{mc11t}\\
 Y_t-(X_t, S_t) -(C, Y_{\sim t},{Z}_{\sim t}, U_t)  \quad &\Longleftrightarrow \quad  Y_t-(X_t, S_t) -(W_t, U_t),  \label{mc21t}\\ 
 V_t-(C,Y_{\sim t},{Z}_{\sim t},Y_t, {Z}_t)-(U_t, {S}_t,X_t) \quad &\Longleftrightarrow \quad  V_t-(W_t,Y_t, {Z}_t)-(U_t, {S}_t,X_t)\label{mc31t}.
\end{align}
}
Then we have
 {\allowdisplaybreaks
\begin{align}
{Z}_T -(U_T, S_T)-(C, X_T, Y_T, Y_{\sim T}, Z_{\sim T}, T) \quad &\Longleftrightarrow \quad  {Z}_T -(U_T, S_T)-(X_T, Y_T, W_T, T), \label{mc11T}\\
Y_T-(X_T, S_T) -(C, Y_{\sim T},{Z}_{\sim T}, U_T, T) \quad &\Longleftrightarrow \quad Y_T-(X_T, S_T) -(W_T, U_T, T), \label{mc21T} \\ 
V_T-(C,Y_{\sim T},{Z}_{\sim T},Y_T, {Z}_T, T)-(U_T, {S}_T,X_T) \quad &\Longleftrightarrow \quad V_T-(W_T,Y_T, {Z}_T, T)-(U_T, {S}_T,X_T).\label{mc31T} 
\end{align}
}
where \eqref{mc11T} and \eqref{mc21T} come from the fact that
{\allowdisplaybreaks
\begin{align*}
\mathbb P \{Z_T=z| S_T=s, U_T=u, X_T=x, Y_T=y, Y_{\sim T}=\tilde {\mathbf y} , Z_{\sim T}=\tilde {\mathbf z}, T=t, C\}&=  \mathbb P \{Z_T=z| S_T=s, U_T=u\},\\
\mathbb P \{Y_T=y| X_T=x, S_T=s, Y_{\sim T}=\tilde {\mathbf y} ,Z_{\sim T}=\tilde {\mathbf z}, U_T=u, T=t, C\}&=  \mathbb P \{Y_T=y| X_T=x, S_T=s\}
\end{align*}}
since the source is i.i.d. and the channel is memoryless.
Then by \eqref{mc31t}, \eqref{mc31T} holds because 
{\allowdisplaybreaks
\begin{align*}
I(V_T;U_T {S}_T X_T| C Y^{n} Z^{n} T)= \sum_{i=1}^n \frac{1}{n} I(V_t;U_t S_t X_t| C Y^{n} Z^{n} T=t)=0.
\end{align*}
}
Since $W=W_t$ when $T=t$, we also have $Z-(U,S)-(X,Y,W)$, $(U,S,X)-(W,Y, Z)-V$ and $Y-(X,S)-(W,U)$. 
The cardinality bound is proved in $\mbox{Appendix \ref{appendix bounds}}$.

\section{Strong coordination region for special cases}\label{sec: special cases}

Although the inner and outer bounds in Theorem \ref{teouv} do not match in general, we characterize the strong coordination region exactly in three special cases: 
perfect channel, lossless decoding and separation between the channel and the source. 

The empirical coordination region for these three settings was derived in \cite{treust2015empirical}. 
In this section we recover the same information constraints as in \cite{treust2015empirical}, 
but we show that for strong coordination a positive rate of common randomness is also necessary.
This reinforces the conjecture, stated in \cite{cuff2010}, that with enough common randomness
the strong coordination capacity region is the same as the empirical coordination capacity region for any specific network setting.

\subsection{Perfect channel}\label{perfect channel}
 \begin{center}
\begin{figure}[ht]
 \centering
 \includegraphics[scale=0.21]{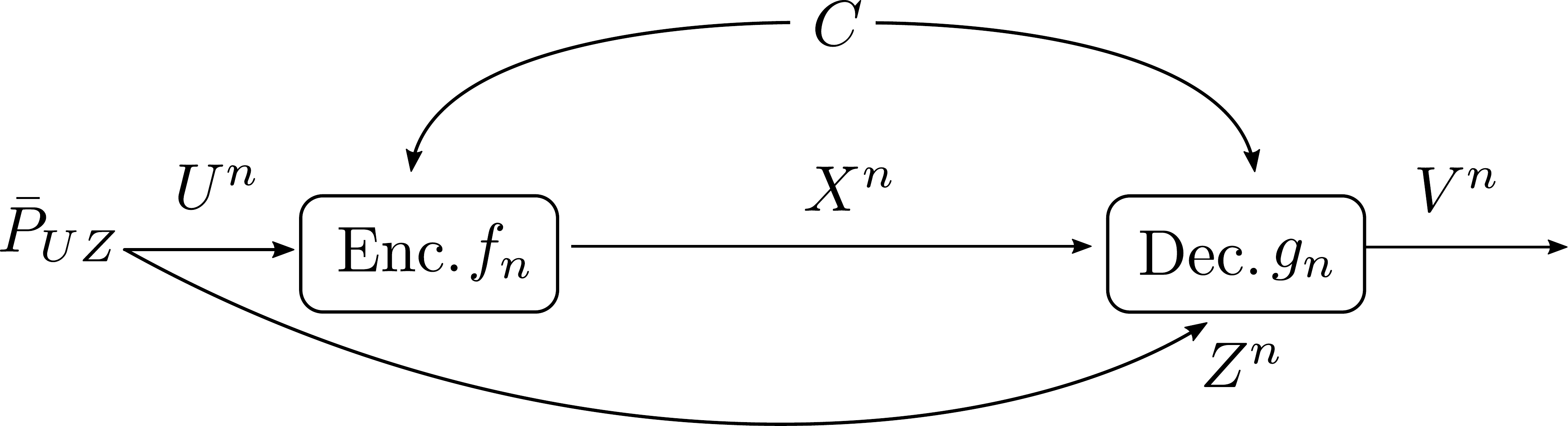}
\caption{Coordination of signals and actions for a two-node network with a perfect channel.}
\label{fig: pc}
\end{figure}
\end{center}
\vspace{-0,7cm}

Suppose we have a perfect channel as in Figure \eqref{fig: pc}. In this case $X^n=Y^n$ and the variable $Z^n$ plays the role of side information at the decoder.
We characterize the strong coordination region $\mathcal R_{\text{PC}}$.
\begin{teo} \label{teopc}
In the setting of Theorem \ref{teouv}, suppose that $\bar P_{Y|XS}( \mathbf{y}|\mathbf{x},\mathbf{s})={\mathds 1}_{X=Y} \{\mathbf{x}= \mathbf{y} \}$. 
Then the strong coordination region is
\begin{equation}\label{eq: regionpc}
\mathcal R_{\text{PC}} := \begin{Bmatrix}[c|l]
 &  \bar P_{UZXV}= \bar P_{UZ} \bar P_{X|U} \bar P_{V|UXZ} \\
 & \exists \mbox{ } W \mbox{ taking values in $\mathcal W$}\\ 
(\bar P_{UZXV}, R_0) & \bar P_{UZWXV}= \bar P_{UZ} \bar P_{W|U} \bar P_{X|UW}  \bar P_{V|WXZ}\\
 & I(WX;U) \leq H(X)+I(W;Z|X)\\
 & R_0 \geq I(W;UV|XZ)\\ 
 & \lvert \mathcal W \rvert \leq \lvert \mathcal U \times \mathcal Z \times \mathcal X  \times {\mathcal V} \rvert+4 \\
\end{Bmatrix}
\end{equation}
\end{teo}

\begin{oss}
 Observe that the decomposition of the joint distributions $\bar P_{UZXV}$ and $\bar P_{UZWXV}$
 is equivalently characterized in terms of Markov chains: 
 {\allowdisplaybreaks
\begin{equation}\label{markov chain pc}
   Z-U-X,
   \quad \quad
 \begin{cases}
Z-U-(X,W)\\
V-(X,Z,W)-U
   \end{cases}.
\end{equation}
}

\end{oss}

\subsubsection{Achievability}
We  show that $\mathcal R_{\text{PC}}$ is contained in the region $\mathcal R_{\text{in}}$ defined in \eqref{eq: region inn}  and thus it is achievable. 
We note $\mathcal R_{\text{in}}(W)$ the subset of $\mathcal R_{\text{in}}$ for a fixed $W \in \mathcal W$ that satisfies:
{\allowdisplaybreaks
\begin{align*}
&\bar P_{USZWXYV}=\bar P_{USZ} \bar P_{W|U} \bar P_{X|UW} \bar P_{Y|XS} \bar P_{V|WYZ}\\
&I(W;U) \leq I(W;YZ) \stepcounter{equation}\tag{\theequation}\label{constrgen} \\
& R_0 \geq I(W;USXV|YZ)
\end{align*}
}Then the set $\mathcal R_{\text{in}}$ is the union over all the possible choices for $W$ that satisfy the \eqref{constrgen}. Similarly, $\mathcal R_{\text{PC}}$ is the union of all $\mathcal R_{\text{PC}}(W)$ with $W$ that satisfies
{\allowdisplaybreaks
\begin{align*}
 &\bar P_{UZWXV}= \bar P_{UZ} \bar P_{W|U} \bar P_{X|UW}  \bar P_{V|WXZ}\\
 & I(W,X;U) \leq H(X)+I(W;Z|X) \stepcounter{equation}\tag{\theequation}\label{constrpc} \\
 & R_0 \geq I(W;UV|XZ)
\end{align*}
}
Let $(\bar P_{UZXV}, R_0) \in \mathcal R_{\text{PC}}(W)$ for some $W \in \mathcal W$.
Then $W$ verifies the Markov chains 
$Z-U-(X,W)$ and $V-WXZ-U$ and the 
information constraints for  $ \mathcal R_{\text{PC}}$. 
Note that $(\bar P_{UZXV}, R_0) \in \mathcal R_{\text{in}}(W')$, where $W'=(W,X)$.
The Markov chains are still valid and the information constraints in \eqref{constrpc} imply 
the information constraints for $\mathcal R_{\text{in}}(W')$ since:
{\allowdisplaybreaks
\begin{align*}
 I(W';U)&=I(W,X;U) \leq H(X)+I(W;Z|X)\\
 &=I(W,X;X)+I(W,X;Z|X) = I(W,X;XZ) = I(W';XZ), \stepcounter{equation}\tag{\theequation}\label{mael2015pc} \\
 R_0& \geq I(W;UV|XZ)=I(WX;UV|XZ) .
\end{align*}
}
Then $(\bar P_{UZXV}, R_0) \in \mathcal R_{\text{in}}(W')$ and if we consider the union over all suitable $W$, 
we have $$\bigcup_{W} \mathcal R_{\text{PC}}(W) \subseteq \bigcup_{(W,X)}  \mathcal R_{\text{in}}(W,X) \subseteq \bigcup_{W}  \mathcal R_{\text{in}}(W).$$
Finally, $  \mathcal R_{\text{PC}} \subseteq  \mathcal R_{\text{in}}$.

\begin{oss}
In the case of perfect channel, \cite[Section IV.A]{treust2015empirical} characterizes the 
empirical coordination region and the information constraint is $I(W,X;U) \leq H(X)+I(W;Z|X)$ as in \eqref{eq: regionpc}.
\end{oss}

\vspace{0,2cm}
\subsubsection{Converse}

Consider a code $(f_n,g_n)$ that induces a distribution $P_{U^{n} S^{n} Z^{n} X^{n}  V^{n}}$ that is $\varepsilon$-close in total variational distance to the i.i.d. distribution $\bar P_{U S Z X V}^{\otimes n}$.
Let $T$ be the random variable defined in Section \ref{outer}.

We would like to prove that 
\begin{equation*}
 0 \leq H(X)+I(W;Z|X)-I(W,X;U)=I(W,X;X Z)-I(W,X;U).
\end{equation*}
The following proof is inspired by \cite{treust2015empirical}.
We have
{\allowdisplaybreaks
\begin{align*}
 & 0  =  H(X^{n}  ,Z^{n}) - I(X^{n} Z^{n} ; U^{n} C) - H(X^{n}  Z^{n}|U^{n} C)\\
 & \overset{(a)}{\leq} \sum_{t=1}^{n} H(X_t ,{Z}_t) - \sum_{t=1}^{n} I(X^{n} Z^{n}; U_t|U_{t+1}^n C)  - H(X^{n} Z^{n}|U^{n} C)\\
 & \overset{(b)}{=} \sum_{t=1}^{n} I(X^{n} Z^{n} C;X_t {Z}_t) -\sum_{t=1}^{n} I(X^{n} Z^{n} U_{t+1}^n C; U_t)  +\sum_{t=1}^{n} I(U_{t+1}^n C; U_t)- H(X^{n} Z^{n}|U^{n} C)\\
 & \overset{(c)}{=} \sum_{t=1}^{n} I(X^{n} Z^{n} C;X_t {Z}_t) -\sum_{t=1}^{n} I(X^{n} Z^{n} U_{t+1}^n C; U_t)  - H(X^{n} Z^{n}|U^{n} C)\\
 & \leq \sum_{t=1}^{n} I(X^{n} Z^{n} C;X_t {Z}_t) -\sum_{t=1}^{n} I(X^{n} Z^{n} C; U_t) - H(X^{n} Z^{n}|U^{n} C)\\
 & \overset{(d)}{=}  \sum_{t=1}^{n} I(X^{n} {Z}_{\sim t} C;X_t {Z}_t)+ \sum_{t=1}^{n} I (Z_t;X_t {Z}_t| X^{n} {Z}_{\sim t} C) -\sum_{t=1}^{n} I(X^{n} {Z}_{\sim t} C; U_t) \\
 & \quad  -\sum_{t=1}^{n} I (Z_t; U_t| X^{n} {Z}_{\sim t} C)- H(X^{n} Z^{n}|U^{n} C)\\
 & = \sum_{t=1}^{n} I(X^{n} {Z}_{\sim t} C;X_t {Z}_t)-\sum_{t=1}^{n} I(X^{n} {Z}_{\sim t} C; U_t)  - H(X^{n} Z^{n}|U^{n} C) + \sum_{t=1}^{n} H(Z_t|X^{n} {Z}_{\sim t} C) \\
 &\quad - \sum_{t=1}^{n} H(Z_t| X^{n} Z^{n} C)  - \sum_{t=1}^{n} H(Z_t|X^{n} {Z}_{\sim t} C)  + \sum_{t=1}^{n} H(Z_t|U_t X^{n} {Z}_{\sim t} C)\\
 & = \sum_{t=1}^{n} I(X^{n} {Z}_{\sim t} C;X_t {Z}_t)-\sum_{t=1}^{n} I(X^{n} {Z}_{\sim t} C; U_t) + \sum_{t=1}^{n} H(Z_t|U_t X^{n} {Z}_{\sim t} C)- H(X^{n} Z^{n}|U^{n} C)\\
 & \overset{(e)}{\leq} \sum_{t=1}^{n} I(X^{n} {Z}_{\sim t} C;X_t {Z}_t)-\sum_{t=1}^{n} I(X^{n} {Z}_{\sim t} C; U_t)+ \sum_{t=1}^{n} H(Z_t|U_t C)- H(Z^{n}|U^{n} C)\\
 & \overset{(f)}{=}  \sum_{t=1}^{n} \!I(X^{n} {Z}_{\sim t} C; X_t {Z}_t)\!-\!\sum_{t=1}^{n} I(X^{n} {Z}_{\sim t} C; U_t) \\
 & = n I(X^{n} {Z}_{\sim T} C;X_T {Z}_T|T) -n I(X^{n} {Z}_{\sim T} C; U_T|T) \\
 & \leq n I(X^{n} {Z}_{\sim T} C T; X_T {Z}_T)  -n I(X^{n} {Z}_{\sim T} C T; U_T) + n  I(T; U_T)\\
 & \overset{(g)}{=} n I(X_T {X}_{\sim T} {Z}_{\sim T} C T; X_T {Z}_T)  - n I(X_T {X}_{\sim T} {Z}_{\sim T} C T; U_T)
 \end{align*}}where $(a)$ and $(b)$ follow from the properties of the mutual information and $(c)$ comes from
the independence between $U^n$ and $C$ and
the i.i.d. nature of the source.
Then $(d)$ comes from the chain rule, $(e)$ from the properties of conditional entropy, $(f)$ from 
the independence between $(U^n,Z^n)$ and $C$ and
the i.i.d. nature of the source.
Finally, $(g)$ comes from the fact that $I(T; U_T)$ is zero due to the i.i.d. nature of the source.

We identify the auxiliary random variable $W_t$ with $(C,X_{\sim t}, {Z}_{\sim t})$ for each $t \in \llbracket 1,n\rrbracket$ and $W$ with $(W_T, T)=$ $(C,X_{\sim T}, {Z}_{\sim T}, T)$. 
Observe that with this identification of $W$ the bound for $R_0$ follows from Section \ref{genconvpart1} with the substitution $Y=X$. 
Moreover, the following Markov chains are verified for each $t \in \llbracket 1,n\rrbracket$:
{\allowdisplaybreaks
\begin{align*}
& {Z}_t-U_t-(W_t, X_t)\\
& V_t-(W_t, X_t, {Z}_t)-U_t
\end{align*}}
The first one holds because the source is i.i.d. and $Z_t$ does not belong to $W_t$. The second Markov chain follows from the fact that 
$V$ is generated using $C$, $X^{n}$ and $Z^{n}$ that are included in $(W_t, X_t, {Z}_t)= (C,X_{\sim t}, {Z}_{\sim t}, X_t, {Z}_t)$.
With a similar approach as in Section \ref{identification} and Section \ref{identification2}, the Markov chains with $T$ hold.
Then since $W=W_t$ when $T=t$, we also have $ {Z}-U-(W, X)$ and $V-(W, X, {Z})-U $. 
The cardinality bound is proved in $\mbox{Appendix \ref{appendix bounds}}$.

\subsection{Lossless decoding}\label{lossless decoding}
 \begin{center}
\begin{figure}[ht]
 \centering
 \includegraphics[scale=0.21]{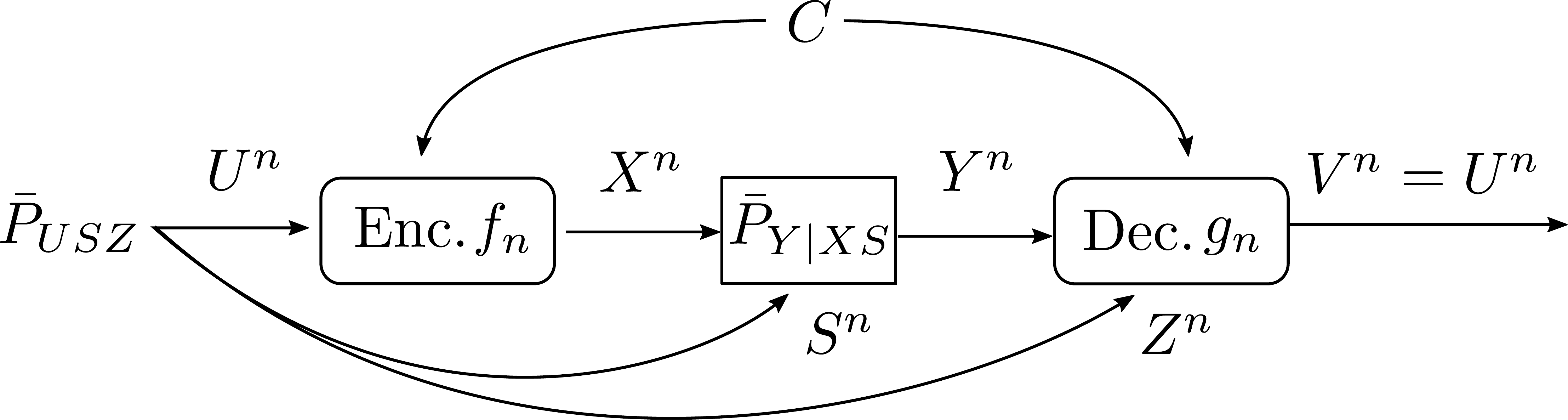}
\caption{Coordination of signals and actions for a two-node network with a noisy channel and a lossless decoder.}
\label{fig: ld}
\end{figure}
\end{center}
\vspace{-0,7cm}

Suppose that the decoder wants to reconstruct the source losslessly, i.e., $V=U$ as in Figure \ref{fig: ld}. 
Then, we characterize the strong coordination region $\mathcal R_{\text{LD}}$.

\begin{teo} \label{teolossless}
Consider the setting of Theorem \ref{teouv} and suppose that $\bar P_{V|USXYZ}( \mathbf{v}|\mathbf{u},\mathbf{s},\mathbf{x},\mathbf{y}, \mathbf{z})=\mathds 1_{ V =U } \{\mathbf{u} =\mathbf{v}  \}$. Then the strong coordination region is
\begin{equation}\label{eq: regionld}
\mathcal R_{\text{LD}} := \begin{Bmatrix}[c|l]
 & \bar P_{USZXYV}= \bar P_{USZ} \bar P_{X|U} \bar P_{Y|XS} \mathds 1_{ V =U }\\
 &\exists \mbox{ } W \mbox{ taking values in $\mathcal W$}\\ 
 (\bar P_{USZXY}, R_0) &\bar P_{USZWXYV}=\bar P_{USZ} \bar P_{W|U} \bar P_{X|UW}  \bar P_{Y|XS} \mathds 1_{ V =U   }\\
 & I(W;U) \leq I(W;YZ)\\
 & R_0 \geq I(W;USX|YZ)\\
 & \lvert \mathcal W \rvert \leq \lvert \mathcal U \times \mathcal S  \times \mathcal Z \times \mathcal X \times \mathcal Y  \rvert+3
\end{Bmatrix}
\end{equation}
\end{teo}

\vspace{0,3cm}
\begin{oss}
 Observe that the decomposition of the joint distributions  $\bar P_{USZXYV}$  and $\bar P_{USZWXYV}$
 is equivalently characterized in terms of Markov chains: 
 {\allowdisplaybreaks
\begin{equation}\label{markov chain lossless}
 \begin{cases}
   Z-(U,S)-(X,Y)\\
   Y-(X,S)-U
   \end{cases},\quad
 \begin{cases}
Z-(U,S)-(X,Y,W)\\
Y-(X,S)-(U,W)
   \end{cases}.
\end{equation}
}

\end{oss}
\vspace*{0.3cm}
\subsubsection{Achievability}

We show that $\mathcal R_{\text{LD}} \subseteq \mathcal R_{\text{in}}$ and thus it is achievable. 
Similarly to the achievability proof in Theorem \ref{teopc} , 
let $(\bar P_{UZXV}, R_0) \in \mathcal R_{\text{LD}}(W)$ for some $W \in \mathcal W$.
Then, $W$ verifies the Markov chains $Z-(U,S)-(X,Y,W)$ and $Y-(X,S)-(U,Z,W)$ and the 
information constraints for  $ \mathcal R_{\text{LD}}$. 
We want to show that  $(\bar P_{UZXV}, R_0) \in \mathcal R_{\text{in}}(W)$.
Observe that the Markov chains are still valid.
Hence, the only difference is the bound on $R_0$, but  $I(W;USXV|YZ)$ $=$ $ I(W;USX|YZ)$ when $U=V$.
Then, $(\bar P_{UZXV}, R_0) \in \mathcal R_{\text{in}}(W)$ and if we consider the union over all suitable $W$, 
we have $$\bigcup_{W} \mathcal R_{\text{LD}}(W) \subseteq \bigcup_{W} \mathcal R_{\text{in}}(W).$$
Finally, $ \mathcal R_{\text{LD}} \subseteq  \mathcal R_{\text{in}}$.

\vspace*{0.2cm}
\subsubsection{Converse}
Consider a code $(f_n,g_n)$ that induces a distribution $P_{U^{n} S^{n} Z^{n} X^{n} Y^{n} V^{n}}$ that is $\varepsilon$-close in total variational
distance to the i.i.d. distribution $\bar P_{U S Z X Y}^{\otimes n} \mathds 1_{ V =U }^{\otimes n}$.
Let $T$ be the random variable defined in Section \ref{outer}.

 We have
{\allowdisplaybreaks
\begin{align*} 
 nR_0 & \geq H(C) \geq H(C|Y^{n} Z^{n}) =H(C U^{n}|Y^{n} Z^{n}) - H( U^{n}|C Y^{n} Z^{n}) \\
 & \overset{(a)}{\geq} H(C U^{n}|Y^{n} Z^{n}) -  n f(\varepsilon)
\geq  I(U^{n} S^{n} X^{n};C U^{n}|Y^{n} Z^{n})  -  n f(\varepsilon)\\
&=\sum_{t=1}^{n}   I(U_t S_t X_t;C U^{n} \!| U^{t-1}  S^{t-1} X^{t-1} Y_{\sim t}  Z_{\sim t} Y_t  Z_t)    -  n f(\varepsilon)\\
& \overset{(b)}{\geq}\! \sum_{t=1}^{n} I(U_t S_t X_t;C U^{n} Y_{\sim t}  Z_{\sim t} U^{t-1}  S^{t-1} X^{t-1} | Y_t  Z_t) -2nf(\varepsilon)\\
& \geq \sum_{t=1}^{n} I(U_t S_t X_t; C U^{n} Y_{\sim t}  Z_{\sim t} | Y_t  Z_t) -2nf(\varepsilon)
  = n I(U_T S_T X_T;   C U^{n} Y_{\sim T}  Z_{\sim T}  | Y_T  Z_T T)  -2nf(\varepsilon) \\
& = n I(U_T S_T X_T;   C U^{n} Y_{\sim T}  Z_{\sim T}  T | Y_T  Z_T) - n I(U_T S_T X_T;   T | Y_T  Z_T ) -2nf(\varepsilon)\\
& = n I(U_T S_T X_T;   C U^{n} Y_{\sim T}  Z_{\sim T}  T | Y_T  Z_T) - n I(U_T S_T X_T Y_T  Z_T;  T ) + n I(Y_T  Z_T;  T )-2nf(\varepsilon)\\
& \overset{(c)}{\geq} n I(U_T S_T X_T;   C U^{n} Y_{\sim T}  Z_{\sim T}  T | Y_T  Z_T) -3nf(\varepsilon)
\end{align*}}where $(a)$ follows Fano's inequality which implies that
\begin{equation}\label{byfano}
  H( U^{n}|C Y^{n} Z^{n}) \leq n f(\varepsilon)
\end{equation}
as proved in Appendix \ref{appendix fano}. 
To prove $(b)$ observe that  
\begin{align*}
 I(U_t S_t X_t;C U^{n} \!| U^{t-1}  S^{t-1} X^{t-1} Y_{\sim t}  Z_{\sim t} Y_t  Z_t) =& I(U_t S_t X_t;C U^{n} Y_{\sim t}  Z_{\sim t} U^{t-1}  S^{t-1} X^{t-1} | Y_t  Z_t)\\
 &- I(U_t S_t X_t;Y_{\sim t}  Z_{\sim t}  U^{t-1}  S^{t-1} X^{t-1}|  Y_t  Z_t) 
\end{align*}and  $I(U_t S_t X_t;Y_{\sim t}  Z_{\sim t}  U^{t-1}  S^{t-1} X^{t-1}|  Y_t  Z_t) \leq f(\varepsilon) $ by Lemma \ref{lemab}. 
Finally, $(c)$ comes from  the fact, proved in \cite[Lemma VI.3]{cuff2013distributed}, that
 $I(U_T S_T X_T Y_T  Z_T;  T )$ vanishes since the distribution is $\varepsilon$-close to i.i.d. by hypothesis.
With the identifications  $W_t=(C,U^{n}, Y_{\sim t}, {Z}_{\sim t})$ for each $t \in \llbracket 1,n\rrbracket$ and $W=(W_T, T)=(C,U^{n},Y_{\sim T},{Z}_{\sim T}, T)$, we have
$R_0 \geq I(W;USX|YZ)$.

For the second part of the converse, we have
{\allowdisplaybreaks
\begin{align*}
  n I(U; W ) & \leq n H(U)= H(U^{n}) = H(U^{n}|C)= I(U^{n} ; Y^{n} Z^{n} C )+ H(U^{n} | Y^{n} Z^{n} C) \\
&\overset{(d)}{\leq}\sum_{t=1}^n I(U^{n} ; Y_t Z_t | Y^{t-1} Z^{t-1} C) + n f(\varepsilon)
 \leq \sum_{t=1}^n I(U^{n} Y^{t-1} Z^{t-1} C ; Y_t Z_t) +n f(\varepsilon) \\
 & \leq \sum_{t=1}^n I(U^{n} Y_{\sim t} Z_{\sim t} C; Y_t Z_t) +n f(\varepsilon)
 = n I(U^{n} Y_{\sim T} Z_{\sim T} C; Y_T Z_T|T) +n f(\varepsilon)\\
& \leq n I(U^{n} Y_{\sim T} Z_{\sim T} C T; Y_T Z_T) +n f(\varepsilon)
 \overset{(e)}{=} n I( W; Y Z)+n f(\varepsilon)
 \end{align*}}where  $(d)$ comes from Fano's inequality and $(e)$ comes 
 from the identification $W=(C,U^{n},Y_{\sim T},{Z}_{\sim T}, T)$.

In order to complete the converse, we show that the following Markov chains hold for each $t \in \llbracket 1,n\rrbracket$:
{\allowdisplaybreaks
\begin{align*}
 & Y_t-(X_t,S_t)-(U_t,Z_t,W_t),\\
 & Z_t-(U_t,S_t)-(X_t,Y_t,W_t).
\end{align*}
}The first one is verified because the channel is memoryless and $Y_t$ does not belong to $W_t$ and the second one holds because 
of the i.i.d. nature of the source and because $Z_t$ does not belong to $W_t$.
With a similar approach as in Section \ref{identification} and Section \ref{identification2}, the Markov chains with $T$ hold.
Then, since $W=W_t$ when $T=t$, we  also have $ Y-(X,S)-(U,Z,W)$ and $Z-(U,S)-(X,Y,W) $. 
The cardinality bound is proved in $\mbox{Appendix \ref{appendix bounds}}$.

\begin{oss}
An equivalent characterization of the region is:
\begin{equation}\label{eq: regionldmael}
\mathcal R_{\text{LD}} := \begin{Bmatrix}[c|l]
& \bar P_{USZXY}= \bar P_{USZ} \bar P_{X|U} \bar P_{Y|XS} \\
& \exists \mbox{ } W \mbox{ taking values in $\mathcal W$}\\ 
(\bar P_{USZXY}, R_0) & \bar P_{USZWXY}= \bar P_{USZ} \bar P_{W|U} \bar P_{X|UW}  \bar P_{Y|XS} \\
& H(U)\leq I(WU;YZ)\\
& R_0 \geq I(W;USX|YZ)+H(U|WYZ)\\
& \lvert \mathcal W \rvert \leq \lvert \mathcal U \times \mathcal S  \times \mathcal Z \times \mathcal X \times \mathcal Y  \rvert+1 \\
\end{Bmatrix}
\end{equation}The region in \eqref{eq: regionldmael} is achievable since with the choice of the auxiliary random variable $W''=(W,U)$, the constraints in \eqref{eq: regionld} become 
 {\allowdisplaybreaks
\begin{align*}
& I(WU;U)=H(U) \leq I(WU;YZ) \stepcounter{equation}\tag{\theequation}\label{mael2015uv}\\
& R_0 \geq I(WU;USX|YZ) =I(W;USX|YZ)+I(U;USX|WYZ) \\
& \phantom{R_0}=I(W;USX|YZ)+H(U|WYZ)-H(U|USXWYZ) \stepcounter{equation}\tag{\theequation}\label{mael2015ro}\\
& \phantom{R_0}=I(W;USX|YZ)+H(U|WYZ).
\end{align*}
}
Moreover, the converse in the proof of Theorem \ref{teolossless} is still valid with the identification $W=(C,U_{\sim T}, Y_{\sim T},{Z}_{\sim T}, T)$.

Note that \cite[Section IV.B]{treust2015empirical} 
gives a characterization of the empirical coordination region and the constraint for 
the mutual information is
\begin{equation*}
0 \leq I(WU;YZ)-H(U)= I(WU;YZ)-H(U)-I(W;S|U)
\end{equation*}
which is the same as in \eqref{mael2015uv} because of the Markov chain $SZ-U-W$.
\end{oss}

\subsection{Separation between source and channel}\label{separation}
\begin{center}
\begin{figure}[ht]
\centering
\includegraphics[scale=0.21]{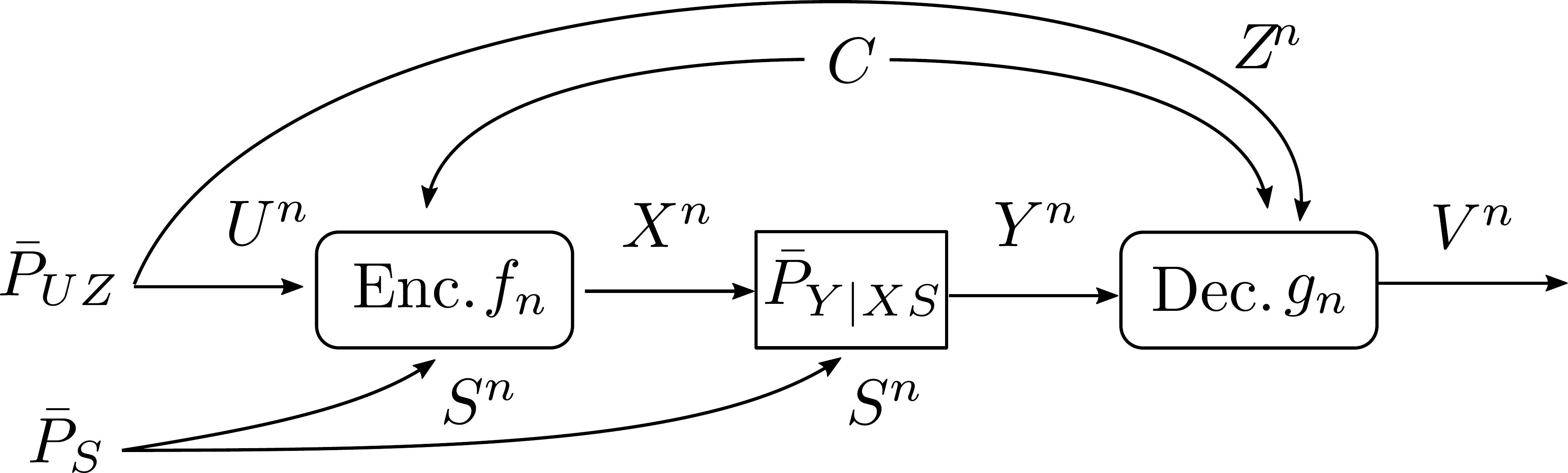}
\caption{Coordination of signals and actions for a two-node network with a noisy channel where the source is separated from the channel.}
\label{fig: sep}
\end{figure}
\end{center}

Suppose that the channel state $P_{S^n}$ is independent of the source and side information $P_{U^nZ^n}$, 
and that the target joint distribution is of the form $\bar{P}_{UZV}^{\otimes n}\bar{P}_{SXY}^{\otimes n}$.   
For simplicity, we will suppose that the encoder has perfect state information (see Figure \ref{fig: sep}). 
Then we characterize the strong coordination region $\mathcal R_{\text{SEP}}$.

Note that in this case the coordination requirements are three-fold: the random variables  $(U^n,Z^n,V^n)$
should be coordinated, the random variables $(S^n,X^n,Y^n)$ should be coordinated and finally  $(U^n,Z^n,V^n)$ should be independent of  $(S^n,X^n,Y^n)$.
We introduce two auxiliary random variables $W_1$ and $W_2$, where $W_2$ is used to accomplish the coordination of $(U^n,Z^n,V^n)$, 
while $W_1$ has the double role of ensuring the independence of source and state as well as coordinating $(S^n,X^n,Y^n)$.
\vspace{-0,7cm}

\begin{teo} \label{teoseparation}
Consider the setting of Theorem \ref{teouv} and suppose that $\bar P_{USXYZV}= \bar P_{UZV} \bar P_{SXY}$. Then, the strong coordination region is
\begin{equation}\label{eq: regionsep}
\mathcal R_{\text{SEP}} := \begin{Bmatrix}[c|l]
&\bar P_{USZXYV}= \bar P_{UZ} \bar P_{V|UZ} \bar P_{S} \bar P_{X|S} \bar P_{Y|XS} \\
&\exists \mbox{ } (W_1,W_2) \mbox{ taking values in $\mathcal W_1 \times \mathcal W_2$}\\ 
(\bar P_{USZXY}, R_0) &\bar P_{USZW_1 W_2XYV}=\bar P_{UZ} \bar P_{W_2|U} \bar P_{V|Z W_2} \bar P_{S} \bar P_{X|S} \bar P_{W_1|SX}  \bar P_{Y|XS} \\
& I(W_1;S) + I(W_2;U) \leq  I(W_1;Y) + I(W_2;Z)\\
&R_0 \geq I(W_1;SX|Y)+I(W_2;UV|Z)\\
&(\lvert \mathcal W_1 \rvert , \lvert \mathcal W_2 \rvert)\leq \lvert \mathcal U \times \mathcal S \times \mathcal Z \times  \mathcal X \times \mathcal Y  \times \mathcal V \rvert +3 .
\end{Bmatrix}
\end{equation}
\end{teo}

\begin{oss}
 Observe that the decomposition of the joint distribution $\bar P_{USZW_1 W_2XYV}$  
 is equivalently characterized in terms of Markov chains: 
 {\allowdisplaybreaks
\begin{equation}\label{markov chain separation}
 \begin{cases}
Z-U-W_2\\
Y-(X,S)-W_1\\
V-(Z,W_2)-U
   \end{cases}.
\end{equation}
}
\end{oss}

\subsubsection{Achievability}
We show that $\mathcal R_{\text{SEP}}$ is contained in the achievable region $\mathcal R_{\text{in}}$ in \eqref{eq: region inn} specialized to this specific setting. 
In this case we are also supposing that the encoder has perfect state information, 
i.e. the input of the encoder is the pair $(U^{n}, S^{n})$ as in Figure \ref{fig: sep} as well as common randomness $C$. 
The joint distribution $\bar P_{USZXYV}$ becomes $\bar P_{UZ} \bar P_{V|UZ} \bar P_{S} \bar P_{X|S} \bar P_{Y|XS}$
since $(U,Z,V)$ is independent of $(S,X,Y)$ and the Markov chains are still valid.

Observe that the set  $\mathcal R_{\text{in}}$ is the union over all the possible choices for $W$ that satisfy 
the joint distribution, rate and  information constraints in \eqref{eq: region inn}.
Similarly, $\mathcal R_{\text{SEP}}$ is the union of all $\mathcal R_{\text{SEP}}(W_1,W_2)$ with $(W_1,W_2)$ that satisfies
the joint distribution, rate and  information constraints in \eqref{eq: regionsep}.
Let $(\bar P_{USZXY}, R_0) \in \mathcal R_{\text{SEP}}(W_1,W_2)$ for some $(W_1,W_2)$ taking values in $\mathcal W_1 \times \mathcal W_2$.
Then, $(W_1,W_2) $ verifies the Markov chains 
$Z-U-W_2$, $V-( W_2, Z)- U$ and $Y-(S,X)-W_1$, and the 
information constraints for  $ \mathcal R_{\text{SEP}}$.
We will show that $(\bar P_{USZXY}, R_0) \in \mathcal R_{\text{in}}(W')$, where $W'=(W_1,W_2)$.
The information constraints in \eqref{eq: regionsep} imply 
the information constraints for $\mathcal R_{\text{in}}(W')$ since:
{\allowdisplaybreaks
\begin{align*}
 &I(W_1 W_2;YZ)- I(W_1W_2;US)\\
&= I(W_1; YZ)+ I(W_2;YZ|W_1) - I(W_1;US)- I(W_2; US|W_1)\\
&= I(W_1; Y)  + I(W_2;Y Z W_1) - I(W_1;S)- I(W_2;U S W_1)\\
 &= I(W_1; Y) + I(W_2;Z) - I(W_1;S)- I(W_2;U) \geq 0\\
& I(W_1W_2;USXV|YZ)\\
& =I(W_1;USXV|YZ) + I(W_2;USXV|YZW_1)\\
& = I(W_1;USXVZ|Y)+ I(W_2;USXVY|ZW_1)\\
&= I(W_1;SX|Y)+  I(W_2;USXVY|Z)\\
&=I(W_1;SX|Y)+  I(W_2;UV|Z) \leq R_0
\end{align*}}because by construction $W_1$ and $W_2$ are independent 
of each other and $W_1$ is independent of $(U,Z,V)$ and $W_2$ is independent of $(S,X,Y)$ .

Then $(\bar P_{USZXY}, R_0) \in \mathcal R_{\text{in}}(W_1, W_2)$ and if we consider the union over all suitable $(W_1, W_2)$, 
we have $$\bigcup_{(W_1, W_2)} \mathcal R_{\text{SEP}}(W_1, W_2) \subseteq \bigcup_{(W_1, W_2)}  \mathcal R_{\text{in}}(W_1, W_2) \subseteq \bigcup_{W}  \mathcal R_{\text{in}}(W).$$
Finally, $  \mathcal R_{\text{SEP}} \subseteq  \mathcal R_{\text{in}}$.

\vspace{0,2cm}
\subsubsection{Converse}
Let $T$ be the random variable defined in Section \ref{outer}.
Consider a code $(f_n,g_n)$ that induces a distribution $P_{U^{n} S^{n} Z^{n} X^{n} Y^{n} V^{n}}$ that
is $\varepsilon$-close in total variational distance to the i.i.d. distribution $\bar P_{U Z V}^{\otimes n} \bar P_{SXY}^{\otimes n}$.
Then, we have
\begin{equation*}
 \mathbb V (P_{S^{n} X^{n} Y^{n} Z^{n} U^{n} V^{n}}, \bar P_{UZV}^{\otimes n} \bar P_{SXY}^{\otimes n}) < \varepsilon.
\end{equation*}
If we apply Lemma \ref{lemmit} to $A=S^{n} X^{n} Y^{n}$ and $B=Z^{n} U^{n} V^{n}$, we have 
\begin{equation}\label{lemmitsep}
I(S^{n} X^{n} Y^{n}; Z^{n} U^{n} V^{n}) < f(\varepsilon).
\end{equation}
Then, we have
{\allowdisplaybreaks
\begin{align*} 
 nR_0 &\geq H(C) \overset{(a)}{\geq} I(U^{n} S^{n} X^{n} V^{n};C|Y^{n} Z^{n})= I(S^{n} X^{n} ;C|Y^{n} Z^{n} U^{n} V^{n})  + I(U^{n}  V^{n};C|Y^{n} Z^{n}) \\
& =  I(S^{n} X^{n} ;C Z^{n} U^{n} V^{n} |Y^{n})  - I(S^{n} X^{n} ; Z^{n} U^{n} V^{n} |Y^{n}) + I(U^{n}  V^{n};C Y^{n}| Z^{n}) - I(U^{n}  V^{n}; Y^{n}| Z^{n})  \\
& \overset{(b)}{\geq}  I(S^{n} X^{n} ;C Z^{n} U^{n} V^{n} |Y^{n})  + I(U^{n}  V^{n};C Y^{n}| Z^{n})- 2 f(\varepsilon)\\
& \overset{(c)}{=}  \sum_{t=1}^n I(S_t X_t ;C Z^{n} U^{n} V^{n} |S_{t+1}^n X_{t+1}^n Y_t Y_{\sim t})+  \sum_{t=1}^n I(U_t  V_t ;C Y^{n}| U^{t-1} V^{t-1} Z_t Z_{\sim t}) - 2 f(\varepsilon)\\
& = \sum_{t=1}^n I(S_t X_t ;C Z^{n} U^{n} V^{n} S_{t+1}^n X_{t+1}^n  Y_{\sim t} | Y_t) - \sum_{t=1}^n I(S_t X_t ; S_{t+1}^n X_{t+1}^n  Y_{\sim t} | Y_t) \\
&\quad  +  \sum_{t=1}^n I(U_t  V_t ;C Y^{n} U^{t-1} V^{t-1}  Z_{\sim t} |Z_t) - \sum_{t=1}^n I(U_t  V_t ; U^{t-1} V^{t-1}  Z_{\sim t} |Z_t) - 2n f(\varepsilon)\\
& \overset{(d)}{\geq}   \sum_{t=1}^n I(S_t X_t ;C Z^{n} U^{n} V^{n} S_{t+1}^n X_{t+1}^n  Y_{\sim t} | Y_t) + \sum_{t=1}^n I(U_t  V_t ;C Y^{n} U^{t-1} V^{t-1}  Z_{\sim t} |Z_t) - 2 f(\varepsilon)- 2n f(\varepsilon)\\
& = n I(S_T X_T ;C  U^{n}  S_{T+1}^n  Y_{\sim T} | Y_T T) + n I(U_T  V_T ; C Y^{n} U^{T-1} V^{T-1}  Z_{\sim T} |Z_T T) - 2(n+1) f(\varepsilon)\\
&\geq n I(S_T X_T ;C  U^{n}  S_{T+1}^n  Y^{T-1} | Y_T T) + n I(U_T  V_T ; C Y^{n} U^{T-1} Z_{\sim T} |Z_T T) - 2(n+1) f(\varepsilon)\\
& = n I(S_T X_T ;C  U^{n}  S_{T+1}^n  Y^{T-1} T| Y_T ) - n I(S_T X_T ; T| Y_T ) \\
& \quad + n I(U_T  V_T ; C Y^{n} U^{T-1} Z_{\sim T} T |Z_T ) - n I(U_T  V_T ;  T |Z_T ) - 2(n+1) f(\varepsilon)\\
& = n I(S_T X_T ;C  U^{n}  S_{T+1}^n  Y^{T-1} T| Y_T ) - n I(S_T X_T  Y_T; T ) + n I( Y_T; T )\\
& \quad + n I(U_T  V_T ; C Y^{n} U^{T-1} Z_{\sim T} T |Z_T ) - n I(U_T  V_T Z_T ;  T ) + n I( Z_T; T ) - 2(n+1) f(\varepsilon)\\
& \overset{(e)}{\geq} n I(S_T X_T ;C  U^{n}  S_{T+1}^n  Y^{T-1} T| Y_T ) + n I(U_T  V_T ; C Y^{n} U^{T-1} Z_{\sim T} T |Z_T ) - 2(2n+1) f(\varepsilon)
\end{align*}}where $(a)$ follows from basic properties of entropy and mutual information. To prove $(b)$, note that  
{\allowdisplaybreaks
\begin{align*}
  I(S^{n} X^{n} ; Z^{n} U^{n} V^{n} |Y^{n})  & \leq I(S^{n} X^{n} Y^{n}; Z^{n} U^{n} V^{n})\\
  I(U^{n}  V^{n}; Y^{n}| Z^{n})  & \leq I(S^{n} X^{n} Y^{n}; Z^{n} U^{n} V^{n})
\end{align*}}and  $I(S^{n} X^{n} Y^{n}; Z^{n} U^{n} V^{n}) < f(\varepsilon)$ by \eqref{lemmitsep}.
Then $(c)$ comes from the chain rule for mutual information, $(d)$ follows from Lemma \ref{lemab} and $(e)$ from  \cite[Lemma VI.3]{cuff2013distributed}
since the distributions are close to i.i.d. by hypothesis.
The lower bound on $R_0$ follows from the identifications
{\allowdisplaybreaks
\begin{align*} 
 & W_{1,t} = (C , U^{n} , S_{t+1}^n,  Y^{t-1}) \quad t \in \llbracket 1,n\rrbracket \\
 & W_{2,t} = (C, Y^{n}, U^{t-1}, Z_{\sim t} ) \quad t \in \llbracket 1,n\rrbracket\\
 & W_1 = (W_{1,T}, T)=(C , U^{n} , S_{T+1}^n,  Y^{T-1}, T)\\
 & W_2 = (W_{2,T}, T)=(C, Y^{n}, U^{T-1}, Z_{\sim T}, T ).
\end{align*}}
Following the same approach as \cite{treust2015empirical, treusttech}, we divide the second part of the converse in two steps.
First, we have the following upper bound:
{\allowdisplaybreaks
\begin{align*} 
I( C U^{n} ;Y^{n})  & =\sum_{t=1}^n I(C U^{n} ; Y_t|Y^{t-1}) \leq \sum_{t=1}^n I(C U^{n} Y^{t-1}; Y_t)\\
 & = \sum_{t=1}^n I(C U^{n}  Y^{t-1}  S_{t+1}^n; Y_t)-\sum_{t=1}^n I(S_{t+1}^n; Y_t| C U^{n}  Y^{t-1})\\
 & \overset{(f)}{=} \sum_{t=1}^n I(C U^{n}  Y^{t-1}  S_{t+1}^n; Y_t)-\sum_{t=1}^n I(S_t;  Y^{t-1}| C U^{n} S_{t+1}^n) \stepcounter{equation}\tag{\theequation}\label{upb}\\
 & = \sum_{t=1}^n I(C U^{n} Y^{t-1}  S_{t+1}^n; Y_t)-\sum_{t=1}^n I(S_t; C Y^{t-1} U^{n}  S_{t+1}^n) + \sum_{t=1}^n I(S_t;  C U^{n}  S_{t+1}^n)\\
 & \overset{(g)}{=} \sum_{t=1}^n I(C U^{n} Y^{t-1}  S_{t+1}^n; Y_t)-\sum_{t=1}^n I(S_t; C Y^{t-1} U^{n}  S_{t+1}^n) \\
 & \overset{(h)}{=} \sum_{t=1}^n I( Y_t; W_{1,t})-\sum_{t=1}^n I(S_t; W_{1,t}) \\
 \end{align*}}where $(f)$ comes from Csisz{\'a}r's Sum Identity \cite{elgamal2011nit}, $(g)$ from the fact that $I(S_t;  C U^{n}  S_{t+1}^n)$ is zero because the source and the common randomness are independent of the state, which 
is  i.i.d. by hypothesis. Finally, $(h)$ comes from the identification of the auxiliary random variable $W_{1,t}$ for $t \in \llbracket 1,n\rrbracket$. 

Then, we show a lower bound:
{\allowdisplaybreaks
\begin{align*} 
 I( C U^{n} ;Y^{n}) & \geq I(U^{n} ;Y^{n}|C) \overset{(i)}{=}  I(U^{n} ;C Y^{n} ) \overset{(j)}{=} I(U^{n} Z^{n};C Y^{n} ) \\
 &\geq I(U^{n};C Y^{n}|Z^{n})= \sum_{t=1}^n I(U_t;C Y^{n}|Z^{n} U^{t-1}) \\
 &= \sum_{t=1}^n I(U_t; C Y^{n} Z_{\sim t} U^{t-1}|Z_t) - \sum_{t=1}^n I(U_t; Z_{\sim t} U^{t-1}|Z_t) \\
 & \overset{(k)}{=} \sum_{t=1}^n I(U_t; C Y^{n} Z_{\sim t} U^{t-1}|Z_t) \stepcounter{equation}\tag{\theequation}\label{lob}\\
 & = \sum_{t=1}^n I(U_t Z_t;C Y^{n} Z_{\sim t} U^{t-1}) - \sum_{t=1}^n I(Z_t;C Y^{n} Z_{\sim t} U^{t-1})  \\
 & \overset{(l)}{=} \sum_{t=1}^n I(U_t; C Y^{n} Z_{\sim t} U^{t-1}) - \sum_{t=1}^n I(Z_t;C Y^{n} Z_{\sim t} U^{t-1}) \\
 & \overset{(m)}{=} \sum_{t=1}^n I(U_t; W_{2,t}) - \sum_{t=1}^n I(Z_t; W_{2,t}) 
 \end{align*}}where $(i)$ comes from the fact that $I(U^{n};C)$ is zero because $U^{n}$ and $C$ are independent, $(j)$ from
the Markov chain $Z^{n}-U^{n}-Y^{n} C$, $(k)$ from the fact that $U^{n}$ and $Z^{n}$ are i.i.d. by hypothesis, 
$(l)$ follows from the the Markov chain $ Z_t-U_t-(Y^{n},Z_{\sim t}, U^{t-1}, C )$ for $t \in \llbracket 1,n\rrbracket$ and finally
$(m)$ comes from the identification of the auxiliary random variable $W_{2,t}$ for $t \in \llbracket 1,n\rrbracket$.

By combining upper and lower bound, we have
{\allowdisplaybreaks
\begin{align*} 
 0 & \overset{(n)}{\leq} \sum_{t=1}^n I( Y_t, W_{1,t})-\sum_{t=1}^n I(S_t; W_{1,t})  + \sum_{t=1}^n I(Z_t; W_{2,t}) - \sum_{t=1}^n I(U_t; W_{2,t})\\
 & = n  I( Y_T, W_{1,T}|T) - n I(S_T; W_{1,T}|T) +n I(Z_T; W_{2,T}|T) - n I(U_T; W_{2,T}|T)\\
 & \leq  n  I( Y_T, W_{1,T} T) - n I(S_T; W_{1,T} T) + n I(S_T; T) + n I(Z_T; W_{2,T} T) - n I(U_T; W_{2,T} T)+n I(U_T; T)\\
 & \overset{(o)}{=} n  I( Y_T, W_{1,T} T)- n I(S_T; W_{1,T} T) + n I(Z_T; W_{2,T} T) - n I(U_T; W_{2,T} T)\\
 & \overset{(p)}{=} n  I( Y; W_1) - n I(S; W_1)+ n I(Z; W_2) - n I(U; W_2)
\end{align*}}where $(n)$ comes from \eqref{upb} and \eqref{lob} and $(o)$ follows from the i.i.d. nature of the source and state.
Finally $(p) $ follows from the identifications for $W_1$ and $W_2$.

With the chosen identification, the Markov chains are verified for each $t \in \llbracket 1,n\rrbracket$:
{\allowdisplaybreaks
\begin{align*}
& Y_t- (X_t, S_t)-W_{1,t} \\
& Z_t-U_t- W_{2,t}\\
& V_t-( W_{2,t}, Z_t)- U_t.
\end{align*}
}The first Markov chain holds  because the channel is memoryless and $Y_t$ does not belong to $W_{1,t}$. The second one holds because $Z^{n}$ 
is i.i.d. and $Z_t$ does not belong to $W_{2,t}$. 
Finally, the third one is verified because the decoder is non causal and $V_t$ is a function of $(Y^{n}, Z^{n})$ that is included in 
$ (W_{2,t}, Z_t)=(Y^{n}, U^{t-1}, Z_{\sim t}, Z_t)$.
With a similar approach as in Section \ref{identification} and Section \ref{identification2}, the Markov chains with $T$ hold.
Then since $W_1=W_{1,t}$  and $W_2=W_{2,t}$ when $T=t$,  we also have $ Y- (X, S)-W_1 $, $ Z-U- W_2$ and $V-( W_2, Z)- U$.
The cardinality bound is proved in $\mbox{Appendix \ref{appendix bounds}}$.

\begin{oss}
Note that even if in the converse proof $W_1$ and $W_2$ are correlated, from them 
we can define two new variables $W'_1$ and $W'_2$  independent of each other, with the same marginal distributions 
$P_{W'_1 SXY} =P_{W_1 SXY}$ and $P_{W'_2 UVZ}=P_{W_2 UVZ}$, such that the joint distribution $P_{W'_1 W'_2 SXYUVZ}$ splits as $P_{W'_1 SXY} P_{W'_2 UVZ}$.
Since we are supposing $(U,V,Z)$ and $(S,X,Y)$ independent of each other and the constraints only depend on the marginal distributions $P_{W_1 SXY}$ and $P_{W_2 UVZ}$, 
the converse is still satisfied with the new auxiliary random variables $W'_1$ and $W'_2$.
Moreover the new variables still verify the the cardinality bounds since they also depend only on the marginal distributions (as shown in $\mbox{Appendix \ref{appendix bounds}}$). 
\end{oss}
\vspace{-0,3cm}
\subsection{Coordination under secrecy constraints }\label{secrecy}

\begin{center}
\begin{figure}[ht]
\centering
\includegraphics[scale=0.21]{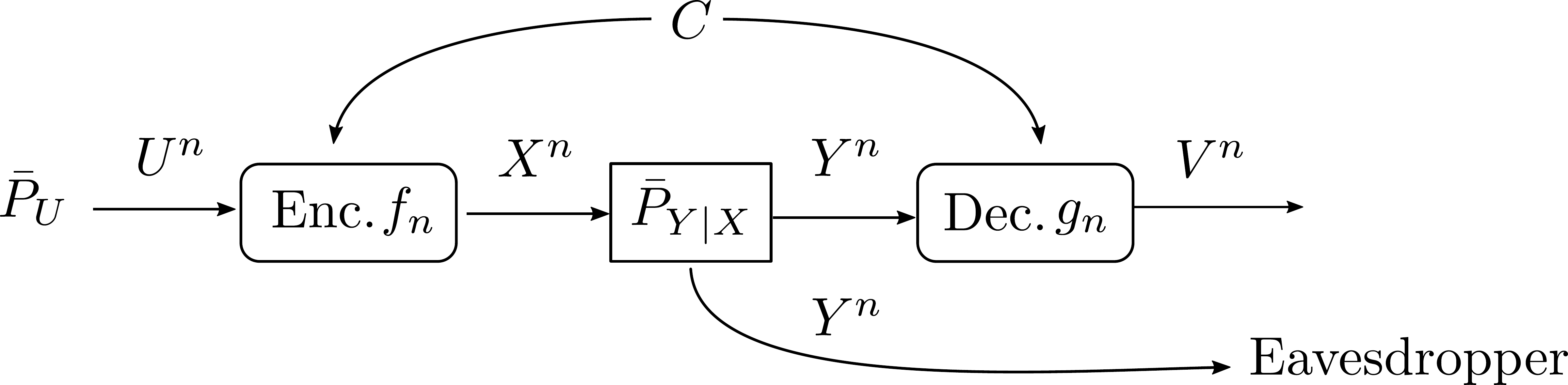}
\caption{Wiretap channel: strong coordination implies secrecy.}
\label{fig: wtap}
\end{figure}
\end{center}
\vspace{-0,8cm}

In this section we briefly discuss how in the separation setting of Section \ref{separation}, strong coordination offers additional security guarantees \vv{for free}.
In this context, the common randomness is not only useful to coordinate signals and actions of the nodes but plays the role of a secret key shared between the two legitimate users.

For simplicity, we do not consider channel state and side information at the decoder.
Suppose there is an eavesdropper who observes the signals sent over the channel. 
We will show  that not knowing the common randomness, the eavesdropper cannot infer any information about the actions.

\begin{lem}
In the setting of Theorem \ref{teoseparation}, without state and side information at the decoder, 
suppose that there is an eavesdropper that receives the same sequence $Y^n$ as the decoder but has no knowledge of the common randomness. 
There exists a sequence $(f_n,g_n)$  of strong coordination codes achieving the pair 
$(\bar P_{UV} \bar P_{X}, R_0) \in \mathcal{R}_{SEP}$ such that 
the induced joint distribution $P_{U^n V^n X^n Y^n}$ satisfies the \emph{strong secrecy condition} \cite{bloch2011physical}:
\begin{equation}\label{strong secrecy}
\lim_{n \to \infty} \mathbb D(P_{U^nV^nY^n} \Arrowvert P_{U^nV^n} P_{Y^n})=\lim_{n \to \infty} I(U^nV^n;Y^n)=0.
\end{equation}

\begin{IEEEproof}
 Observe that in this setting the target joint distribution is of the form $\bar{P}_{UV}^{\otimes n}\bar{P}_{XY}^{\otimes n}$. Therefore achieving strong coordination means that 
$\mathbb V(P_{U^nV^nY^n}, \bar{P}_{UV}^{\otimes n}\bar{P}_{Y}^{\otimes n})$ vanishes.
By the upperbound on the mutual information in Lemma \ref{lem1csi},  we have secrecy if 
$\mathbb V(P_{U^nV^nY^n}, \bar{P}_{UV}^{\otimes n}\bar{P}_{Y}^{\otimes n})$ goes to zero exponentially.
But we have proved in Section \ref{inner} that there exists a sequence of codes such that
$\mathbb V (\bar P_{U^{n} S^{n} Z^{n} X^{n} Y^{n} V^{n}}, P_{U^{n} S^{n} Z^{n} X^{n} Y^{n} V^{n}}) $ 
goes to zero exponentially \eqref{finaleq}. Hence, so does $\mathbb V(P_{U^nV^nY^n}, \bar{P}_{UV}^{\otimes n}\bar{P}_{Y}^{\otimes n})$.
\end{IEEEproof}
\end{lem}

\vspace{-0,3cm}
\subsection{Is separation optimal?}

\begin{center}
\begin{figure}[ht]
\centering
\includegraphics[scale=0.21]{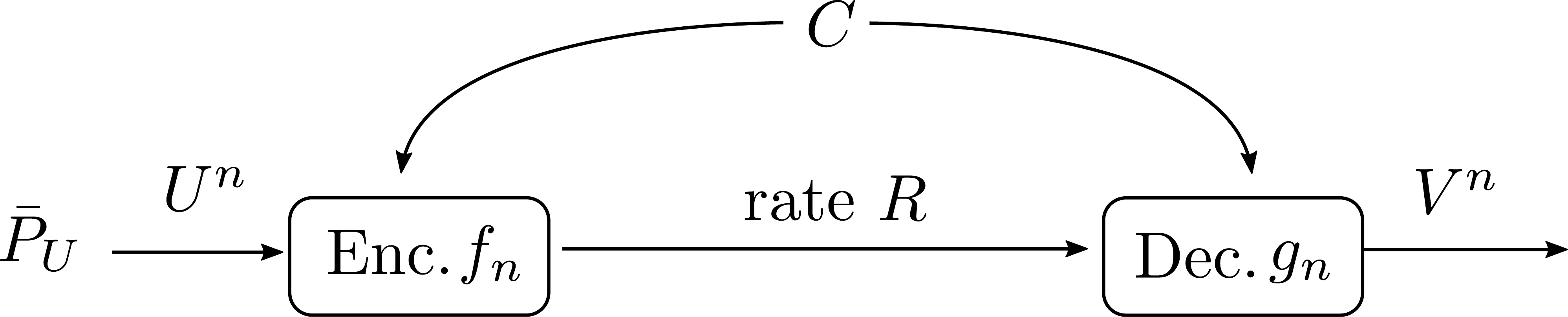}
\caption{Coordination of the actions $U^n$ and $V^n$ for a two-node network with an error-free link of rate $R$.}
\label{fig: cuff}
\end{figure}
\end{center}
\vspace{-0,7cm}

Strong coordination over error-free channels was investigated in \cite{cuff2010, cuff2008communication}. 
When extending this analysis to noisy channels, it is natural to ask whether some form of separation theorem holds between coding for coordination and channel coding.
In this section, we show that unlike the case of empirical coordination, separation does not hold for strong coordination.

If the separation principle is still valid for strong coordination, by concatenating the strong coordination of 
the source and its reconstruction with the strong coordination of the input and output of the channel we should retrieve the 
same mutual information and rate constraints. 
In order to prove that separation does not hold, first we consider the optimal result for coordination 
of actions in \cite{cuff2010, cuff2008communication} and than we compare it with our result on joint coordination of signals and actions.
In particular, since we want to compare the result in \cite{cuff2010, cuff2008communication} with an exact region, 
we consider the case in which the channel is perfect and the target joint distribution is of the form 
$\bar{P}_{UV}^{\otimes n}\bar{P}_{X}^{\otimes n}$. 
The choice of a perfect channel might appear counterintuitive but 
it is motivated by the fact that we are trying to find a counterexample. 
As a matter of fact, if the separation principle holds for any noisy link, it should in particular hold for a perfect one.

We start by considering the two-node network with fixed source $\bar P_U$ and an error-free link of rate $R$ 
(Figure \ref{fig: cuff}). For this setting, \cite{cuff2010, cuff2008communication} 
characterize the strong coordination region as
\begin{equation}\label{region cuff}
\mathcal{R}_{\text{Cuff}}:=\begin{Bmatrix}[c|l]
& \bar P_{UV}= \bar P_{U}  \bar P_{V|U}  \\
&\exists \mbox{ } W \mbox{ taking values in $\mathcal W$}\\ 
(\bar P_{UWV}, R, R_0)  & \bar P_{UWV}= \bar P_{U} \bar P_{W|U} \bar P_{V|UW}\\
& R \geq I(U;W) \\
& R+R_0 \geq I(UV;W)\\
& \lvert \mathcal W \rvert \leq \lvert \mathcal U \times   {\mathcal V} \rvert+1 
\end{Bmatrix}.
\end{equation}The result in \cite{cuff2010,cuff2008communication} characterizes the trade-off between the rate $R_0$ of available common randomness and 
the required description rate $R$ for simulating a discrete memoryless channel for a fixed input distribution.
We compare this region to our results when the requirement to coordinate the signals $X^n$ and $Y^n$ in addition 
to the actions $U^n$ and $V^n$ is relaxed.
We consider, in the simpler scenario with no state and no side information, the intersection  $\mathcal{R}_{UV\otimes X}:= \mathcal R_{\text{PC}}\cap \mathcal R_{\text{SEP}}$.
The following result characterizes the strong coordination region (proof in Appendix \ref{appendix UVotimesX}).

\begin{prop} \label{teoUVotimesX}
Consider the setting of Theorem \ref{teoisit} and suppose that 
$\bar P_{Y|X}( \mathbf{y}|\mathbf{x})={\mathds 1}_{X=Y} \{\mathbf{x}= \mathbf{y} \}$ and 
$\bar P_{UXV}=$ $\bar P_{UV} \bar P_{X}$. Then, the strong coordination region is

\begin{equation}\label{UVotimesX}
 \mathcal R_{UV\otimes X} := \begin{Bmatrix}[c|l]
&\bar P_{UXV}= \bar P_{U} \bar P_{V|U}  \bar P_{X}\\
&\exists \mbox{ } W \mbox{ taking values in $\mathcal W$}\\ 
(\bar P_{UXV}, R_0) &\bar P_{UW XV}=\bar P_{U} \bar P_{W|U} \bar P_{V| W}  \bar P_{X}\\
& I(W;U) \leq H(X)\\
&R_0 \geq I(UV;W)\\
& \lvert \mathcal W \rvert\leq \lvert \mathcal U  \times \mathcal V \rvert +1
 \end{Bmatrix}
\end{equation}
\end{prop}

To compare $\mathcal{R}_{\text{Cuff}}$ and $\mathcal R_{UV\otimes X}$, suppose that in the setting of Figure \ref{fig: cuff} we use a codebook to send a message to coordinate $U^n$ and $V^n$. 
In order to do so we introduce an i.i.d. source $X^n$ with uniform distribution $P_X$ in the model and we use the entropy typical sequences of $X^n$ as a codebook.
Note that in the particular case where $X^n$ is generated according to the uniform distribution, all the sequences in $\mathcal X^n$ are entropy-typical
and $P_{X^n}$ is equal in total variational distance to the i.i.d. distribution $\bar P_{X}^{\otimes n}$.
Hence, we identify $R = H(X)$ and we rewrite the information contraints in \eqref{region cuff} as 
\begin{equation*}
H(X) \geq I(U;W) \quad R_0 \geq I(UV;W)-H(X).
\end{equation*}Since in \cite{cuff2008communication} the request is to induce a joint distribution $P_{U^{n} V^{n}}$ 
that is $\varepsilon$-close in total variational distance to the i.i.d. distribution $\bar P_{UV}^{\otimes n}$, by 
imposing $X^n$ generated according to the uniform distribution, we have coordinated separately $X^n$ and $(U^n,V^n)$.

Observe that, while the information constraint is the same in the two regions \eqref{region cuff} and \eqref{UVotimesX}, the rate of common randomness $R_0$ required for strong coordination region in  \eqref{UVotimesX} is larger than the rate of common randomness in \eqref{region cuff}.
In fact, in the setting of Figure \ref{fig: cuff} both $X^n$ and the pair $(U^n,V^n)$ achieve coordination separately 
(i.e. $P_X^n$ is close to $\bar P_X^{\otimes n}$ and $P_{U^n V^n}$ is close to $\bar P_{UV}^{\otimes n}$ in total variational distance), but there is no extra constraint on the joint distribution $P_{U^{n} X^{n} V^{n}}$.
On the other hand, the structure of our setting in \eqref{UVotimesX} is different and requires the control of the joint distribution $P_{U^{n} X^{n} V^{n}}$ 
which has to be $\varepsilon$-close in total variational distance to the i.i.d. distribution $\bar P_{UV}^{\otimes n} \bar P_{X}^{\otimes n}$.
Since we are imposing a more stringent constraint, it requires more common randomness. 

\begin{oss}
We found $\mathcal R_{UV\otimes X}$ as the intersection of two regions, 
but we can give it the following interpretation starting from  $\mathcal{R}_{\text{Cuff}}$.
By identifying $R = H(X)$  in $\mathcal{R}_{\text{Cuff}}$, we find that the rate of common randomness has to be greater than $I(UV;W)-H(X)$.
But this is not enough to ensure that $X^n$ is independent of $(U^n,V^n)$. In order to guarantee that, we apply a one-time pad on $X^n$ 
(which requires an amount of fresh randomness equal to $H(X)$) and  we have
$$R_0 \geq I(UV;W)-H(X)+H(X)= I(UV;W)$$
which is the condition on the rate of common randomness in \eqref{UVotimesX}.
\end{oss}

The following example shows that, unlike the case of empirical coordination \cite{treust2017joint},
separation does not hold for strong coordination.

\begin{center}
\begin{figure}[ht!]
\centering
\includegraphics[scale=0.44]{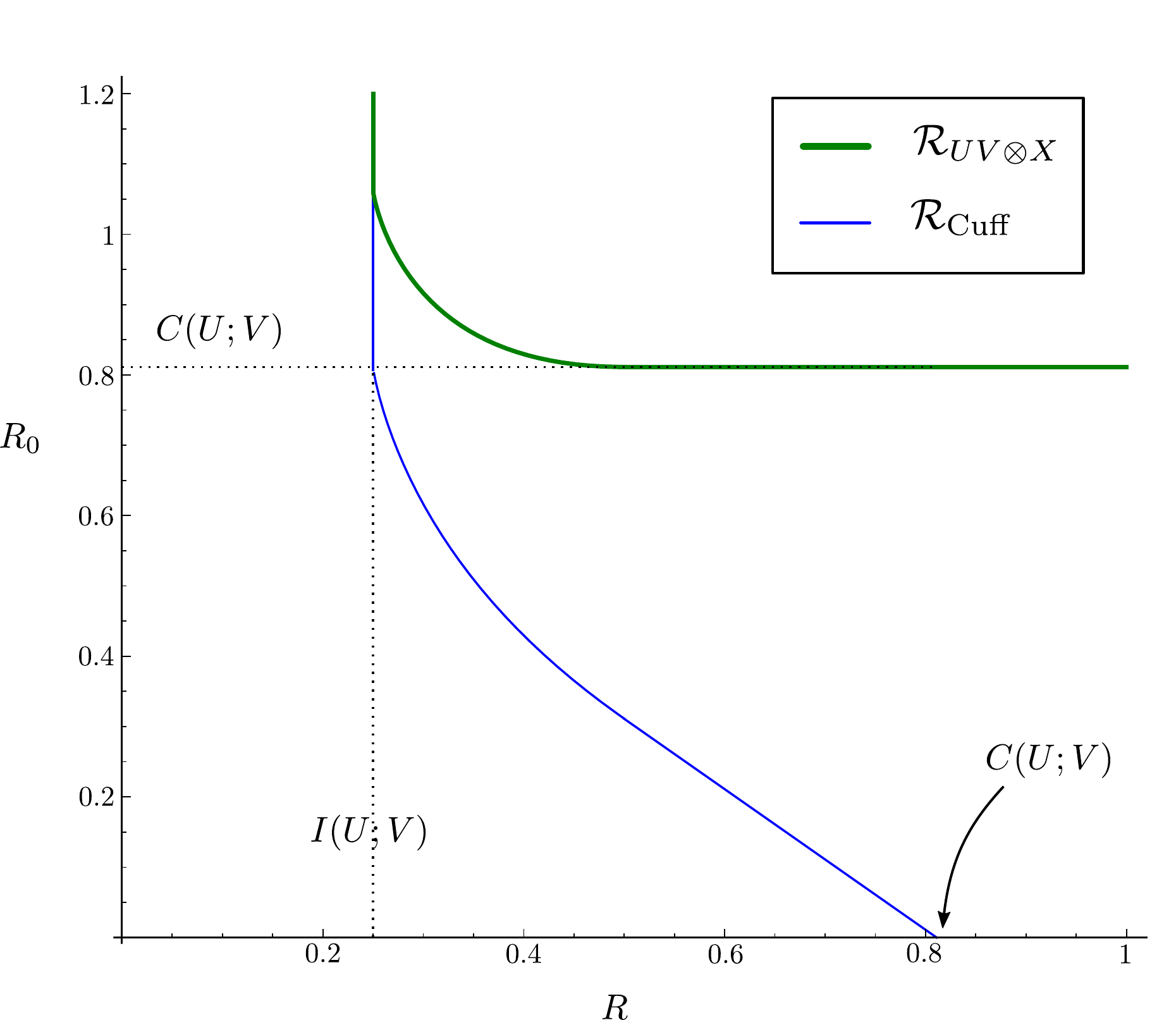}
\caption{Comparison of the joint coordination region $\mathcal{R}_{UV \otimes X}$ with $\mathcal{R}_{\text{Cuff}}$  \cite{cuff2013distributed, cuff2008communication}: 
boundaries of the regions for a binary erasure
channel with erasure probability $p_e = 0.75$ and a Bernoulli-half input.}
\label{fig: plotcuff}
\end{figure}
\end{center}

\vspace{-0,6cm}
\begin{exe}
The difference in terms of rate of common randomness $R_0$ is better shown in an example: 
when separately coordinating the two blocks $X^n$ and $(U^n,V^n)$ without imposing a joint behavior 
$P_{U^n V^n X^n}$, the same bits of common randomness can be reused for both purposes, and the required rate $R_0$ is lower.
We consider the case, already analyzed in \cite{cuff2013distributed, cuff2008communication}, of a Bernoulli-half source $U$, 
and $V$ which is an erasure with probability $p_e$ and is equal to $U$ otherwise. 
In \cite{cuff2013distributed} the authors prove that the optimal choice for  the joint distributed $P_{UWV}$
is the concatenation of two erasure channels $\bar P_{W|U}$ and $\bar P_{V|W}$ 
with erasure probability $p_1$ and $p_2$ respectively.
Then we have
\begin{equation*}
p_2 \in [0, \min\{1/2; p_e\}],\quad
p_1= 1-\frac{1-p_e}{1-p_2}
\end{equation*}
and therefore we obtain
{\allowdisplaybreaks
\begin{align*}
 & I(U;W)=1-p_1, \qquad I(UV;W)=h(p_e) + (1 - p_1 )(1 - h(p_2 ))
\end{align*}}where $h$ is the binary entropy function.
Figure \ref{fig: plotcuff} shows the boundaries of the regions \eqref{region cuff} (blue) and \eqref{UVotimesX} (green) for $p_e = 0.75$ and a Bernoulli-half input. 
The dotted bound $R \geq I(U;V)$ comes directly from combining $R \geq I(U;W)$ with the Markov chain $U-W-V$.
At the other extreme, if $R_0 = 0$ in \eqref{region cuff}, $R+R_0 \geq I(UV;W)\geq C(U;V) $ where 
 $C(U;V):= \min_{U-W-V} I(U,V;W)$ is Wyner's common information   \cite{cuff2008communication}. 
On the other hand, in our setting \eqref{UVotimesX}, $R_0\geq I(UV;W)\geq C(U;V)$ for any value of $R=H(X)$.

Moreover, note that as $R=H(X)$ tends to infinity, there is no constraint on the auxiliary random variable $W$ (aside from the Markov chain $U-V-W$) 
and similarly to \cite{lapidoth2016conditional} the minimum rate of common randomness $R_0$ needed for strong coordination is Wyner's common information $C(U;V)$.
In particular to achieve joint strong coordination of $(U,X,V)$ a positive rate of common randomness is required.
The boundaries of the rate regions only coincide on one extreme, and  $\mathcal R_{UV\otimes X}$ is strictly contained in $\mathcal{R}_{\text{Cuff}}$.
\end{exe}

\section{Polar coding schemes for strong coordination with no state and side information}\label{sec: polarcoding}

Although our achievability results shed some light on the fundamental limits of coordination over noisy channels, 
the problem of designing practical codes for strong coordination in this setting is still open.  
In this section we focus on channels without state and side information for simplicity, 
and we show that the coordination region of Theorem \ref{teoisit} is achievable using polar codes, if 
an error-free channel of negligible rate is available between the encoder and decoder. 

We note that polar codes have already been proposed for coordination in other settings:  
\cite{blasco-serrano2012} proposes polar coding schemes for point-to-point empirical coordination with error free links and uniform actions,
while \cite{chou2015coordination} generalizes the polar coding scheme to the case of non uniform actions.
Polar coding for strong point-to-point coordination has been presented in  \cite{bloch2012strong, chou2016soft}.
In \cite{obead2017joint} the authors construct  a joint coordination-channel polar coding scheme for strong coordination of actions.
We present a joint source-channel polar coding scheme for strong coordination and we require joint coordination of signals and actions over a noisy channel.

For brevity, we only focus on the set of achievable distributions in $\mathcal R'_{\text{in}}$ for which the auxiliary variable $W$ is binary. 
The scheme can be extended to the case of a non-binary random variable $W$ using non-binary polar codes \cite{csacsouglu2012polar}.

\begin{teo}\label{polarregion}
The subset of the region $\mathcal R'_{\text{in}}$ defined in \eqref{eq: regionisit} for which 
the auxiliary random variable $W$ is binary is achievable using polar codes, 
provided there exists an error-free channel of negligible rate between the encoder and decoder. 
\end{teo}
\vspace{0,2cm}

To convert the information-theoretic achievability proof of Theorem \ref{teoisit} into a polar coding proof, 
we use source polarization \cite{arikan2010source} to induce the desired joint distribution. 
Inspired by \cite{chou2015polar}, we want to translate the random binning scheme into a polar coding scheme. 
The key idea is that the information contraints and rate conditions found in the random binning proof directly convert into the definition of the polarization sets.
In the random binning scheme we reduced the amount of common randomness $F$ by having the nodes to agree on an instance of $F$, 
here we recycle some common randomness using a chaining construction as in \cite{hassani2014universal,mondelli2015achieving}.  

Consider  random vectors $U^{n}$, $W^{n}$, $X^{n}$, $Y^{n}$ and $V^{n}$
generated i.i.d. according to $\bar P_{UWXYV}$ that satisfies the inner bound of \eqref{eq: regionisit}.
For $n=2^m$, $m \in \mathbb N$, we note $G_n:= \begin{footnotesize}
\begin{bmatrix}
1 & 0\\
1 & 1
\end{bmatrix}^{\otimes m}
\end{footnotesize}$ the source polarization transform defined in \cite{arikan2010source}.
Let $R^{n}:=W^{n}G_n$ be the polarization of $W^{n}$.
For some $0<\beta<1/2$, let $\delta_n = 2^{-n^ {\beta}}$ and define the very high entropy and high entropy sets:
 {\allowdisplaybreaks
 \begin{align}\label{eq: hv}
 \begin{split}
\V_{W}: & =\left\{j\in\llbracket 1,n\rrbracket:H(R_j|R^{j-1})>1-\delta_n \right\},\\
\V_{W | U}: & =\left\{j\in\llbracket 1,n\rrbracket:H(R_j|R^{j-1} U^{n})>1-\delta_n \right\}, \\
\V_{W | Y}: & =\left\{j\in\llbracket 1,n\rrbracket : H(R_j|R^{j-1} Y^{n})>1-\delta_n \right\},\\
\h_{W | Y}: & =\left\{j\in\llbracket 1,n\rrbracket:H(R_j|R^{j-1} Y^{n})>\delta_n \right\} .
\end{split}
 \end{align}}

Now define the following disjoint sets:
{\allowdisplaybreaks
\begin{align*}
A_1 := \V_{W|U} \cap \h_{W|Y} , \quad & A_2 : = \V_{W|U} \cap \h_{W|Y}^c,\\
A_3 := \V_{W|U}^c \cap \h_{W|Y} ,\quad
& A_4 := \V_{W|U}^c \cap \h_{W|Y}^c.
\end{align*}}

\begin{oss}\label{oss card}
We have:
\begin{itemize}
\item $\V_{W | Y} \subset \h_{W | Y}$ and $ \lim_{n \rightarrow \infty} \frac{\lvert \h_{W | Y} \setminus \V_{W|Y} \rvert}{n} = 0$ \cite{arikan2010source},
 \item  $\lim_{n \rightarrow \infty} \frac{\lvert \V_{W|U} \rvert}{n}  = H(W|U)$  \cite {chou2015secretkey},
 \item  $ \lim_{n \rightarrow \infty} \frac{\lvert \h_{W | Y} \rvert}{n} = H(W|Y)$ \cite{arikan2010source}.
\end{itemize}
Since  $H(W|U)- H(W|Y) = I(W;Y)- I(W;U),$ for sufficiently large $n$ 
the assumption $I(W;Y) \geq I(W;U)$ directly implies that $\lvert A_2 \rvert \geq \lvert A_3 \rvert$.
\end{oss}

\vspace{0,2cm}
\paragraph*{Encoding}
The encoder observes $k$ blocks of the source $U^{n}_{(1:k)}:=(U^{n}_{(1)}, \ldots, U^{n}_{(k)})$ and 
generates for each block $i\in \llbracket 1,k\rrbracket$ a random variable $\widetilde R^{n}_{(i)}$ following the procedure described in Algorithm \ref{algcnc}. Similar to \cite{Cervia2016}, the chaining construction proceeds as follows:
\begin{itemize}
\item let $A'_1:=\mathcal V_{W|UXYV}$, observe that $A'_1$ is a subset of $ A_1$ since
 $\mathcal V_{W|UXYV} \subset \mathcal V_{W|U}$ and 
 $\mathcal V_{W|UXYV} \subset \mathcal V_{W|Y} $ $\subset \mathcal H_{W|Y}$. 
 The bits in  $A'_1 \subset \V_{W|U}$ in block $i \in \llbracket 1,k\rrbracket$ are chosen with uniform probability using a uniform randomness 
 source $\bar C'$ shared with the decoder, and  their
value is reused over all blocks; 
\item the bits in  $A_1 \setminus A'_1 \subset \V_{W|U}$ in block $i \in \llbracket 1,k\rrbracket$ are chosen with uniform probability using a
uniform randomness source $\bar C_{i}$ shared with the decoder;
\item in the first block the bits in  $A_2 \subset \V_{W|U}$ are chosen with uniform probability using a local randomness source $M$;
\item for the  following blocks, let $A'_3$ be a subset of $A_2$ such that $\lvert A'_3 \rvert= \lvert A_3 \rvert$.  The bits of $A_3$ in block $i$ are sent to $A'_3$ in the block $i+1$ using a one time pad with key $C_{i}$. Thanks to the Crypto Lemma 
 \cite[Lemma 3.1]{bloch2011physical}, if we choose $C_{i}$ of size $\lvert A_3 \rvert$ to be a uniform random key, the bits in $A'_3$ in the block $i+1$ are uniform. The bits in $A_2 \setminus A'_3$ are chosen with uniform probability using the local randomness source $M$;
\item the bits in $A_3$ and in $A_4$ are generated according to the previous bits using successive cancellation 
encoding as in  \cite{arikan2010source}. Note that it is possible to sample efficiently from 
$\bar P_{R_i \mid R^{i-1} U^{n}}$ given $U^{n}$  \cite{arikan2010source}.
\end{itemize}

As in \cite{chou2015polar}, 
to deal with unaligned indices, chaining also requires in the last encoding block to transmit $R_{(k)}[A_{3}]$ to the decoder.
Hence the coding scheme requires an error-free channel between the encoder and decoder which has negligible rate 
since $\lvert R_{(k)}[A_{3}] \rvert \leq \lvert \h_{W | Y} \rvert$ and
\begin{equation*}
\lim_{n \rightarrow \infty \atop k \rightarrow \infty} \frac{\lvert \h_{W|Y}\rvert}{kn} = \lim_{k \rightarrow \infty} \frac{H(W|Y)}{k} =0.
\end{equation*}

The encoder then computes $\widetilde W^{n}_{(i)}= \widetilde  R^{n}_{(i)} G_n$ for $i=1, \ldots, k$ and generates $X^{n}_{(i)}$
symbol by symbol from $\widetilde W^{n}_{(i)}$ and $U^{n}_{(i)}$ using the conditional distribution 
$$\bar P_{X_{j,(i)} |\widetilde W_{j,(i)}  U_{j,(i)}}(x|\widetilde w_{j,(i)}, u_{j,(i)})=\bar P_{X|WU} ( x | w_{j,(i)}, u_{j,(i)})$$ and sends $X^{n}_{(i)}$ 
over the channel.

\begin{center}\label{fig:kblorecc}
\begin{figure}[ht!]
 \centering
 \includegraphics[scale=0.9]{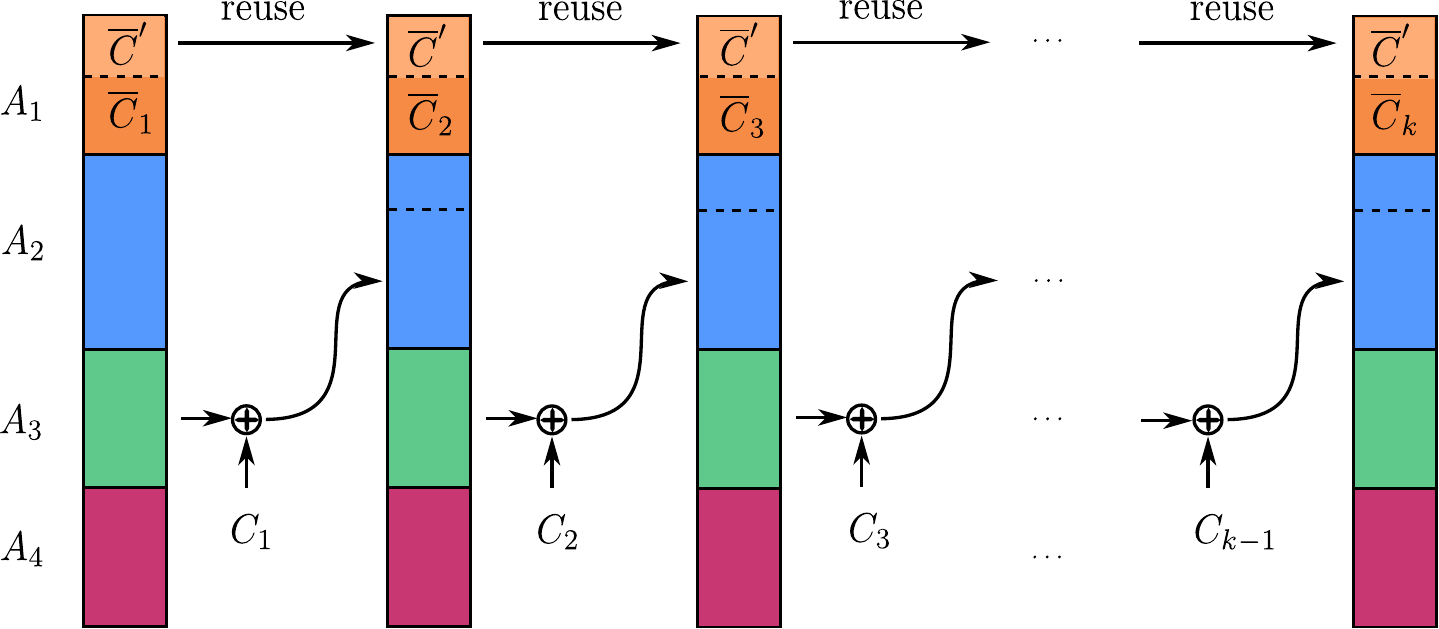}
\caption{Chaining construction for block Markov encoding}
\end{figure}
\end{center}

\vspace{-0,7cm}
\begin{algorithm}[ht!]\label{algcnc}
\begin{small}
\DontPrintSemicolon
\SetAlgoLined
\SetKwInOut{Input}{Input}
\SetKwInOut{Output}{Output}
\Input{ $(U^{n}_{(1)}, \ldots, U^{n}_{(k)})$, $M$ local randomness (uniform random bits), 
common randomness  $(\bar C', \bar C_{1:k},C_{1:k-1})$  shared with the decoder, where 
$\bar C'$ has size $\lvert A'_1 \rvert$,
$\bar C_{1:k}$ has size $k \lvert A_1 \setminus A'_1 \rvert$,
$C_{1:k-1}$ has size $(k-1) \lvert A_3 \rvert$.}

\Output{$\left( \widetilde R_{(1)}^n, \ldots, \widetilde R_{(k)}^n \right)$}
\If{$i=1$}{
 $\widetilde R_{(1)}[A'_1] \longleftarrow \bar C' \qquad \widetilde R_{(1)}[A_1 \setminus A'_1] \longleftarrow \bar C_{1} \qquad  \widetilde R_{(1)}[A_2]   \longleftarrow M $\;
\For{$j \in A_{3} \cup A_{4}$}{
Given $U^{n}_{(1)}$, successively choose the bits $\widetilde R_{j,(1)}$ according to
\begin{equation} \label{eq: p1}
\bar P_{R_j \mid R^{j-1} U^{n}} \left(\widetilde R_{j,(1)}\mid \widetilde R^{j-1}_{(1)} U^{n}_{(1)} \right)
\end{equation}
}
}

\For{$i=2, \ldots, k$}{
 $\widetilde R_{(i)}[A'_1] \longleftarrow \bar C' \qquad \widetilde R_{(i)}[A_1 \setminus A'_1] \longleftarrow \bar C_i $\;
 $ \widetilde R_{(i)}[A'_3] \longleftarrow \widetilde R_{(i-1)}[A_3]  \oplus C_{i-1} \quad \widetilde R_{(i)}[A_2 \setminus A'_3] \longleftarrow M $ \;
\For{$j \in A_{3} \cup A_{4}$}{
Given $U^{n}_{(i)}$, successively choose the bits $\widetilde R_{j,(i)}$ according to
\begin{equation} \label{eq: p2}
\bar P_{R_j \mid R^{j-1}U^{n}} \left(\widetilde R_{j,(i)} \mid \widetilde R_{(i-1)}^j U^{n}_{(i)} \right)
\end{equation} 
}
}

\BlankLine
\caption{Encoding}
\end{small}
\end{algorithm}

\paragraph*{Decoding}

The deconding procedure  described in Algorithm \ref{algdecnc} proceeds as follows.
The decoder observes $(Y^{n}_{(1)}, \ldots, Y^{n}_{(k)})$  and $R_{(k)}[A_3]$
which allows it to decode in reverse order. 
We note $\widehat R^n_{(i)}$ the estimate of $R^n_{(i)}$ at the decoder, for $i \in \llbracket 1,k\rrbracket$.
In block $i \in \llbracket 1,k\rrbracket$, the decoder has access to $\widehat R_{(i)}[A_{1} \cup A_{3}]= \widehat R_{(i)}[\h_{W \mid Y}]$:
\begin{itemize}
\item the bits in $A_1$ in block $i$ correspond to shared randomness $\bar C'$ and $\bar C_i$ for $A'_1$ and 
$A_1 \setminus A'_1$ respectively;
\item in block $i \in [1, k-1]$ the bits in $A_3$ are obtained by successfully recovering $A_2$ in block $i+1$.
\end{itemize}

\paragraph*{Rate of common randomness}
The rate of common randomness is $I(W;UXV|Y)$ since:
{\allowdisplaybreaks
\begin{align*}
 &\lim_{n \to \infty} \frac{ k \lvert A_1 \rvert - (k-1) \lvert A'_1 \rvert + (k-1) \lvert A_3  \rvert}{kn} = \lim_{n \to \infty}\frac{ \lvert A_1 \rvert + \lvert A_3  \rvert - \lvert A'_1 \rvert}{n}\\
 &= H(W|Y)- H(W|UXYV) =I(W;UXV|Y).
\end{align*}
}
\begin{algorithm}[ht!]\label{algdecnc}
\begin{small}
\DontPrintSemicolon
\SetAlgoLined

\SetKwInOut{Input}{Input}
\SetKwInOut{Output}{Output}
\Input{$(Y^{n}_{(1)}, \ldots, Y^{n}_{(k)})$, $R_{(k)}[A_3]$ shared with the encoder, 
common randomness $(\bar C', \bar C_{1:k-1},C_{1:k-1})$  shared with the encoder, where $\bar C'$ has size $\lvert A'_1 \rvert$,  
$\bar C_{1:k}$ has size $k \lvert A_1 \setminus A'_1 \rvert$  and  $C_{1:k-1}$ has size $(k-1) \lvert A_3 \rvert$.}
\Output{$(\widehat R_{(1)}^n, \ldots, \widehat R_{(k)}^n)$}
\For{$i=k, \ldots, 1$}{
 $\widehat R_{(i)}[A'_1] \longleftarrow \bar C' \qquad \widehat R_{(i)}[A_1 \setminus A'_1] \longleftarrow \bar C_i$\;
 \If{$i=k$}{$\widehat R_{(i)}[A_3]$ shared with the decoder}
 \Else{$\widehat R_{(i)}[A_3] \longleftarrow \widehat R_{(i+1)}[A'_3]$\;}
\For{$j \in A_{2} \cup A_{4}$}{ Successively choose the bits according to
$\widehat R_{j,(i)} = \begin{cases}
 0 \quad \mbox{if } L_n(Y_{(i)}^n,R_{(i-1)}^j) \geq 1\\
 1 \quad \mbox{else} 
 \end{cases}$\;
where
$$L_n(Y_{(i)}^n,R_{(i-1)}^j) = \frac{\bar P_{R_{j,(i)} \mid R_{(i-1)}^j  Y_{(i)}^n}\left(0 \mid \widehat R_{(i-1)}^j Y_{(i)}^n \right) }{\bar P_{R_{j,(i)} \mid R_{(i-1)}^j  Y_{(i)}^n}\left(1 \mid \widehat R_{(i-1)}^jY_{(i)}^n \right)}$$
}
}
\BlankLine
\caption{Decoding}
\end{small}
\end{algorithm}

\paragraph*{Proof of Theorem \ref{polarregion}}

We note with $\tilde{P}$ the joint distribution induced by the encoding and decoding algorithm of the previous sections. 
The proof requires a few steps, here presented as different lemmas. The proofs are in Appendix \ref{appendix polar}.
First, we want to show that we have strong coordination in each block.
\begin{lem}\label{scblock}
In each block $i \in \llbracket 1,k\rrbracket$, we have 
\begin{equation}\label{1block}
\tv\left(\widetilde P_{U^{n}_{(i)} \widetilde W_{(i)}^n X_{(i)}^n Y_{(i)}^n V_{(i)}^n}, \bar P_{UWXYV}^{\otimes n}\right) \leq \delta_n^{(1)}
\end{equation}
where $\delta_n^{(1)}:= 2 \mathbb P \left\{\widehat W_{(i)}^n  \neq \widetilde W_{(i)}^n \right\}+ \sqrt{2 \log 2} \sqrt{n \delta_n}$.
\end{lem}

Now, we want to show that two consecutive blocks are almost independent.
To simplify the notation, we set 
 {\allowdisplaybreaks
 \begin{equation*}
\begin{matrix}[ll]
  L:=U^{n} X^{n} Y^{n} W^{n} &\\
  L_{i}:=U^{n}_{(i)} X_{(i)}^n Y_{(i)}^n V_{(i)}^n & i \in \llbracket 1,k\rrbracket\\
  L_{a:b}:=U^{n}_{(a:b)} X^{n}_{(a:b)} Y^{n}_{(a:b)} V^{n}_{(a:b)} & \llbracket a,b\rrbracket \subset \llbracket 1,k\rrbracket
\end{matrix}
\end{equation*}
}

\begin{lem} \label{step2a'}
For $i \in \llbracket 2,k\rrbracket$, we have
$$\tv \left( \widetilde P_{L_{i-1:i} \bar C'} ,  \widetilde P_{ L_{i-1}\bar C'}  \widetilde P_{L_{i}}\right) \leq  \delta_n^{(3)}$$
where $ \delta_n^{(3)}:= \sqrt{2 \log 2} \sqrt{n \delta_n  + 2 \delta_n^{(1)} (\log{\lvert \mathcal U  \times  \mathcal X  \times   \mathcal W  \times   \mathcal Y  \times  \mathcal V \rvert}-\log{\delta_n^{(1)}})}$ 
and $\delta_n^{(1)}$ is defined in Lemma \ref{scblock}.
\end{lem}

Now that we have proven the asymptotical independence of two consecutive blocks, 
we use Lemma \ref{step2a'} to prove the asymptotical independence of all blocks. First we need an intermediate step.
\vspace*{0.3cm}
\begin{lem}\label{step2b'}
We have
$$ \tv \left( \widetilde P_{ L_{1:k}}, \prod_{i=1}^k  \widetilde P_{ L_{i}} \right) \leq \sqrt{k-1} \delta_n^{(3)}$$
where $\delta_n^{(3)} $ is defined in Lemma \ref{step2a'}.
\end{lem}

\vspace*{0.3cm}
Finally, we prove the asymptotical independence of all blocks.
\begin{lem}\label{step2c'}
We have
$$ \tv \left( \widetilde P_{L_{1:k}}, \bar P_{UXYV}^{\otimes nk}\right) \leq \delta_n^{(5)}$$
where $ \delta_n^{(5)}:=\sqrt{k} (\delta_n^{(3)} + \delta_n^{(2)})$ and  
$\delta_n^{(2)}$ and $\delta_n^{(3)}$ are defined in \eqref{emp2} and Lemma \ref{step2a'} respectively.
\end{lem}

\section{Conclusions and perspectives}\label{sec: conclusions}
In this paper we have developed an inner and an outer bound for the strong coordination region when the input and output signals 
have to be coordinated with the source and reconstruction. 
Despite the fact that we have fully characterized the region in some special cases in Section \ref{sec: special cases}, 
inner and outer bound differ in general on the information constraint. Closing this gap is left for future study.

The polar coding proof in Section \ref{sec: polarcoding}, though it provides an explicit coding scheme, relies on a chaining construction over several blocks, 
which is not practical for delay-constrained applications. This is another issue that may be studied further.

Some important questions have not been addressed in this study and are left for future work. By coordinating signals and 
actions, the synthesized sequences would appear to be statistically indistinguishable from i.i.d. to an outside observer. 
As suggested in the example in Section \ref{secrecy}, this property could be exploited in a more general setting where two legitimate nodes wish to coordinate while concealing their actions from an eavesdropper who observes the signals sent over the channel.

Moreover, our results could be extended to a strategic coordination setting. This represents a scenario where the objectives of the two agents are not necessarily aligned, and has been investigated for empirical coordination in \cite{treust2016information, LeTreustTomala17}.

\appendices

\section{Proof of preliminary results} \label{appendix prel}

\begin{IEEEproof}[Proof of Lemma \ref{lemmit}]
We have $$I(A_t;A_{\sim t})= H(A_t)-H(A) + H(A)-H(A_t|A_{\sim t})$$ and we prove separately that 
{\allowdisplaybreaks
\begin{align*}
 & H(A)-H(A_t|A_{\sim t})  \leq f(\varepsilon),\\
 & H(A_t)-H(A) \leq f(\varepsilon).
\end{align*}}

First, we need two results.
\begin{lem}[$\mbox{\cite[Lemma 2.7]{csiszar2011information}}$]\label{csi2.7}
Let $P$ and $Q$  two distributions on $\mathcal A $  
such that 
$\tv(P,Q) = \varepsilon$ and $\varepsilon \leq 1/2$, then
\vspace{-0.3cm}
\begin{equation*}
 \lvert H(P)-H(Q) \rvert \leq  \varepsilon \log{\frac{\lvert \mathcal A \rvert}{\varepsilon}}.
\end{equation*}
\end{lem}

\begin{lem}[$\mbox{\cite[Lemma 3.2]{yassaee2014achievability}}$]\label{yas3.2'}
If $\tv(P_{A}P_{B|A}, Q_{A}Q_{B|A})\leq \varepsilon$ then
\begin{equation*}
 \mathbb P\{A \in \mathcal A | \tv(P_{B|A=a}, Q_{B|A=a}) \leq \sqrt{\varepsilon}\} \geq 1-2 \sqrt{\varepsilon}.
\end{equation*}
\end{lem}
\vspace{0,3cm}

Now, consider the set
$\mathcal E:=\{ \mathbf a \in  \mathcal A^{n-1}| \tv(P_{A_t |A_{\sim t}=\mathbf a}, \bar P_{A}) \leq \sqrt{\varepsilon}\}.$
By Lemma \ref{yas3.2'}, $\mathbb  P \{\mathcal E\} \geq 1-2 \sqrt{\varepsilon}$.

Then, we have
{\allowdisplaybreaks
\begin{align*}
& H(A)-H(A_t|A_{\sim t}) =  H(A) -    \!\!\!  \!\!\!\sum_{\mathbf a \in \mathcal A^{n-1}} \!\! \!\! P_{A_{\sim t}}(\mathbf a)  H(A_t|A_{\sim t}= \mathbf a) = \!\!\!\! \sum_{\mathbf a \in \mathcal A^{n-1}} \!\!\!\! \left(P_{A_{\sim t}}(\mathbf a) H(A)\!  -\!   P_{A_{\sim t}}(\mathbf a)  H(A_t|A_{\sim t}= \mathbf a) \right)\\
& =  \sum_{\mathbf a \in \mathcal E} \left( P_{A_{\sim t}}(\mathbf a) H(A)- P_{A_{\sim t}}(\mathbf a)  H(A_t|A_{\sim t}= \mathbf a)\right)  +  \sum_{\mathbf a \in \mathcal E^c} \left( P_{A_{\sim t}}(\mathbf a) H(A)- P_{A_{\sim t}}(\mathbf a) H(A_t|A_{\sim t}= \mathbf a)\right) \stepcounter{equation}\tag{\theequation}\label{part1}\\
& \overset{(a)}{\leq} \sum_{\mathbf a \in \mathcal E} P_{A_{\sim t}}(\mathbf a) \delta + \mathbb P \{\mathcal E^c\} \left(H(A_t)+H(A)\right)  \leq \delta + 2 \sqrt{\varepsilon} \left(2 H(A) + \delta \right)
\end{align*}}where $(a)$ comes from the fact that by Lemma \ref{csi2.7} for $\mathbf a \in \mathcal E$  
\begin{equation*}
\lvert H(A_t|A_{\sim t}=\mathbf a)-H(A) \rvert \leq \varepsilon \log{\frac{\lvert \mathcal A \rvert}{\varepsilon}}:=\delta.
\end{equation*}

Lemma \ref{csi2.7} also implies that
\begin{equation}\label{part2}
\lvert H(A_t)-H(A) \rvert \leq \delta.
\end{equation}
Hence by \eqref{part1} and \eqref{part2}, we have $I(A_t;A_{\sim t}) \leq 2 \sqrt{\varepsilon} (2H(A)+\delta)+ 2\delta$.
\end{IEEEproof}

\vspace{0,2cm}
\begin{IEEEproof}[Proof of Lemma \ref{lemab}]
The proof of \eqref{lemab1} comes directly from Lemma \ref{lemmit}: 
\begin{equation}\label{eqab}
 \sum_{t=1}^{n} I(A_t;A^{t-1} B_{\sim t}|B_t)
\leq  \sum_{t=1}^{n} I(A_t;A_{\sim t} B_{\sim t}|B_t) 
 \leq \sum_{t=1}^{n} I(A_t B_t;A_{\sim t} B_{\sim t}) \leq n f(\varepsilon).
\end{equation}

To prove \eqref{lemab2}, we have
{\allowdisplaybreaks
\begin{align*}
  H(C|B^{n})  &\geq   I(A^{n}; C|B^{n}) = \sum_{t=1}^{n} I(A_t;C|A^{t-1} B_{\sim t}B_t)\\
& = \sum_{t=1}^{n} I(A_t;CA^{t-1}B_{\sim t}|B_t)- \sum_{t=1}^{n} I(A_t; A^{t-1} B_{\sim t}|B_t) \\
& \geq \sum_{t=1}^{n} I(A_t;CB_{\sim t}|B_t) - \sum_{t=1}^{n} I(A_t;A^{t-1} B_{\sim t}|B_t)\\
& \overset{(a)}{\geq} \sum_{t=1}^{n} I(A_t;CB_{\sim t}|B_t)-n f(\varepsilon) 
= n I(A_T;CB_{\sim T}|B_T T) -n f(\varepsilon)  \\
&= n I(A_T;CB_{\sim T} T|B_T ) -n I(A_T; T|B_T)  -n f(\varepsilon)\\
& \geq  n I(A_T;CB_{\sim T} T|B_T ) -n I(A_T  B_T;T)  -n f(\varepsilon)
\end{align*}}where $(a)$ comes from Lemma \ref{lemab}.

\end{IEEEproof}

\section{Proof of Remark \ref{binnings2}} \label{appendix bin}
We want to prove that there exists a fixed binning that satisfy both the conditions in Section \ref{rc gen} and Section \ref{rf gen}.
If we denote with $\mathbb E_{\varphi_1 \varphi_2 }$ and $\mathbb E_{\varphi_2}$ the  expected value with respect to the random binnings, 
for all $\varepsilon$, there exists $\bar n$  such that $\forall n \geq \bar n$
{\allowdisplaybreaks
\begin{align*}
& \mathbb E_{\varphi_1 \varphi_2 } \left[\tv \left(\bar P^{\varphi_1 \varphi_2}_{ U^{n} S^{n} Z^{n} FC}, Q_{F} Q_{C}  \bar P_{U^{n} S^{n} Z^{n}} \right)\right]  < \frac{\varepsilon}{2}\\
& \mathbb E_{ \varphi_2} [\tv (\bar P_{U^{n} S^{n} Z^{n} X^{n} Y^{n} V^{n} F}^{ \varphi_2}, Q_{F} \bar P_{U^{n} S^{n} Z^{n} X^{n} Y^{n} V^{n} } )] < \frac{\varepsilon}{2}
\end{align*}}which implies by Markov's inequality
{\allowdisplaybreaks
\begin{align*}
&\mathbb P_{\varphi_1 \varphi_2} \left\{ \tv \left(\bar P^{\varphi_1 \varphi_2}_{U^{n} S^{n} Z^{n} FC}, Q_{F} Q_{C}  \bar P_{U^{n} S^{n} Z^{n}}\right) <  \varepsilon \right\} > \frac{1}{2} \\
& \mathbb P_{\varphi_2}\{\tv (\bar P_{U^{n} S^{n} Z^{n} X^{n} Y^{n} V^{n} F}^{\varphi_2},  Q_{F} \bar P_{U^{n} S^{n} Z^{n} X^{n} Y^{n} V^{n}} )  < \varepsilon \} > \frac{1}{2}. \stepcounter{equation}\tag{\theequation}\label{min122}
\end{align*}
}

In Section \ref{rc gen} and \ref{rf gen} we have chosen the binnings  $(\varphi'_1, \varphi'_2)$  and $\varphi''_2$ respectively such that 
{\allowdisplaybreaks
\begin{align*}
&\lim_{n \to \infty} \tv \left(\bar P^{\varphi'_1 \varphi'_2}_{U^{n} S^{n} Z^{n} FC}, Q_{F} Q_{C}  \bar P_{U^{n} S^{n} Z^{n}}  \right)=0\\
& \lim_{n \to \infty} \tv (\bar P^{\varphi''_2}_{U^{n} S^{n} Z^{n} X^{n} Y^{n} V^{n} F}, Q_{F} \bar P_{U^{n} S^{n} Z^{n} X^{n} Y^{n} V^{n}})=0.
\end{align*}
}
It follows from \eqref{min122} that the intersection of the two sets is non-empty, therefore there exists a binning 
$\varphi_2^{*}$ that satisfies both conditions.
\hfill \IEEEQED

\section{Comparison between $I(XUS;YZ)$ and $I(XS;Y)+I(U;Z)$}\label{appendix compare}

Observe that
{\allowdisplaybreaks
\begin{align*}
  I(XUS; YZ)& = I(XS;YZ)+I(U; YZ|XS)\\
  &\overset{(a)}{=}  I(XS;Y) +I(U;Z) + I(SXY;Z) -I(Y;Z) +I(UZ;SXY) - I(SXY;Z)-I(U;XS)\\
  & \overset{(b)}{=} I(XS;Y) +I(U;Z)   -I(Y;Z) +I(UZ;SX) -I(U;XS)\\
  & = I(XS;Y) +I(U;Z) - I(Y;Z) +I(Z;SX|U) +I(U;XS) -I(U;XS)\\
  & \overset{(c)}{=} I(XS;Y) +I(U;Z)  -I(Y;Z) +I(Z;S|U) 
\end{align*}}where $(a)$ follows from basic properties of the mutual information, $(b)$ and $(c)$ from the
Markov chains $ Y-XS-UZ$ and  $X-US-Z$ respectively.
If we note $\Delta:=I(Z;S|U)-I(Y;Z) $, then  $I(XUS; YZ)= I(XS;Y) +I(U;Z) + \Delta$ where 
$\Delta$ may be either positive or negative, for instance:
\begin{itemize}
 \item in the special case where $S-U-Z$ holds and $ Y=Z$, $\Delta=-H(Y) \leq 0$,
 \item if we suppose $Y$ independent of $Z$, $\Delta=I(Z;S|U) \geq 0$.
\end{itemize}
\hfill \IEEEQED

\section{Proof of Lemma \ref{ossmc1}.}\label{appendix ossmc1}
 To prove that $I(Y_t {Z}_t; C, X_{\sim t} U_{\sim t} S_{\sim t} Y_{\sim t}{Z}_{\sim t}|X_t U_t S_t)=0$, we have
 {\allowdisplaybreaks
\begin{align*}
 I(Y_t {Z}_t ; C X_{\sim t} S_{\sim t} U_{\sim t}& Y_{\sim t} {Z}_{\sim t} |X_t S_t U_t)\\
& = I({Z}_t; C X_{\sim t} S_{\sim t} U_{\sim t} Y_{\sim t} {Z}_{\sim t}|X_t S_t U_t) + I( Y_t; C X_{\sim t} S_{\sim t} U_{\sim t} Y_{\sim t} {Z}_{\sim t}|X_t S_t U_t {Z}_t )\\
& = I({Z}_t; C X_t X_{\sim t} S_{\sim t} U_{\sim t} Y_{\sim t} {Z}_{\sim t}|S_t U_t)-I({Z}_t; X_t|S_t U_t)\\
&\quad  + I(Y_t; C U_t {Z}_t X_{\sim t} S_{\sim t} U_{\sim t} Y_{\sim t} {Z}_{\sim t}|X_t S_t)-I(Y_t;U_t {Z}_t|X_t S_t)\\
& \leq I({Z}_t; C X^{n} S_{\sim t} U_{\sim t} Y_{\sim t} {Z}_{\sim t}|S_t U_t) + I(Y_t; C U^{n} Z^{n} X_{\sim t} S_{\sim t}  Y_{\sim t} |X_t S_t)\\
& \leq I({Z}_t; C X^{n} Y^{n} S_{\sim t} U_{\sim t}  {Z}_{\sim t}|S_t U_t) + I(Y_t; C U^{n} Z^{n} X_{\sim t} S_{\sim t}  Y_{\sim t} |X_t S_t) 
\end{align*}}where both $I({Z}_t; C X^{n} Y^{n} S_{\sim t} U_{\sim t}  {Z}_{\sim t}|S_t U_t)$ and $I(Y_t; C U^{n} Z^{n} X_{\sim t} S_{\sim t}  Y_{\sim t} |X_t S_t)$ are equal to zero because 
by \eqref{markov chain general} the following Markov chains hold:
{\allowdisplaybreaks
\begin{align*}
 {Z}_t -(U_t, S_t)-(C, X^{n}, Y^{n}, U_{\sim t}, S_{\sim t}, {Z}_{\sim t}), \qquad Y_t-(X_t,S_t)-(C, Z^{n}, U^{n}, X_{\sim t}, S_{\sim t}, Y_{\sim t} ).
\end{align*}
}\hfill \IEEEQED

\vspace{-0,3cm}
\section{Proof of \eqref{byfano}} \label{appendix fano}

Define the  event of error $E$ as follows:
\begin{equation*}
 E := \begin{cases}
      0 \quad \mbox{ if } \quad U^{n} = V^{n}\\
      1 \quad \mbox{ if } \quad U^{n} \neq V^{n}
     \end{cases}.
\end{equation*}

We note $p_e:= \mathbb{P} \{U^{n}\neq V^{n}\}$ and recall that by hypothesis the distribution $P_{U^{n} S^{n} Z^{n} X^{n} Y^{n} V^{n}}$ is $\varepsilon$-close 
in total variational distance to the i.i.d. distribution 
$\bar P_{U S Z X Y V}^{\otimes n}$ where the decoder is lossless. 
Then $\mathbb V (P_{V^{n}},  \mathds 1_{ V =U }^{\otimes n}) < \varepsilon$
and therefore $p_e$ vanishes.
By Fano's inequality \cite{fano1961transmission}, we have
\begin{equation}\label{eqfano}
H( U^{n}|C, Y^{n}, Z^{n}) \leq H_2(p_e)+p_e \log{(\lvert \mathcal U^n\rvert -1)}.
\end{equation}Since $p_e$ vanishes, $H_2(p_e)$ is close to zero and the right-hand side of \eqref{eqfano} goes to zero.
Hence, we have that $H( U^{n}|C, Y^{n}, Z^{n}) \leq n f(\varepsilon)$, where $f(\varepsilon)$ 
denotes a function which tends to zero as $\varepsilon$ does.
\hfill \IEEEQED

\section{Proof of Proposition \ref{teoUVotimesX}} \label{appendix UVotimesX}
\subsection{Achievability}
We show that $\mathcal R_{UV \otimes X}$ is contained in the region $\mathcal R_{\text{PC}}\cap \mathcal R_{\text{SEP}}$ and thus it is achievable. 

We consider the subset of $\mathcal R_{\text{SEP}}$ when $\bar P_{Y|X}( \mathbf{y}|\mathbf{x})={\mathds 1}_{X=Y} \{\mathbf{x}= \mathbf{y} \}$ as the union of all
$\mathcal R_{\text{SEP}}(W)$ with $W=(W_1,W_2)$ that satisfies
{\allowdisplaybreaks
\begin{align*}
 &\bar P_{UW_1W_2XV}= \bar P_{U} \bar P_{W_2|U} \bar P_{V| W_2}  \bar P_{X} \bar P_{W_1|X},\\
 & I(W_1;X) \geq  I(W_2;U),  \stepcounter{equation}\tag{\theequation}\label{sepmi} \\
 & R_0 \geq I(W_2;UV).
\end{align*}
}
\vspace{-0,2cm}
Similarly, $\mathcal R_{\text{PC}}$ is the union of all 
$\mathcal R_{\text{PC}}(W)$ with $W$ that satisfies
{\allowdisplaybreaks
\begin{align*}
 &\bar P_{UWXV}= \bar P_{U} \bar P_{W|U} \bar P_{X|UW}  \bar P_{V|WX},\\
 & H(X) \geq I(WX;U),   \stepcounter{equation}\tag{\theequation}\label{pcmi} \\
 & R_0 \geq I(W;UV|X).
\end{align*}
}If we choose $W=(W_1,W_2)$ and we
add the hypothesis that $(W_2, U,V)$ is independent of $(W_1,X)$, \eqref{pcmi} becomes
{\allowdisplaybreaks
\begin{align*}
 &\bar P_{UW_1W_2XV}= \bar P_{U} \bar P_{W_2|U} \bar P_{V| W_2}  \bar P_{W_1} \bar P_{X|W_1},\\
 & H(X) \geq I(W_1W_2X;U)=I(W_2;U), \stepcounter{equation}\tag{\theequation}\label{pcm2} \\
 & R_0 \geq I(W_1W_2;UV|X)=I(W_2;UV).
\end{align*}}
Note that if we identify $X=W_1$, we have
$H(X) = I(W_1;X)$ and $\bar P_{W_1} \bar P_{X|W_1}=\bar P_{X} \bar P_{W_1|X}= \bar P_{X} {\mathds 1}_{X=W_1}$.
Then, there exists a subset of $\mathcal R_{\text{SEP}}$  and  $\mathcal R_{\text{PC}}$ defined as 
the union over all $W_2$ such that
{\allowdisplaybreaks
\begin{align*}
 &\bar P_{UW_2XV}= \bar P_{U} \bar P_{W_2|U} \bar P_{V| W_2}  \bar P_{X} ,\\
 & H(X) \geq I(W_2;U), \stepcounter{equation}\tag{\theequation}\label{inter} \\
 & R_0 \geq I(W_2;UV).
\end{align*}}Finally, observe that, 
by definition of the region \eqref{UVotimesX}, 
$\mathcal R_{UV \otimes X}$ is the union over all the possible choices for $W_2$ that satisfy \eqref{inter} and 
therefore $ \mathcal R_{UV \otimes X} \subseteq  \mathcal R_{\text{PC}} \cap \mathcal R_{\text{SEP}}$.

\subsection{Converse}
Consider a code $(f_n,g_n)$ that induces a distribution $P_{U^{n} X^{n} V^{n}}$ that is $\varepsilon$-close in total variational
distance to the i.i.d. distribution $\bar P_{U V}^{\otimes n} \bar P_{X}^{\otimes n} $.
Let $T$ be the random variable defined in Section \ref{outer}.

Then, we have
{\allowdisplaybreaks
\begin{align*} 
 nR_0 &\geq H(C) \overset{(a)}{\geq} I(U^{n}  V^{n};C|X^n)  = I(U^{n}  V^{n};C X^{n}) - I(U^{n}  V^{n}; X^{n})  \\
 & \overset{(b)}{\geq}  I(U^{n}  V^{n};C X^{n})- n f(\varepsilon) =  
 \sum_{t=1}^n I(U_t  V_t ;C X^{n}| U^{t-1} V^{t-1} ) - n f(\varepsilon)\\
& = \sum_{t=1}^n I(U_t  V_t ; C X^{n} U^{t-1} V^{t-1} ) - \sum_{t=1}^n I(U_t  V_t ; U^{t-1} V^{t-1} ) - n f(\varepsilon)\\
& \overset{(c)}{\geq}\sum_{t=1}^n I(U_t  V_t ; C X^{n} U^{t-1} V^{t-1} ) - 2nf(\varepsilon) 
\geq \sum_{t=1}^n I(U_t  V_t ; C X^{n} U^{t-1} ) - 2nf(\varepsilon)\\
&= n  I(U_T  V_T ; C X^{n} U^{T-1}|T ) - 2nf(\varepsilon)=
n  I(U_T  V_T ; C X^{n} U^{T-1} T ) -  n  I(U_T  V_T ;  T ) - 2nf(\varepsilon)\\
& \overset{(d)}{\geq} n  I(U_T  V_T ; C X^{n} U^{T-1} T ) - 3nf(\varepsilon)
\end{align*}}where $(a)$ follows from basic properties of entropy and mutual information and $(b)$ from
the upperbound on the mutual information in Lemma \ref{lem1csi} since we assume $\mathbb V(P_{U^nV^nX^n}, \bar{P}_{UV}^{\otimes n}\bar{P}_{X}^{\otimes n}) \leq \varepsilon$
and $ \lvert \mathcal U \times \mathcal V \rvert \geq 4$.
Finally, since the distributions are close to i.i.d. by hypothesis, $(c)$ and $(d)$ come from Lemma \ref{lemmit} and 
\cite[Lemma VI.3]{cuff2013distributed} respectively. 

For the second part of the converse, observe that
{\allowdisplaybreaks
\begin{align*}
 & 0  =  H(X^{n}) - I(X^{n}; U^{n} C) - H(X^{n}|U^{n} C) \leq \sum_{t=1}^{n} H(X_t) - \sum_{t=1}^{n} I(X^{n}; U_t|U^{t-1} C)  - H(X^{n}|U^{n} C)\\
 & \leq \sum_{t=1}^{n} H(X_t) - \sum_{t=1}^{n} I(X^{n}; U_t|U^{t-1} C) = \sum_{t=1}^{n} H(X_t) - \sum_{t=1}^{n} I(X^{n} U^{t-1} C; U_t) + \sum_{t=1}^{n} I(U^{t-1} C; U_t)\\
 & \overset{(e)}{=} \sum_{t=1}^{n} H(X_t) - \sum_{t=1}^{n} I(X^{n} U^{t-1} C; U_t) = n H(X_T|T)  - n I(X^{n} U^{T-1} C; U_T|T) \\
 &\leq n H(X_T) - n I(X^{n} U^{T-1} C T; U_T) + n I(T; U_T)  = n H(X_T) - n I(X^{n} U^{T-1} C T; U_T).
 \end{align*}
}where $(e)$ follows from the i.i.d. nature of the source $\bar P_{U}$.

Then, we identify the auxiliary random variable $W_t$ with $(C,X^n, U^{t-1})$ for 
each $t \in \llbracket 1,n\rrbracket$ and $W$ with $(W_T, T)=$ $(C,X^n, U^{T-1}, T)$.
\hfill \IEEEQED

\section{Proof of cardinality bounds}\label{appendix bounds}

Here we prove separately the cardinality bound for all the outer bounds in this paper. 
Note that since the proofs are basically identical we will prove it in the first case and then omit most details in all the other cases.
First, we state the Support Lemma \cite[Appendix C]{elgamal2011nit}.
\begin{lem}\label{support lemma}
 Let $\mathcal{A}$ a finite set and $\mathcal W$ be an arbitrary set. Let $\mathcal P$ be a connected compact subset of probability mass functions on $\mathcal A$ and $P_{A|W}$ be 
 a collection of conditional probability mass functions on $\mathcal A$. Suppose that $h_i(\pi)$, $i=1, \ldots, d$, are real-valued continuous functions of $\pi \in \mathcal P$. Then for every $W$ defined 
 on $\mathcal W$ there exists a random variable $W'$ with $\lvert \mathcal W' \rvert \leq d$ and a collection of conditional probability mass functions $P_{A|W'} \in \mathcal P$ such that
 \begin{equation*}
  \sum_{w \in \mathcal W} P_W(w) h_i(P_{A|W}(a|w))=\sum_{w \in \mathcal W'} P_{W'}(w)h_i(P_{A|W'}(a|w)) \quad i=1, \ldots, d.
 \end{equation*}
\end{lem}

\begin{IEEEproof}[Proof of cardinality bound of Theorem \ref{teoisit}]
 We consider the probability distribution 
$\bar P_{U}  \bar P_{W|U}  \bar P_{X|UW} \bar P_{Y|X}  \bar P_{V|WY}$
that is $\varepsilon$-close in total variational distance to the i.i.d. distribution. 
We identify $\mathcal{A}$ with $\{1,\ldots,\lvert \mathcal{A}\rvert \}$ 
and we consider $\mathcal P$  a connected compact subset of probability mass functions on $\mathcal A=\mathcal U \times \mathcal X \times \mathcal Y \times \mathcal V$.
Similarly to \cite{treust2017joint}, suppose that $h_i(\pi)$, $i=1, \ldots, \lvert \mathcal A \rvert +4$, 
 are real-valued continuous functions of $\pi \in \mathcal P$ such that:
{\allowdisplaybreaks
\begin{align*}
 h_i (\pi) =
 \begin{cases}
   \pi(i) &\mbox{for } i= 1, \ldots, \lvert \mathcal A \rvert -1  \\
   H(U) &\mbox{for } i= \lvert \mathcal A  \rvert  \\
   H(UXV|Y)& \mbox{for } i= \lvert \mathcal  A \rvert +1\\
   H(Y|UX)& \mbox{for } i= \lvert \mathcal  A \rvert +2\\
   H(V|Y)& \mbox{for } i= \lvert \mathcal  A \rvert +3\\
   H(V|UXY)& \mbox{for } i= \lvert \mathcal  A \rvert +4\\
 \end{cases}.
\end{align*}}
Then by Lemma \ref{support lemma} there exists an auxiliary random variable $W'$ taking at most 
$ \lvert \mathcal U \times \mathcal X \times \mathcal Y \times \mathcal V  \rvert +4$ values such that:
{\allowdisplaybreaks
\begin{align*}
& H(U|W)= \sum_{w \in \mathcal W} P_W(w) H(U|W\!=\!w)= \sum_{w \in \mathcal W'} P_{W'}(w) H(U|W'\!=\!w)=H(U|W'),\\
& H(UXV|YW)= \sum_{w \in \mathcal W} P_W(w) H(UXV|YW\!=\!w)= \sum_{w \in \mathcal W'} P_{W'}(w) H(UXV|YW'\!=\!w)=H(UXV|YW'),\\
& H(Y|UXW)= \sum_{w \in \mathcal W} P_W(w) H(Y|UXW\!=\!w)= \sum_{w \in \mathcal W'} P_{W'}(w) H(Y|UXW'\!=\!w)=H(Y|UXW'),\\
& H(V|YW)= \sum_{w \in \mathcal W} P_W(w) H(V|YW\!=\!w)= \sum_{w \in \mathcal W'} P_{W'}(w) H(V|YW'\!=\!w)=H(V|YW'),\\
& H(V|UXYW)= \sum_{w \in \mathcal W} P_W(w) H(V|UXYW\!=\!w)= \sum_{w \in \mathcal W'} P_{W'}(w) H(V|UXYW'\!=\!w)=H(V|UXYW').
\end{align*}}

Then the constraints on the conditional distributions, the information constraints and the Markov chains are still verified 
since we can rewrite the inequalities in \eqref{eq: regionisit2} and the Markov chains in \eqref{markov chain isit} as 
{\allowdisplaybreaks
\begin{align*}
& H(U)-H(U|W) \leq I(X;Y),\\
& R_0 \geq H(UXV|Y)-H(UXV|WY),\\
& I(Y;UW|X)= H(Y|X)- H(Y|UXW)=0,\\
& I(V;UX|YW)= H(V|YW)-H(V|UXYW)=0.
\end{align*}}
Note that we are not forgetting any constraints: to preserve $H(U)-H(U|W) \leq I(X;Y)$ we only need to fix $H(U|W)$ because the other quantities 
depend only on the joint distribution $P_{UXYV}$ (which is preserved). 
Similarly, once the distribution $\bar P_{UXYV}$ is preserved, the difference $H(UXV|Y)-H(UXV|WY)$ 
only depends on the conditional entropy $H(UXV|WY)$ and the difference $H(Y|X)- H(Y|UXW)$ only 
depends on $H(Y|UXW)$.
\end{IEEEproof}

\vspace{0.3cm}

\begin{IEEEproof}[Proof of cardinality bound of Theorem \ref{teouv}]
 Here let $\mathcal A=\mathcal U \times \mathcal S \times \mathcal Z \times \mathcal X \times \mathcal Y \times \mathcal V $ 
and suppose that $h_i(\pi)$, $i=1, \ldots, \lvert \mathcal A \rvert +5$, are real-valued continuous functions of $\pi \in \mathcal P$ such that:
{\allowdisplaybreaks
\begin{align*}
 h_i (\pi) =
 \begin{cases}
   \pi(i) & \mbox{for } i= 1, \ldots, \lvert \mathcal A  \rvert -1  \\
   H(U) & \mbox{for } i= \lvert \mathcal A \rvert  \\
   H(USXV|YZ) & \mbox{for } i= \lvert \mathcal A \rvert +1\\
   H(Y|USX) & \mbox{for } i= \lvert \mathcal A \rvert +2\\
   H(V|YZ) & \mbox{for } i= \lvert \mathcal A \rvert +3\\
   H(V|YZUSX) & \mbox{for } i= \lvert \mathcal A \rvert +4\\
   H(Z|YUSX) & \mbox{for } i= \lvert \mathcal A \rvert +5
 \end{cases}.
\end{align*}}
By the Markov chain $Z-(U,S)-(X,Y,W)$, the mutual information $I(Z;XYW|US)$ is zero
and once the distribution $\bar P_{USZXYV}$ is preserved, the mutual information
$I(Z;XYW|US)= H(Z|US)- H(Z|USXYW)$ only depends on $H(Z|YUSX)$.
Therefore there exists an auxiliary random variable $W'$ taking at most
$ \lvert \mathcal U \times \mathcal S \times \mathcal Z \times  \mathcal X \times \mathcal Y \times \mathcal V  \rvert +5 $ 
values such that the constraints on the conditional distributions and the information constraints are still verified.
\end{IEEEproof}

\vspace{0.3cm}

\begin{IEEEproof}[Proof of cardinality bound of Theorem \ref{teopc}]
 Similarly, here let $\mathcal A=\mathcal U \times \mathcal Z \times \mathcal X \times \mathcal V $ and suppose that $h_i(\pi)$, $i=1, \ldots, \lvert \mathcal A \rvert +4$, 
 are real-valued continuous functions of $\pi \in \mathcal P$ such that:
 {\allowdisplaybreaks
\begin{align*}
 h_i (\pi) =
 \begin{cases}
   \pi(i) &  \mbox{for } i= 1, \ldots, \lvert \mathcal A  \rvert -1  \\
   H(U|XZ) &\mbox{for } i= \lvert \mathcal A \rvert  \\
   H(UV|XZ)  & \mbox{for } i= \lvert \mathcal A \rvert +1\\
   H(V|XZ) & \mbox{for } i= \lvert \mathcal A \rvert +2\\
   H(V|ZUX) & \mbox{for } i= \lvert \mathcal A \rvert +3\\
   H(Z|UX) & \mbox{for } i= \lvert \mathcal A \rvert +4
 \end{cases}.
\end{align*}}
The information constraint in Theorem \ref{teopc} can be written as 
{\allowdisplaybreaks
\begin{align*}
  &H(X)+I(W;Z|X) - I(WX;U) = H(X)+I(WX;Z)-I(Z;X)- I(WX;U) \\
  &{\overset{{(a)}}{=}} H(X)-I(Z;X)+I(WX;Z)-I(WX;UZ)=H(X)-I(Z;X)+I(WX;U|Z)\\
  &=H(X)-I(Z;X)+H(U|Z)-H(U|WXZ) \geq 0
\end{align*}}
where $(a)$ follows from the fact that $I(WX;UZ)=I(WX;U)$ by the Markov chain $Z-U-(W,X)$.
By fixing $H(UV|XZ)$ the constraint on the bound for $R_0$ is satisfied and similarly to the previous cases
the Markov chains are still verified. 
Thus there exists an auxiliary random variable $W'$ taking at most 
$ \lvert \mathcal U  \times \mathcal Z \times  \mathcal X \times \mathcal V  \rvert +4 $ values.
\end{IEEEproof}

\vspace{0.3cm}

\begin{IEEEproof}[Proof of cardinality bound of Theorem \ref{teolossless}]
 For the lossless decoder, we rewrite the constraints in the equivalent characterization of the region \eqref{eq: regionldmael} as:
{\allowdisplaybreaks
 \begin{align*}
H(U)&\leq H(YZ)- H(YZ|UW),\\
R_0 &\geq I(W;USX|YZ)+H(U|WYZ)= H(USX|YZ)- H(USX|WYZ)+H(U|WYZ)\\
&=H(USX|YZ)-H(USX)+H(U)+H(SX|U)-H(SX|UWYZ).
\end{align*}}
Then let $\mathcal A=\mathcal U \times \mathcal S \times \mathcal Z \times \mathcal X \times \mathcal Y$
and suppose that $h_i(\pi)$, $i=1, \ldots, \lvert \mathcal A \rvert +3$, 
 are real-valued continuous functions of $\pi \in \mathcal P$ such that:
 {\allowdisplaybreaks
\begin{align*}
 h_i (\pi) =
 \begin{cases}
   \pi(i) &  \mbox{for } i= 1, \ldots, \lvert \mathcal A  \rvert -1  \\
   H(YZ|U) & \mbox{for } i= \lvert \mathcal A \rvert  \\
   H(SX|UYZ)  & \mbox{for } i= \lvert \mathcal A \rvert +1\\
   H(Y|USX) & \mbox{for } i= \lvert \mathcal A \rvert +2\\
   H(Z|YUSX) & \mbox{for } i= \lvert \mathcal A \rvert +3
 \end{cases}.
\end{align*}}and therefore there exists an auxiliary random variable $W'$ taking at most $\lvert \mathcal U \times \mathcal S \times \mathcal Z \times  \mathcal X \times \mathcal Y   \rvert +3 $ values.
\end{IEEEproof}

\vspace{0.3cm}

\begin{IEEEproof}[Proof of cardinality bound of Theorem \ref{teoseparation}]
 For case of separation between channel and source, we consider the following equivalent characterization of the information constraints:
{\allowdisplaybreaks
 \begin{align*}
& 0 \leq H(YZ)-H(YZ|W_1 W_2) -H(US)+H(US|W_1 W_2),\\
& R_0 \geq H(USXV|YZ)-H(USXV|YZ W_1 W_2)
\end{align*}}In this case we have $W=(W_1, W_2)$. 
Let $\mathcal A=\mathcal U \times \mathcal S \times \mathcal Z \times \mathcal X \times \mathcal Y \times \mathcal V $ 
and suppose that $h_i(\pi)$, $i=1, \ldots, \lvert \mathcal A \rvert +3$,  
 are real-valued continuous functions of $\pi \in \mathcal P$ such that:
 {\allowdisplaybreaks
\begin{align*}
 h_i (\pi) =
 \begin{cases}
   \pi(i) &  \mbox{for } i= 1, \ldots, \lvert \mathcal A  \rvert -1  \\
   H(US) &\mbox{for } i= \lvert \mathcal A \rvert  \\
   H(USXV|YZ) & \mbox{for } i= \lvert \mathcal A \rvert +1\\
   H(V|Z)& \mbox{for } i= \lvert \mathcal  A \rvert +2\\
   H(V|UZ)& \mbox{for } i= \lvert \mathcal  A \rvert +3\\ 
 \end{cases}.
\end{align*}}
Then there exists an auxiliary random variable $W'=(W'_1, W'_2)$ taking at most
$ \lvert \mathcal U \times \mathcal S \times \mathcal Z \times  \mathcal X \times \mathcal Y  \times \mathcal V \rvert +3$ values.
\end{IEEEproof}

\section{Polar coding achievability proofs}\label{appendix polar}
Here we prove the results used in the achievability proof of Theorem \ref{polarregion}.

\begin{IEEEproof}[Proof of Lemma \ref{scblock}]
 Similarly to \cite[Lemma 5]{Cervia2016}, we first prove that in each block $i \in \llbracket 1,k\rrbracket$ 
\begin{equation}\label{emp1}
 \D\left(  \bar P_{UW}^{\otimes n} \Big\Arrowvert  \widetilde P_{U^{n}_{(i)} \widetilde W_{(i)}^n }\right) =n \delta_n.
\end{equation}
In fact, we have
{\allowdisplaybreaks
  \begin{align*}
 \D\left(   \bar P_{UW}^{\otimes n} \Big\Arrowvert  \widetilde P_{U^{n}_{(i)} \widetilde W_{(i)}^n }\right) 
 & {\overset{{(a)}}{=}} \mathbb{D}\left(  \bar P_{UR}^{\otimes n}  \Big\Arrowvert \widetilde P_{U^{n}_{(i)} \widetilde R_{(i)}^n } \right)
    {\overset{{(b)}}{=}} \mathbb{D}\left( \bar P_{R^n|U^n} \Big\Arrowvert \widetilde P_{\widetilde R_{(i)}^n |U^{n}_{(i)}} \Big| \bar P_{U^n}\right)\\
   &  {\overset{{(c)}}{=}} \sum_{j =1}^n \mathbb D \left( \bar P_{R_j|R^{j-1} U^{n} }\Big\Arrowvert \widetilde P_{\widetilde R_{(i),j}| \widetilde R_{(i)}^{j-1} U_{(i)}^{n}} \Big| \bar P_{R^{j-1} U^{n}}  \right)\\
   &  {\overset{{(d)}}{=}}\sum_{j \in A_1 \cup A_2} \mathbb D \left(  P_{R_j|R^{j-1} U^{n} }\Big\Arrowvert \widetilde P_{\widetilde R_{(i),j}| \widetilde R_{(i)}^{j-1} U_{(i)}^{n} } \Big| \bar P_{R^{j-1} U^{n}}\right)\\
   &  {\overset{{(e)}}{=}} \sum_{j \in A_1 \cup A_2} \left( 1- H(R_j \mid R^{j-1} U^{n}) \right) {\overset{{(f)}}{<}} \delta_n \lvert \mathcal V _{W\mid U}\rvert 
   \leq n \delta_n,
  \end{align*}}where $(a)$ comes from the invertibility of $G_n$, $(b)$ and $(c)$ come from the chain rule, $(d)$ comes from \eqref{eq: p1} and \eqref{eq: p2}, 
$(e)$ comes from the fact that the conditional distribution $\widetilde P_{\widetilde R_{(i),j}| \widetilde R_{(i)}^{j-1} U_{(i)}^{n} } $ is uniform for $j$ in $A_1$ and $A_2$ and $(f)$ from \eqref{eq: hv}. 
Therefore, applying Pinsker's inequality to \eqref{emp1} we have
\begin{equation}\label{emp2}
 \tv\left(\widetilde P_{U^{n}_{(i)} \widetilde W_{(i)}^n }, \bar P_{UW}^{\otimes n}\right)\leq  \sqrt{2 \log 2} \sqrt{n \delta_n}:= \delta_n^{(2)} \to 0.
\end{equation}
Note that $X_{(i)}^n$ is generated symbol by symbol from $U^{n}_{(i)}$ and $\widetilde W_{(i)}^n$ via the 
conditional distribution $\bar P_{X|UW}$ and $Y_{(i)}^n$ is generated symbol by symbol via the channel $\bar P_{Y|X}$.
By Lemma \ref{cuff17}, we add first $X_{(i)}^n$ and then $Y_{(i)}^n$ and we obtain that for each $i \in \llbracket 1,k\rrbracket$,
\begin{equation}\label{oneblockconv}
 \tv \!\left(\!\widetilde P_{U^{n}_{(i)} \widetilde W_{(i)}^n X_{(i)}^n Y_{(i)}^n}, \bar P_{UWXY}^{\otimes n}\right)\!=\!  \tv\left(\!\widetilde P_{U^{n}_{(i)} \widetilde W_{(i)}^n }, \bar P_{UW}^{\otimes n}\right) \! \leq \! \delta_n^{(2)}
\end{equation}
and therefore the left-hand side of \eqref{oneblockconv} vanishes.

Observe that we cannot use Lemma \ref{cuff17} again because $V_{(i)}^n$ is generated using $\widehat W_{(i)}^n$ (i.e. the estimate of $W_{(i)}^n$ at the decoder)
and not $\widetilde W_{(i)}^n$.
By the triangle inequality for all $i \in \llbracket 1,k\rrbracket$ 
\begin{equation}\label{triangle} 
\tv \left( \widetilde P_{U^{n}_{(i)} \widehat W_{(i)}^n X_{(i)}^n Y_{(i)}^n},  \bar P_{UWXY}^{\otimes n} \right) \!\!
 \leq \! \tv \left( \widetilde P_{U^{n}_{(i)} \widehat W_{(i)}^n X_{(i)}^n Y_{(i)}^n},   \widetilde P_{U^{n}_{(i)} \widetilde W_{(i)}^n X_{(i)}^n Y_{(i)}^n} \right) \!\! + \! \tv \left( \widetilde P_{U^{n}_{(i)} \widetilde W_{(i)}^n X_{(i)}^n Y_{(i)}^n}, \bar P_{UWXY}^{\otimes n} \right).
\end{equation}We have proved in  \eqref{oneblockconv}  that the second term of the right-hand side in \eqref{triangle} goes to zero, we show that the first term tends to zero as well. 
To do so,  we apply Proposition \ref{theocoup} to 
$$\begin{matrix}[ll]
  A = U^{n}_{(i)} \widehat W_{(i)}^n  X_{(i)}^n Y_{(i)}^n & A'=U^{n}_{(i)} \widetilde W_{(i)}^n X_{(i)}^n Y_{(i)}^n\\
  P =\widetilde P_{U^{n}_{(i)} \widehat W_{(i)}^n  X_{(i)}^n  Y_{(i)}^n} & P'=\widetilde P_{U^{n}_{(i)} \widetilde W_{(i)}^n X_{(i)}^n Y_{(i)}^n}\\
\end{matrix}$$
on  $\mathcal A= \mathcal U \times \mathcal W \times \mathcal X \times \mathcal Y$.
Since it has been proven in \cite{arikan2010source} that
\begin{equation*}
p_e:= \mathbb P \left\{\widehat W_{(i)}^n  \neq \widetilde W_{(i)}^n \right\}=O(\delta_n)
\end{equation*}we find that 
$\tv \left( \widetilde P_{U^{n}_{(i)} \widehat W_{(i)}^n X_{(i)}^n Y_{(i)}^n},   \widetilde P_{U^{n}_{(i)} \widetilde W_{(i)}^n X_{(i)}^n Y_{(i)}^n} \right) \leq 2 p_e$
and therefore
 {\allowdisplaybreaks
\begin{align*}
&\tv \left( \widetilde P_{U^{n}_{(i)} \widehat W_{(i)}^n X_{(i)}^n Y_{(i)}^n}, \bar P_{UWXY}^{\otimes n} \right)\leq 2p_e+\delta_n^{(2)}=\delta_n^{(1)}\to 0.
\end{align*}}
Since $V^{n}_i$ is generated symbol by symbol from $\widehat W^{n}_i$ and $Y^{n}_i$, we apply Lemma \ref{cuff17} again and find 
 {\allowdisplaybreaks
\begin{align*}
&\tv\left(\widetilde P_{U^{n}_{(i)} \widehat W_{(i)}^n X_{(i)}^n Y_{(i)}^n V_{(i)}^n}, \bar P_{UWXYV}^{\otimes n}\right)\leq \delta_n^{(1)}\to 0. \stepcounter{equation}\tag{\theequation}\label{convblock}
\end{align*} }
\end{IEEEproof}

\vspace{0.3cm}

\begin{IEEEproof}[Proof of Lemma \ref{step2a'}]
 For $i \in \llbracket 2,k\rrbracket$, we have
{\allowdisplaybreaks
\begin{align*}
 \D \left(\widetilde P_{ L_{i-1:i} \bar C'} \Big\Arrowvert \widetilde P_{L_{i-1}\bar C'}  \widetilde P_{L_{i}} \right) &= I (L_{i-1} \bar C';L_{i})  {\overset{{(a)}}{=}}  I (L_{i}; \bar C' ) +  I (L_{i-1};L_{i}| \bar C') \\ 
 & {\overset{{(b)}}{=}}  I (L_{i}; \bar C' ) =  I (L_{i}; \widetilde R_i [A'_1] )  {\overset{{(c)}}{=}} \lvert A'_1 \rvert - H (\widetilde R_i [A'_1] | L_{i}) \\
& {\overset{{(d)}}{=}} \lvert A'_1 \rvert - H (R [A'_1] | W) +  \delta_n^{(4)} 
 {\overset{{(e)}}{\leq}} \lvert A'_1 \rvert - \sum_{j \in A'_1} H (R_j | R^{j-1} L) + \delta_n^{(4)} \stepcounter{equation}\tag{\theequation}\label{eqlem10}\\
&{\overset{{(f)}}{\leq}} \lvert A'_1 \rvert - \lvert A'_1 \rvert (1 - \delta_n) + \delta_n^{(4)}  \leq n \delta_n + \delta_n^{(4)}
\end{align*}
}where $(a)$ comes from the chain rule, $(b)$ from the Markov chain 
$L_{i-1} - \bar C' - L_{i}$, $(c)$ from the fact that the bits in $A'_1$ are uniform. 
To prove $(d)$ observe that
{\allowdisplaybreaks
\begin{align*}
 H (\widetilde R_i [A'_1] | L_{i}) - H (R [A'_1] | L)&= H (\widetilde R_{(i)}[A'_1] L_{i}) - H (R [A'_1] L) - H (L_{i}) + H (L)\\
&{\overset{{(g)}}{\leq}}  \delta_n^{(1)} \log{\frac{\lvert \mathcal U \times \mathcal X \times  \mathcal W \times  \mathcal Y \times \mathcal V \rvert}{\delta_n^{(1)}}} \! + \!
\delta_n^{(1)} \log{\frac{\lvert \mathcal U \times  \mathcal X \times  \mathcal Y \times \mathcal V \rvert}{\delta_n^{(1)}}}\\
&\leq 2 \delta_n^{(1)} (\log{\lvert \mathcal U \times \mathcal X \times  \mathcal W \times  \mathcal Y \times \mathcal V \rvert}-\log{\delta_n^{(1)}}):= \delta_n^{(4)}
\end{align*}}where $(g)$ comes from Lemma \ref{csi2.7} since
\begin{equation*}
 \tv \left( \widetilde P_{L_i }, \bar P_{UXYV}^{\otimes n}\right) \leq \tv \left( \widetilde P_{L_i \widetilde W_{(i)}^n},\bar P_{UWXYV}^{\otimes n}\right) \leq \delta_n^{(1)}
\end{equation*}that  vanishes as $n$ goes to infinity.
Finally $(e)$ is true because conditioning does not increase entropy and 
$(f)$ comes by definition of the set $A'_1$. Then from Pinsker's inequality
\begin{equation}\label{deltan2}
\tv \left( \widetilde P_{L_{i-1:i} \bar C'} ,  \widetilde P_{ L_{i-1}\bar C'}  \widetilde P_{L_{i}}\right) \leq \sqrt{2 \log 2} \sqrt{n \delta_n + \delta_n^{(4)}}=  \delta_n^{(3)} \to 0.  
\end{equation}
\end{IEEEproof} 

\vspace{0.3cm}

\begin{IEEEproof}[Proof of Lemma \ref{step2b'}]
We have
{\allowdisplaybreaks
\begin{align*}
 \D ( \widetilde P_{ L_{1:k}} \Big\Arrowvert \prod_{i=1}^{k}  \widetilde P_{ L_{i}} )  & {\overset{{(a)}}{=}}   \sum_{i=2}^{k} I (L_{i} ;L_{1:i-1})  \leq \sum_{i=2}^{k} I (L_{i} ;L_{1:i-1} \bar C') \\
& = \sum_{i=2}^{k} \left( I (L_{i} ;L_{i-1} \bar C') +  \sum_{j=1}^{i-2} I (L_{i} ;L_{i-j-1} | L_{i-j:i-1} \bar C') \right)\\
& \leq \sum_{i=2}^{k} \left( I (L_{i} ;L_{i-1} \bar C') +  \sum_{j=1}^{i-2} I (L_{i} ;L_{i-j-1:i-2} |L_{i-1} \bar C') \right)\\
&{\overset{{(b)}}{=}}  \sum_{i=2}^{k} I (L_{i} ;L_{i-1} \bar C')  {\overset{{(c)}}{\leq}} (k-1) (n \delta_n + \delta_n^{(4)})
\end{align*}}where $(a)$ comes from \cite[Lemma 15]{chou2016soft}, $(b)$ is true because the dependence structure of the blocks gives the Markov chain $L_{i-j-1:i-2}-L_{i-1} \bar C' -L_{i} $ and $(c)$ follows from \eqref{eqlem10}.
We conclude with Pinsker's inequality.
\end{IEEEproof}

\vspace{0.3cm}

\begin{IEEEproof}[Proof of Lemma \ref{step2c'}]
  By the triangle inequality
 {\allowdisplaybreaks
\begin{align}\label{tri}
\begin{split}
& \tv \left( \widetilde P_{ L_{1:k}}, \bar P_{UXYV}^{\otimes nk}\right) \leq \tv \left( \widetilde P_{ L_{1:k}}, \prod_{i=1}^{k} \widetilde P_{ L_{i}}\right)+ \tv \left( \prod_{i=1}^{k} \widetilde P_{ L_{i}}, \bar P_{UXYV}^{\otimes nk}\right)
\end{split}
\end{align}}where the first term is smaller than $\sqrt{k-1} \delta_n^{(3)}$ by Lemma \ref{step2b'}.
To bound the second term, observe that
{\allowdisplaybreaks
\begin{align}\label{eqdiv}
\begin{split}
 & \D \left(  \prod_{i=1}^{k} \widetilde P_{ L_{i}} \Big\Arrowvert \bar P_{UXYV}^{\otimes nk}  \right)   = \D \left(\prod_{i=1}^{k} \widetilde P_{L_{i}} \Big\Arrowvert \prod_{i=1}^{k} \bar P_{UXYV}^{\otimes n}  \right) =  \sum_{i=1}^{k} \D \left( \widetilde P_{L_{i}} \Big\Arrowvert \bar P_{UXYV}^{\otimes n} \right).
\end{split}
\end{align}}
By the chain rule we have that 
$\D \left( \widetilde P_{ L_{i}} \Big\Arrowvert\bar P_{UXYV}^{\otimes n} \right) \leq \D \left( \widetilde P_{ L_{i} W_{(i)}^n} \Big\Arrowvert \bar P_{UWXYV}^{\otimes n} \right)$.
Since $X_{(i)}^n$, $Y_{(i)}^n$ and $V_{(i)}^n$ are generated symbol by symbol via the conditional distributions $\bar P_{X|UW}$, $\bar P_{Y|X}$ and $\bar P_{V|WY}$ respectively,
by Lemma \ref{lemkl} we have that
\begin{equation}\label{eq div2}
\D \left( \widetilde P_{ L_{i} W_{(i)}^n} \Big\Arrowvert \bar P_{UWXYV}^{\otimes n} \right) = \D \left( \widetilde P_{ U^{n}_{(i)} W_{(i)}^n } \Big\Arrowvert \bar P_{UW} \right).
\end{equation}
Hence, we have
{\allowdisplaybreaks
\begin{align*}
\D \left(  \prod_{i=1}^{k} \widetilde P_{ L_{i}} \Big\Arrowvert \bar P_{UXYV}^{\otimes kn}  \right) = \sum_{i=1}^{k}  \D \left( \widetilde P_{ L_{i}} \Big\Arrowvert \bar P_{UXYV}^{\otimes n} \right) 
{\overset{{(a)}}{\leq}} \sum_{i=1}^{k}  \D \left( \widetilde P_{ U^{n}_{(i)}  W_{(i)}^n} \Big\Arrowvert \bar P_{UW}^{\otimes n} \right){\overset{{(b)}}{=}} kn \delta_n
\end{align*}}where $(a)$ follows from the chain rule and \eqref{eq div2} and $(b)$ comes from \eqref{emp1}. 

Then, by Pinsker's inequality, \eqref{tri} becomes:
\begin{equation*}
 \tv \left( \widetilde P_{ L_{1:k}}, \bar P_{UXYV}^{\otimes nk}\right) \leq  \sqrt{k-1} \delta_n^{(3)} +\sqrt{k} \delta_n^{(2)} \leq \sqrt{k} (\delta_n^{(3)} + \delta_n^{(2)})= \delta_n^{(5)} \to 0.
\end{equation*}
\end{IEEEproof}

\begin{small}
\bibliographystyle{IEEEtran}
\bibliography{mybib}
\end{small}

\end{document}